\documentclass[structabstract]{aa}
\usepackage{natbib}
\usepackage{graphicx}
\usepackage[varg]{txfonts}
\usepackage{longtable,lscape}

\begin{document}

\title{
  Constraining the structure and formation of the Galactic bulge from a
  field in its outskirts\thanks{Based on observations at the Very Large
    Telescope of the European Southern Observatory, Cerro Paranal/Chile under
    Programme 083.D-0046(A). Table~\ref{taba1} is only available in electronic
    form at the CDS via anonymous ftp to cdsarc.u-strasbg.fr (130.79.128.5) or
    via http://cdsweb.u-strasbg.fr/cgi-bin/qcat?J/A+A/.}}
\subtitle{FLAMES-GIRAFFE spectra of about 400 red giants around
  $(l,b)=(0\degr,-10\degr)$}

   \author{S.\ Uttenthaler\inst{1}
     \and
     M.\ Schultheis\inst{2,3}
     \and
     D.\ M.\ Nataf\inst{4}
     \and
     A.\ C.\ Robin\inst{2}
     \and
     T.\ Lebzelter\inst{1}
    \and
     B.\ Chen\inst{2}
   }

   \institute{
     University of Vienna, Department of Astronomy, T\"urkenschanzstra\ss e 17,
     1180 Vienna, Austria\\
     \email{stefan.uttenthaler@univie.ac.at; thomas.lebzelter@univie.ac.at}
     \and
     Institut Utinam, CNRS UMR6213, OSU THETA, Universit\'{e} de
     Franche-Comt\'{e}, 41bis avenue de l'Observatoire, 25000 Besan\c{c}on,
     France\\
     \email{mathias@obs-besancon.fr; annie.robin@obs-besancon.fr}
     \and
     Institut d'Astrophysique de Paris, UMR 7095 CNRS, Universit\'{e} Pierre et
     Marie Curie, 98bis boulevard Arago, 75014 Paris, France
     \and
     Department of Astronomy, Ohio State University, 140 West 18th Avenue,
     Columbus, OH 43210, USA\\
     \email{nataf@astronomy.ohio-state.edu}
   }

   \date{Received 16 February 2012; accepted 30 July 2012}

 
  \abstract
   {The presence of two stellar populations in the Milky Way bulge have been
     reported recently, based on observations of giant and dwarf stars in the
     inner an intermediate bulge.}
   {We aim at studying the abundances and kinematics of stars in the outer
    Galactic bulge, thereby providing additional constraints on formation models
    of the bulge.}
   {Spectra of 401 red giant stars in a field at $(l,b)=(0\degr,-10\degr)$
    were obtained with the FLAMES-GIRAFFE spectrograph at the VLT. Stars of
    luminosities down to below the two bulge red clumps are included in the data
    set. From these spectra we measure general metallicities, abundances of
    iron and the $\alpha$-elements, and radial velocities of the stars. The
    abundances are derived from an interpolation and fitting procedure within a
    grid of COMARCS model atmospheres and spectra. These measurements as well as
    photometric data are compared to simulations with the Besan\c{c}on and
    TRILEGAL models of the Galaxy.}
   {We confirm the presence of two populations among our sample stars: i) a
    metal-rich one at $[{\rm M}/{\rm H}]\sim+0.3$, comprising about 30\% of the
    sample, with low velocity dispersion and low $\alpha$-abundance, and ii) a
    metal-poor population at $[{\rm M}/{\rm H}]\sim-0.6$ with high velocity
    dispersion and high $\alpha$-abundance. The metallicity difference between
    the two populations, a systematically and statistically robust figure, is
    ${\Delta}[{\rm M}/{\rm H}]=0.87\pm0.03$. The metal-rich population could be
    connected to the Galactic bar. We identify this population as the carrier of
    the double red clump feature. We do not find a significant difference in
    metallicity or radial velocity between the two red clumps, a small
    difference in metallicity being probably due to a selection effect and
    contamination by the metal-poor population. The velocity dispersion agrees
    well with predictions of the Besan\c{c}on Galaxy model, but the metallicity
    of the ``thick bulge'' model component should be shifted to lower
    metallicity by 0.2 to 0.3\,dex to well reproduce the observations. We
    present evidence that the metallicity distribution function depends on the
    evolutionary state of the sample stars, suggesting that enhanced mass loss
    preferentially removes metal-rich stars. We also confirm the decrease of
    $\alpha$-element over-abundance with increasing metallicity.}
   {Our sample is consistent with the existence of two populations, one being a
    metal-rich bar, the second one being more like a metal-poor classical bulge
    with larger velocity dispersion.}
   

   \authorrunning{Uttenthaler et al.}
   \titlerunning{The structure and formation of the Galactic bulge}

   \maketitle
%

\section{Introduction}

The bulge of the Milky Way galaxy is such a complex system that its formation
and evolution is still poorly understood. There exist two main scenarios for the
\object{Galactic bulge} (GB) formation. The first one, called the ``classical''
scenario, describes the bulge formation through initial collapse of gas at
early times \citep{Egg62} or through hierarchical merging of sub-clumps
\citep{Nog99,Agu01}. In the monolithic collapse case the bulge formed
{\em before} the disc and the star-formation time-scale was very short
\citep[$\sim0.5$\,Gyr;][]{Tho05}. The resulting stars are old ($\gtrsim10$\,Gyr)
and have enhancements of $\alpha$-elements relative to iron in a large range of
Fe abundances, which are characteristic of classical bulges. This indicates a
very fast bulge formation, where SNe\,Ia did not have time to pollute the gas
with $\alpha$-element-free ejecta. In the hierarchical merging the bulge also
formed before the disc but on a longer timescale
\citep[of the order of a few Gyr;][]{Nog99} and therefore this approach predicts
lower over-abundances of $\alpha$-elements.

The second scenario, called the ``pseudo-bulge'' scenario, describes the bulge
formation by the buckling of the disc, following a disc instability
\citep[also called a bar;][]{Com90,Rah91,Nor96,KK04,Lia05}. In this case the
bulge forms {\em after} the disc and on a much longer timescale than a classical
bulge. After the bar formation the disc is heated in vertical direction
\citep{CS81}, giving rise to the typical boxy/peanut shape. The bulge formed in
this way will be a mixture of disc stars and stars formed in situ by gas which
is likely to be well mixed by the action of the bar. In this scenario, the
enhancement of the $\alpha$-elements in the bulge stars is predicted to be low,
similar to that of the inner disc stars in the models.

It should be noted that, in addition to the classical bulge and pseudo-bulge
classifications, an alternative formation mechanism has been suggested, namely
the ``clump-origin bulge''. In this scenario, first proposed by \citet{Nog98}
and recently elaborated upon by \citet{IS11}, stellar clumps spontaneously form
in high gas-density discy galaxies, as shown by N-body/SPH simulations, and
dynamical friction drags these clumps to the centre of the galaxy where they
aggregate into a bulge-like structure. The models of \citet{IS11} reproduce the
observed boxy shape, rapid star formation, and vertical metallicity gradients
observed in the bulge of the Milky Way. Clumpy galaxies that could be the
analogues of the primordial Milky Way and that are consistent with these models
have been observed in the high-redshift universe \citep{EE05,Genz11}, but also
locally \citep{Elm12}.

Each of these scenarios is supported by at least some of the observational
constraints. The question of formation history is crucial and necessary to
investigate because our Galaxy is a benchmark for understanding the formation of
disc galaxies. Recently, \citet{Bab10} and \citet{Hill11} analysed large samples
of red clump stars in Baade's window and three fields close to the minor axis,
at $b\sim-4\degr$, $-6\degr$, and $-12\degr$, which revealed the presence of
two distinct populations: a metal-poor component around
$[{\rm Fe}/{\rm H}]\approx-0.3$\,dex with a broad distribution in [Fe/H], and
metal-rich component centered around $[{\rm Fe}/{\rm H}]\approx+0.3$ with a
small spread in metallicity. In addition, these two populations show kinematical
differences: the metal-poor component is compatible with an old spheroid,
whereas the metal-rich component is consistent with a bar population. Therefore,
two different formation scenarios have been proposed: A rapid formation
timescale for the metal-poor component, and a formation over a longer timescale
driven by the evolution of the bar (pseudo-bulge) for the metal-rich component.
Similarly, \citet{Bens11} find two populations in their sample of micro-lensed
bulge dwarf and sub-giant stars, centred at $[{\rm Fe}/{\rm H}]\approx-0.6$ and
$[{\rm Fe}/{\rm H}]\approx+0.3$, with a dearth of stars around
$[{\rm Fe}/{\rm H}]=0.0$.

Other observations favoured one or the other scenario, which prevented a clear
picture of bulge formation to emerge. For example, \citet{Zoc08}, from the
observed sample re-analysed later by \citet{Bab10}, found a clear metallicity
gradient in the bulge; this was interpreted as a challenge to the scenario in
which the bulge would result solely from the vertical heating of the bar. On the
other hand, recent radial velocity studies \citep{How08,How09,Shen10,Kun12} find
that the radial velocity dispersion ($\sigma_{\rm RV}$) of bright bulge giants
is fully consistent with the pseudo-bulge scenario. In particular,
\citet{Shen10} show that any classical bulge contribution cannot be larger than
$\sim8$\% of the disc mass. Furthermore, recent spectroscopic studies of the
abundances of the $\alpha$-elements as a function of [Fe/H] revealed that at
least the metal-poor bulge stars are chemically similar to stars in the local
thick disc \citep{Mel08,Alv10,Ryde10} and the inner disc at Galactocentric
distances of 4 to 7\,kpc \citep{Bens10b}.

A comprehensive summary of the current observational picture of the bulge
structure, formation, and evolution cannot be given in this paper. Instead, we
refer to the recent review by \citet{Rich11}. In any case, it is clear that the
current observational evidence is not sufficient to conclusively constrain the
structure and formation history of the Galactic bulge, and that more
observations are required.

In the past, most of the attention has been paid to the intermediate bulge,
e.g.\ Baade's window toward $(l,b)=(+1\degr,-3.9\degr)$, while the outer bulge
has been somewhat neglected. This may explain why an important feature such as
the double red clump (RC), which becomes apparent only at $|b|\gtrsim5\degr$,
was detected only recently \citep{Nat10,MZ10}. This feature is interpreted as
two over-densities of bulge stars at different distances from the sun
\citep{MZ10,Sai11}. It is suggested that the over-densities look like a
three-dimensional X-structure \citep{MZ10}. A first spectroscopic study of stars
located in the two RCs was performed by \citet{DePro10}. These authors could not
discern any difference in kinematics and metal abundance between the two RCs,
which is in agreement with the interpretation that the two RCs represent the
same parent population at different distances from the sun. However, these
studies were based on optical spectra with relatively low resolution and
signal-to-noise ratio (S/N), from which only an abundance index could be
measured. An investigation of the double RC at higher spectral resolution and
S/N is desirable to foster these conclusions.

In this Paper we present medium-resolution optical spectra of $\sim400$ bulge
stars towards a field at $(l,b)=(0\degr,-10\degr)$, i.e.\ $\sim1.4$\,kpc south
of the Galactic plane (assuming a bulge distance of 8.0\,kpc) in the outer part
of the bulge. The spectra were obtained with the FLAMES spectrograph at the VLT
in GIRAFFE multi-object mode.
Our sample includes stars from the tip of the RGB to stars less luminous than
the expected bump in the red giant branch (RGB) luminosity function (RGB bump),
hence it also includes the two RCs. This is the first study of the two RCs with
medium-resolution, high-S/N spectra. We measure from these spectra radial
velocities, general metallicities, and abundances of iron and the
$\alpha$-elements, and combine this information to obtain a clearer picture
of the structure of the outer bulge.

Furthermore, we extensively compare our results with predictions by models of
the Galaxy and use these models to infer selection biases and the contamination
by non-bulge stars. Recently, \citet{Rob12} presented a model where two
populations co-exist in the bulge region: a bar or pseudo-bulge of high
metallicity and small scale height, and a ``thick bulge'' or classical bulge
with a higher scale height, lower metallicity, and higher velocity dispersion.
This ``Besan\c{c}on model of the Galaxy'' (BGM) explains well the apparent
gradient in metallicity that is observed along the minor axis, by a variable
proportion of the two populations of different scale height. Here, we use our
observed sample as a test case for the new version of the BGM.

The paper is structured in the following way: The sample selection and the
observations are presented in Sect.~\ref{SampleObs}; the analysis of the data
with the help of COMARCS atmospheric models and spectral synthesis techniques,
as well as with the Besan\c{c}on and TRILEGAL models of the Galaxy, is
introduced in Sect.~\ref{analysis}; Sect.~\ref{results} presents and discusses
our results on the radial velocities, metallicities, and $\alpha$-element
abundances; in Sect.~\ref{compCMD}, a comparison of the Galaxy models with
2MASS photometry is done; finally, conclusions are drawn in
Sect.~\ref{conclusio}.

\section{Sample and observations}\label{SampleObs}

\subsection{Target selection, observations, and data reduction}\label{targsel}

Initially, the sample of stars was selected for the study of the evolution of
lithium along the bulge RGB \citep{LU12}. For a detailed description of the
sample selection and the observations we refer to that paper, here we
reproduce only the most important points.

\begin{figure}
  \centering
  \includegraphics[width=\linewidth,bb=88 364 536 699,clip]{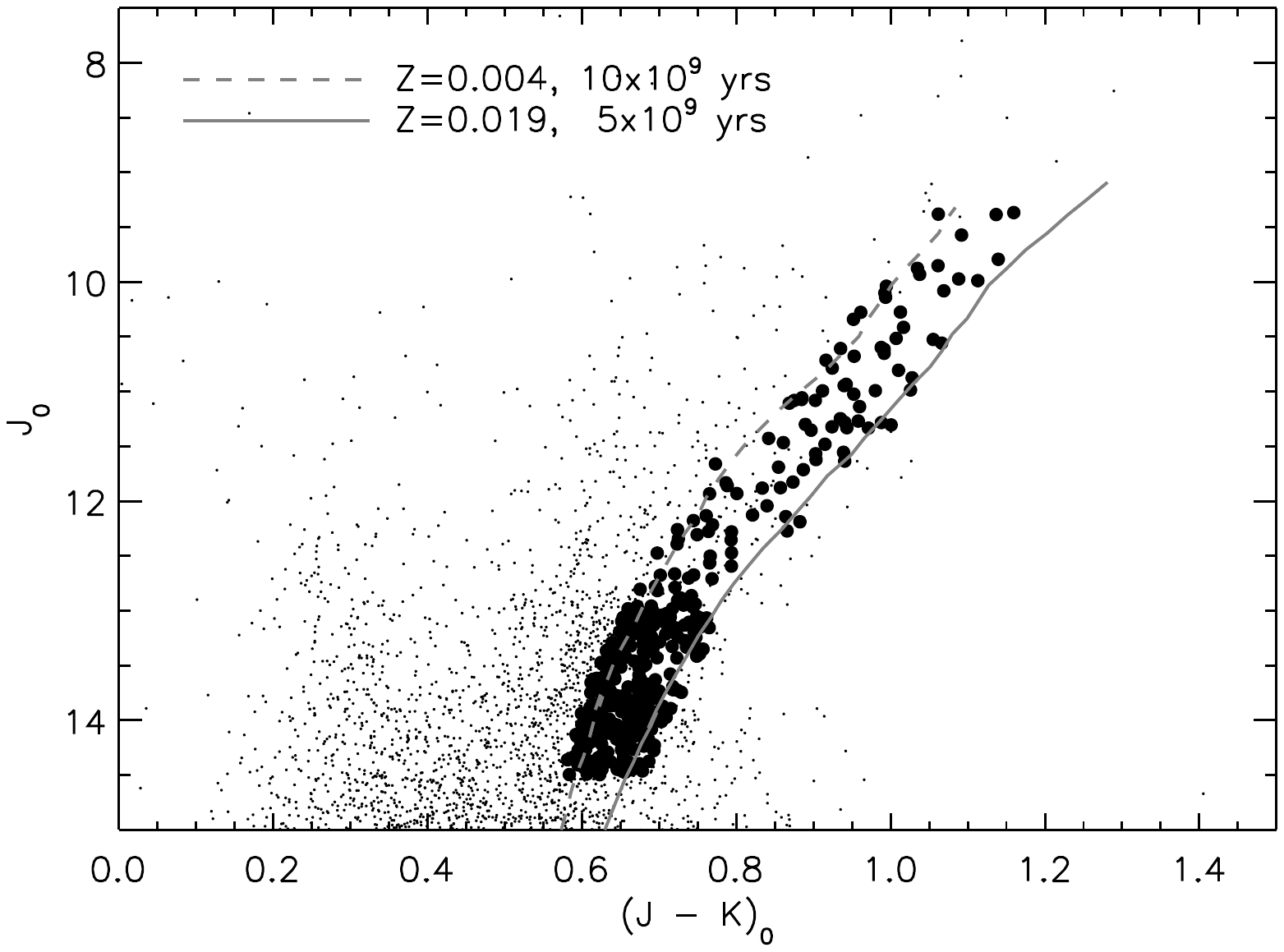}
  \caption{De-reddened 2MASS colour-magnitude diagram of the 25\arcmin\
    dia\-me\-ter field around $(l,b)=(0\degr,-10\degr)$. Targets observed in
    the present programme are plotted as black circles, those that have not
    been observed as black dots. The two RGB isochrones from \citet{Gir00} used
    for the target selection, with ages and metallicities as indicated in the
    legend, are shown as dashed and solid lines. The isochrones are truncated
    at the tip of the RGB. The two RCs are the over-densities at
    $J_0\sim13\fm2$ and $J_0\sim13\fm9$.}
  \label{selcmd}
\end{figure}

The selection of targets was based on data from the 2MASS catalogue
\citep{2MASS} in a 25\arcmin\ diameter circle towards the direction
$(l,b)=(0\degr,-10\degr)$, which is the centre of the Palomar-Groningen field
\#3 (PG3). The sample selection is illustrated in Fig.~\ref{selcmd}, which
shows a colour-magnitude diagram of that circular field. The observed targets
(black circles) were chosen to fall close to two isochrones from \citet{Gir00}:
$Z=0.004$ and age $10\times10^9$ years, and $Z=0.019$
\citep[which is $Z_{\sun}$ on the scale used by][]{Gir00} and age $5\times10^9$
years. These isochrones were used because they cover well the RGB of the bulge.
They are not chosen as to reproduce any of the GB's properties, in particular
not its age spread. A distance modulus of $14\fm5$ to the GB was adopted. The
chosen targets were allowed to have a $(J - K_S)_0$ colour either bracketed by
the isochrones, or $0\fm02$ redder or bluer than them. Dereddening of the
photometry was done as described in \citet{LU12}. We find a mean reddening of
$0\fm129$ in the $J$-band and $0\fm049$ in the $K_S$-band. Besides the colour
criterion, only stars were included that are fainter than $J_0=9\fm0$ to exclude
AGB stars above the tip of the RGB, and stars brighter than $J_0=14\fm5$ to
include the RGB bump, which is expected from isochrones at
$13\fm8 \leqslant J_0 \leqslant 14\fm1$.
This region in the CMD will henceforth be referred to as ``selection region''.
Furthermore, all targets were excluded that had fewer than two quality flags
'A' in the 2MASS $JHK_S$ photometry, and targets that had another source within
3\arcsec\ that was not fainter by at least $2\fm0$ in the $J$-band than the
target itself. Applying all those criteria yielded 514 targets for the
observations.

\begin{figure}
\centering
\includegraphics[width=\linewidth,bb=85 370 538 700,clip]{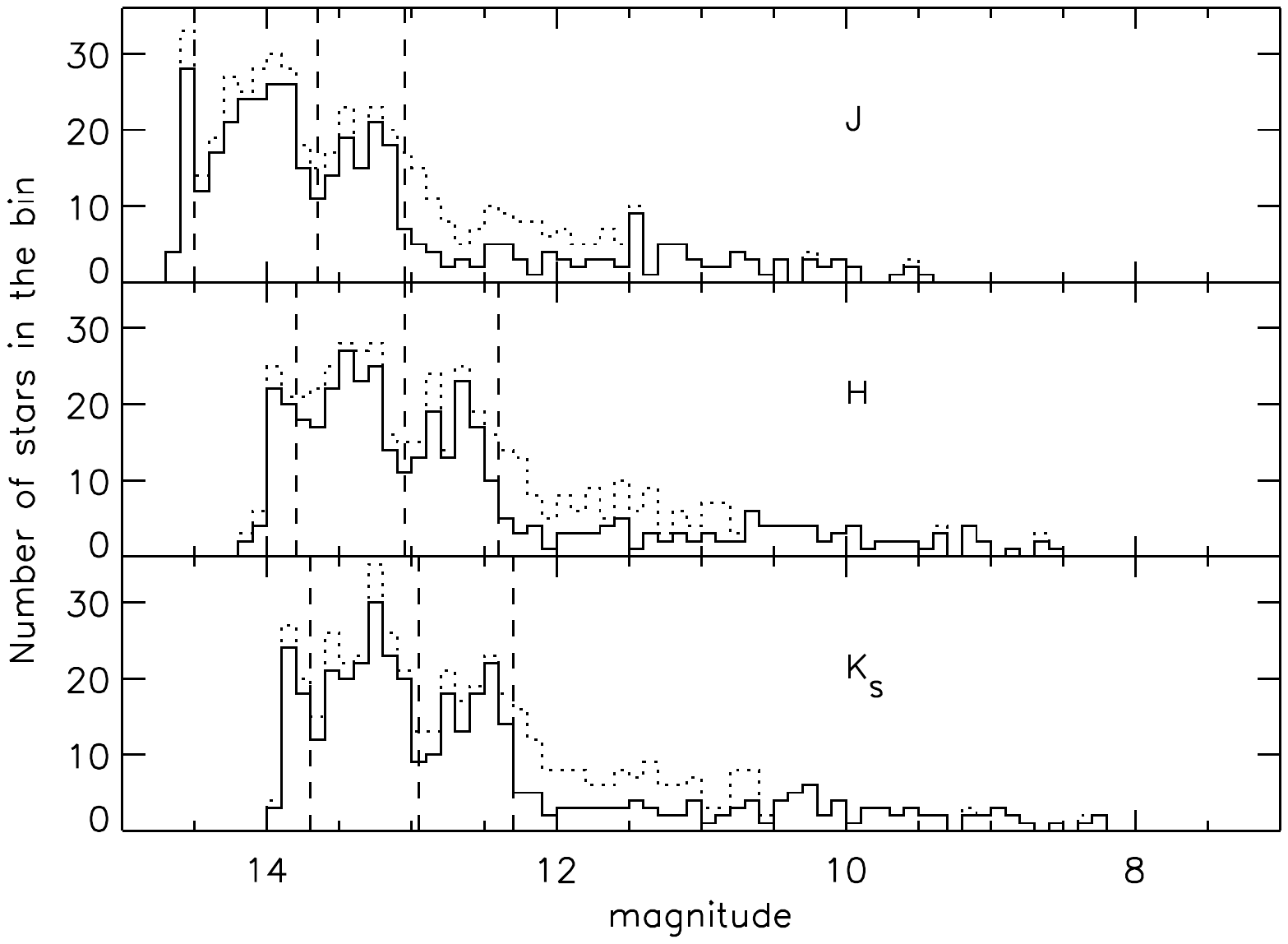}
\caption{Histograms of the observed (not de-reddened) 2MASS magnitudes of our
  sample stars. The dotted line shows the histograms of all stars in the
  selection region (527 stars, including those that did not fulfil the quality
  flag criteria to be selected as target), the solid line shows the histogram
  of the 401 stars for which the radial velocity could be determined. The
  vertical dashed lines indicate the magnitude limits that were adopted to
  define the memberships in the two RCs (Sect.~\ref{maghist2rcs}). The double RC
  is clearly visible in all three bands.}
\label{mag_histo}
\end{figure}

The targets were observed with the FLAMES spectrograph, mounted to the VLT UT2
at Paranal observatory, Chile, in the GIRAFFE configuration. The grating HR15
centred at 665\,nm was used, which gave a spectral resolution of 17\,000 for a
wavelength coverage from 644 to 682\,nm. The observations were obtained in
service mode in June 2009.
Of the initially proposed sample, 401 spectra had sufficiently high quality to
measure their radial velocity (RV, see below).

\subsection{Magnitude histograms and the two RCs}\label{maghist2rcs}

Fig.~\ref{mag_histo} shows histograms of the 2MASS magnitudes of all stars in
the selection region (527 stars, dotted line), and of the stars for which the
RV could be determined (401 stars, solid line). At intermediate magnitudes
($13\fm0 \leq J_0 \leq 11\fm5$), spectra of only about 40\% of the selected
targets are available because one fibre configuration for the stars in this
brightness range was not observed.
The reduced fraction of actually observed stars will be taken into account in
the analysis (Sect.~\ref{MDF_whole}).

Also discernible in Figs.~\ref{selcmd} and \ref{mag_histo} are the two red
clumps (RCs) recently identified by \citet{Nat10} and \citet{MZ10}. These
structures were not known to us in the design phase of the programme. Initially,
the fainter RC at $J_0\sim13\fm9$ was interpreted as to be the RGB bump, and
only the brighter one at $J_0\sim13\fm2$ as the red clump. Inspecting
Fig.~\ref{mag_histo}, we defined membership of stars in the two RCs as follows.
Bright RC: $13\fm65>J>13\fm05$, $13\fm05>H>12\fm40$, and $12\fm95>K>12\fm30$;
faint RC: $14\fm50>J>13\fm65$, $13\fm80>H>13\fm05$, and $13\fm70>K>12\fm95$. We
found 96 and 150 stars, respectively, which fulfill these criteria. Because it
is impossible to ascribe with certainty a given star to either the front or the
back arm of the X-structure, the uncertainty of a bulge star's absolute
magnitude is about 0\fm35.

%

\subsection{Foreground contamination from proper motions}\label{foreground}

\begin{figure}
\centering
\includegraphics[width=\linewidth,bb=86 369 544 700,clip]{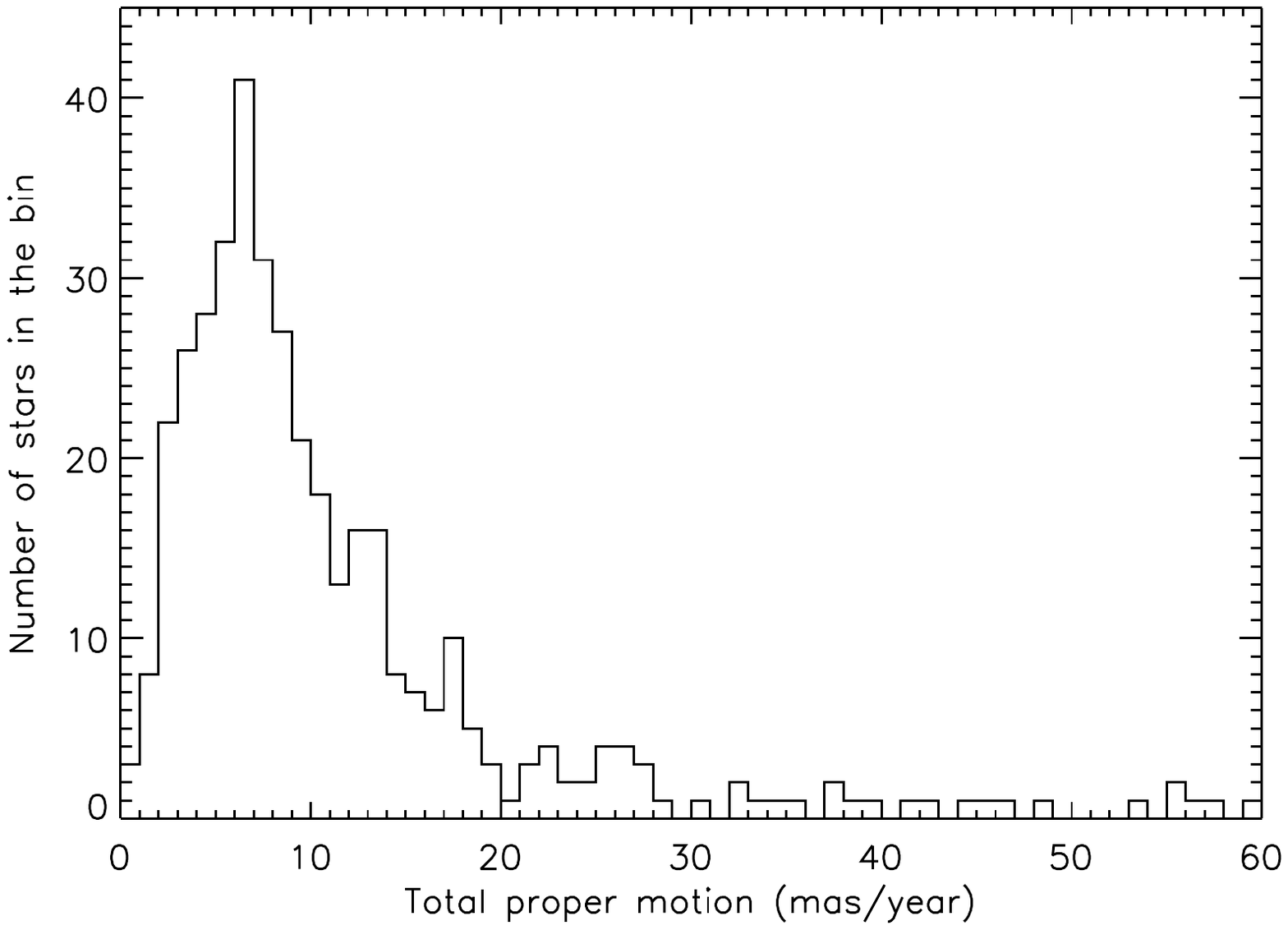}
\caption{Histogram of the total proper motion of our sample stars. Stars
  with a total proper motion larger than 20\,mas/year are defined as foreground
  candidates. Note that four sample stars with a total proper motion larger than
  60\,mas/year fall outside the plot range.}
\label{pm_hist}
\end{figure}

An important issue for this kind of study is the foreground contamination of
the sample. The proper motion can be used to identify possible foreground stars.
We therefore searched the recently published Southern Proper Motion Program
catalogue version 4 \citep[SPM4;][]{Gir11} for sample stars with large proper
motion. All but ten of our sample stars were found in the SPM4 catalogue. In
principle, foreground stars could be identified as those stars whose total
space velocity clearly exceeds the maximum measured RV, assuming that they are
placed at the distance of the GB ($\sim8$\,kpc). The maximum RV found in our
sample is about 300\,km\,s$^{-1}$, which corresponds to a (tangential) proper
motion of 7.91 milli arc-seconds (mas) per year at a distance of 8\,kpc.
Unfortunately, the mean combined error on the proper motion quoted in SPM4
(6.3\,mas/yr) is only barely smaller than this value, so with this catalogue it
is impossible to probe in detail the proper motions out to the GB. Applying the
criterion described above necessarily would return a mixture of bulge and disc
foreground stars. Figure~\ref{pm_hist} shows the distribution of the total
(combined) proper motions of our sample. The distribution essentially consists
of a broad peak centred on $\sim7$\,mas/yr and a long tail of higher proper
motion stars. The tail seems to start at $\sim20$\,mas/yr, which is why we
decided to assign those stars to the foreground whose combined proper motion is
larger than this value. This is also approximately the number that one gets when
two times the mean combined error ($2\sigma$) is added to the most common total
proper motion. Applying this criterion, we found 50 (12.5\%) foreground star
candidates, which roughly agrees with the fraction of non-bulge stars estimated
from the Galaxy models (see Sect.~\ref{selbias}). The foreground star candidates
are marked with an asterisk in Table~\ref{taba1}. One star in our sample (\#300)
has a particularly large proper motion of more than 200\,mas/yr, hence it is
probably a nearby K dwarf. In addition, \#143 is probably a foreground M dwarf.

Unfortunately, there is reason to believe that there are large (systematic)
errors in the SPM4 catalogue because the double RC as well as the two peaks in
the metallicity distribution (Sect.~\ref{MDF_bf}), two features which are
believed to be related to the GB rather than to the disc, are also found among
these 50 foreground candidates. Probably there are still many genuine bulge
stars among these foreground candidates, which is why we refrain from excluding
these stars from the analysis.

\section{Analysis}\label{analysis}

The spectra were analysed with the help of model atmospheres and synthetic
spectra, while the results are interpreted with models of the Milky Way galaxy.
Both these types of models and the associated analysis are described in the
following.

\subsection{COMARCS atmosphere models and COMA synthetic spectra}\label{COMARCS}

For the analysis of the spectra we used model atmospheres calculated with
the COMARCS code \citep{Ari09}, and spectra based on these models synthesised
with COMA \citep[Copenhagen Opacities for Model Atmospheres;][]{Ari09}.


A determination of the stellar temperature and surface gravity with a classical
equivalent width analysis using only the FLAMES spectra was not possible due to
the relatively small spectral range and the dominance of TiO lines in the cooler
targets. Instead, the temperatures of the stars were determined from their 2MASS
$(J-K_S)_0$ colour and the effective temperature calibration based on COMARCS
models presented in \citet{LU12}. Hence, the calculations to derive temperatures
are consistent with the spectral synthesis applied in this work to obtain
abundances in the stars. The surface gravities $(g)$ of the stars were obtained
from the $J_0$ magnitudes of the stars and the isochrones in Fig.~\ref{selcmd},
assuming that the stars belong to the bulge RGB. The uncertainty in the distance
modulus of $\pm0\fm35$ (Sect.~\ref{maghist2rcs}) introduces an uncertainty in
$\log g$ of $\pm0.16$\,dex, which however has only a minor impact on the
uncertainty in the abundances. Furthermore, a generic micro-turbulence of
2.5\,km\,s$^{-1}$, typical of red giant stars, a C/O ratio of 0.3, and a generic
overabundance of the $\alpha$-elements (O, Ne, Mg, Si, S, Ar, Ca, and Ti) of
+0.2\,dex were adopted for the model calculations. This overabundance was
adopted because previous spectral abundance studies of bulge stars showed a
general enhancement of the $\alpha$-elements of this order of magnitude
\citep[e.g.][]{Alv10}.

With the temperature, $\log g$, and micro-turbulence fixed in this way, a grid
of model atmospheres and associated synthetic spectra was calculated. To keep
the grid within reasonable size limits, the RGB was sampled by five different
values of $\log g$. For each of these bins, models at various temperatures, in
a range applicable for this $\log g$ bin, were calculated. For each combination
of $\log g$ and $T_{\rm eff}$, model atmospheres at five different metallicities
$[{\rm M}/{\rm H}] = 1.5$, $-1.0$, $-0.5$, $0.0$, and $+0.5$ were calculated.
The parameters of the model grid are summarised in Table~\ref{COMARCSgrid}.

\begin{table}
\caption{Parameters of the COMARCS model grid.} 
\label{COMARCSgrid}          
\centering                   
\begin{tabular}{ccl}         
\hline\hline                 
$\log g$ & $J_0$ range      & $T_{\rm eff}$ (K) \\ 
\hline                       
0.6      & $ 9\fm0 - 10\fm5$ & 3460, 3625, 3695, 3935 \\
1.1      & $10\fm5 - 11\fm5$ & 3685, 3830, 4000, 4170 \\
1.6      & $11\fm5 - 12\fm5$ & 3950, 4140, 4330, 4520 \\
2.1      & $12\fm5 - 13\fm5$ & 4275, 4450, 4500, 4525,\\
~        &                   &                   4725 \\
2.5      & $13\fm5 - 14\fm5$ & 4425, 4430, 4550, 4645,\\
~        &                   & 4675, 4700, 4750, 4860 \\
\hline                       
\end{tabular}
\end{table}

In the next step, spectra were synthesised for all the models in the grid and
convolved with a Gaussian to the resolving power of the observed spectra
($R = \lambda/\Delta\lambda = 17\,000$). The most important line lists used
are the one for TiO \citep{Schw98}, CN \citep{Jor97}, and the atomic lines
\citep[VALD;][]{Kup99}. This last list was checked and improved by comparing a
model spectrum of \object{Arcturus}
\citep[adopting the stellar parameters and abundances found by][]{Ryde10} with
the observed high-resolution Arcturus spectrum \citep{Hin00}. The solar
abundances listed in \citet{Caf08} were adopted as reference scale. The solar
metallicity on that scale is $Z_{\sun} = 0.0156$.

We emphasise that a particular strength of our method to measure abundances is
the consistency of the temperature determination as well as model atmosphere and
synthetic spectra calculation, for which the same opacity tables and radiative
transfer routines were applied.

\subsubsection{Radial velocity measurement}\label{rvmeasure}

The first quantity that was measured from the spectra was the heliocentric
radial velocity. This was done with the help of a cross-correlation technique
over the whole available wavelength range, using a synthetic spectrum as
template. At this step, a model spectrum with similar temperature as the star
under investigation was used for the cross-correlation. The abundance of the
model spectrum was not adjusted for each star, we simply adopted solar
abundances. The RV was successfully determined for 401 sample stars. We
estimated a typical error of $\sim0.5$\,km\,s$^{-1}$ on the measured RV from a
Gaussian fit to the peak in the cross-correlation function.

\subsubsection{Determination of the metal abundance}\label{mhmeasure}

To determine the general metal abundance from the spectra, we employed a simple
interpolation and fitting procedure. First, the value for $\log g$ was chosen
according to the $J_0$ magnitude of a given star (see Table~\ref{COMARCSgrid}).
Then, all model spectra with that $\log g$ and a given metallicity were
interpolated to the temperature of the star using spline polynomials. This
interpolation was done for every point of the wavelength vector. For the
subsequent fitting procedure, the observed spectrum had to be normalised. For
the hotter stars ($T_{\rm eff} > 3980$\,K, corresponding to $(J-K_S)_0<0.93$) this
was done by adjusting the observed flux to the model spectrum flux around a few
continuum points. In the spectra of the cooler stars, lines of TiO are appearing
or even dominating the spectrum, which prohibit a safe definition of the
continuum; therefore, for these stars the observed spectrum was normalised as to
have the same median flux as the model spectrum over the used wavelength range.
The wavelength range was limited to 649 -- 680\,nm (vacuum wavelength). Finally,
the metal abundance was determined by a $\chi^2$ minimisation method. The figure
of merit to minimise was chosen to be

$$
\chi([{\rm M}/{\rm H}]) = \frac{1}{N} \sum_{i=1}^{N} \frac{\sqrt{\left(f_{{\rm obs},i} - f_{{\rm model},i}([{\rm M}/{\rm H}])\right)^2}}{f_{{\rm obs},i}},
$$

where $N$ is the number of wavelength points, and $f_{{\rm obs},i}$ and
$f_{{\rm model},i}$ are the observed and synthetic model flux at wavelength point
$i$. Hence, $\chi$ is the mean difference between observed and synthetic model
flux, in units of the observed flux. The IDL routine {\tt amoeba.pro} was used
to find the metal abundance [M/H] where $\chi$ reaches a minimum. The general
metal abundance was successfully determined for 383 sample stars.

\begin{figure}
  \centering
  \includegraphics[width=\linewidth,bb=83 369 553 699,clip]{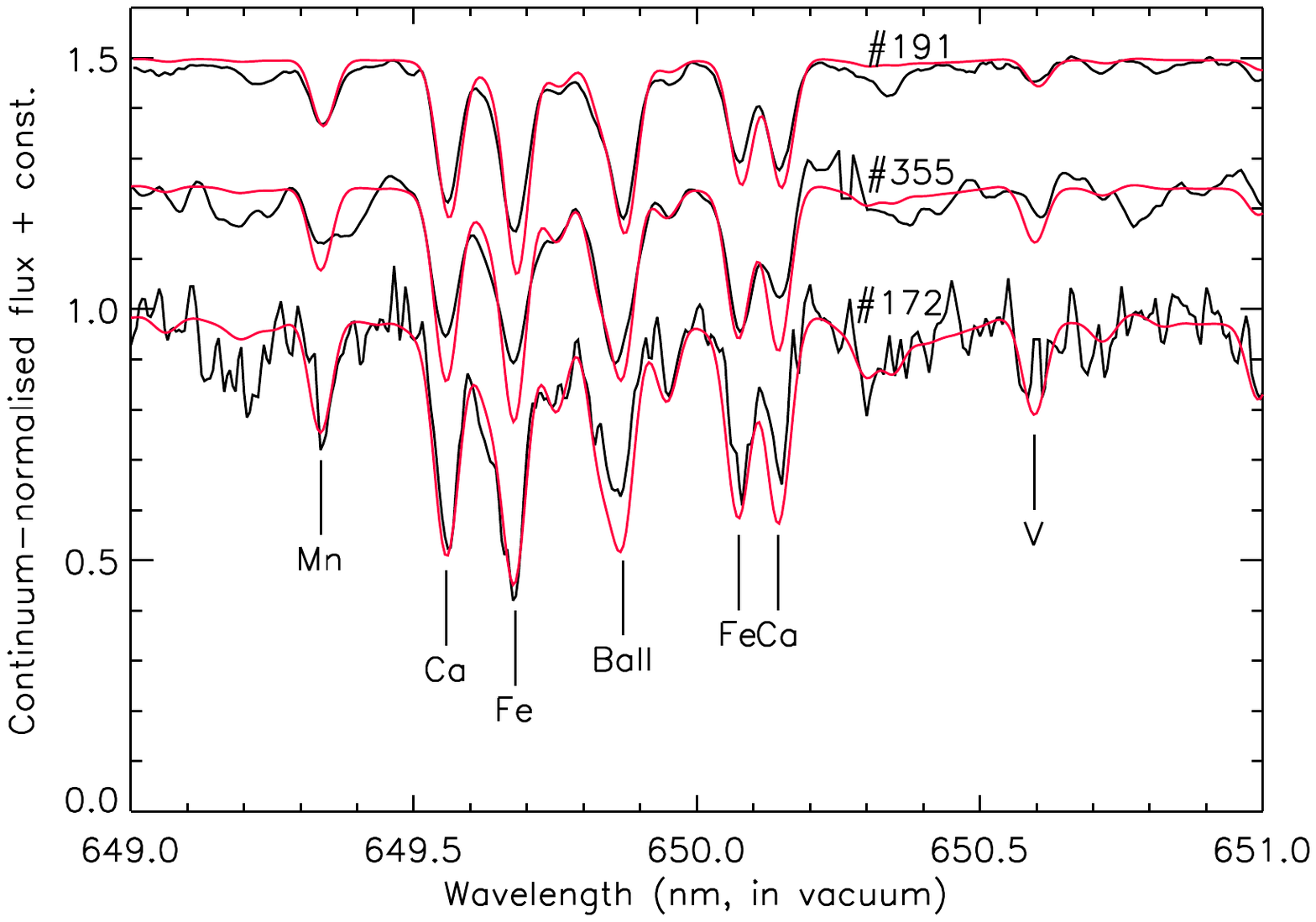}
  \caption{Typical examples of the fit qualities achieved by our grid
    interpolation. Observed spectra are plotted as solid black lines, best-fit
    interpolated model spectra as red lines. From bottom to top: star \#172
    ($T_{\rm eff}=4444$\,K, $\log g = 2.1$, [M/H]=+0.06, $\chi_{\rm min}=0.0446$),
    star \#355 ($T_{\rm eff}=4547$\,K, $\log g = 2.5$, [M/H]=--0.44,
    $\chi_{\rm min}=0.0303$), and star \#191 ($T_{\rm eff}=4569$\,K, $\log g = 2.1$,
    [M/H]=--1.05, $\chi_{\rm min}=0.0166)$. The spectra of \#355 and \#191 are
    shifted vertically by 0.25 and 0.5 for clarity. A few prominent metal lines
    are identified.}
  \label{example_fits}
\end{figure}

A few examples of the typically achieved fit qualities are displayed in
Fig.~\ref{example_fits}. This figure shows a zoom-in on the group of metal lines
located between 649 and 651\,nm. An advantage of our fitting method is that it
also works well at relatively low S/N because all the information in the whole
wavelength range is taken into account. A good example for this is star \#172 in
Fig.~\ref{example_fits}.

\begin{figure}
  \centering
  \includegraphics[width=\linewidth,bb=88 367 409 792,clip]{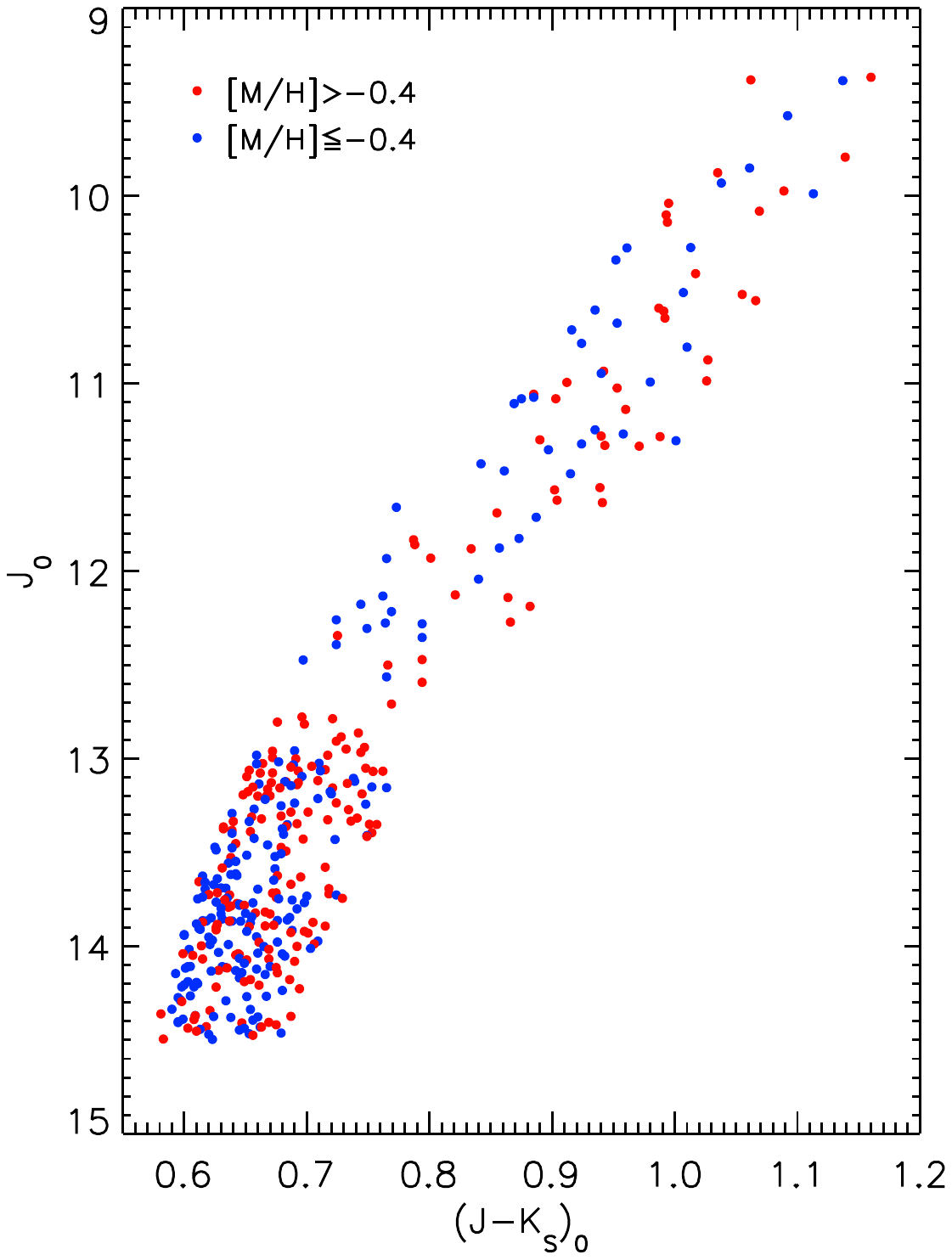}
  \caption{Colour-magnitude diagram of our sample stars, with the metallicity
    colour-coded as indicated in the legend. The sample was divided at
    $[{\rm M}/{\rm H}]=-0.4$ into two approximately equally large parts. In
    particular on the upper RGB, metal-rich stars are preferentially located at
    the red edge of the selection region, whereas metal-poor stars are
    preferentially at the blue edge.}
  \label{CMD_MH}
\end{figure}

An illustrative check of our metallicity determination is displayed in
Fig.~\ref{CMD_MH}, which shows a CMD of our sample stars where the colour of the
plotting symbol codes the metallicity of each star. Metal-rich stars tend to be
more common at the red edge of the selection region, whereas the metal-poor
stars tend to be more abundant on the blue edge. This is what can be expected
from the evolution along the RGB: at a given brightness, a metal-rich star will
be cooler, hence redder in $(J-K_S)_0$, than a metal-poor star. This suggests
that our methods for temperature and metallicity determination work well.
Additional sub-structure around the double RC is hidden by the high density of
stars in that figure; it will be discussed in more detail in Sect.~\ref{MH_RC}.

There are a few sample stars for which the fitting routine extrapolated beyond
the abundance range covered by the COMARCS model grid. The ten metal-poor
outliers clearly have very weak metal lines, indicating a very low metallicity
($[{\rm M}/{\rm H}]\lesssim-1.5$). With such a low metallicity, these stars
possibly belong to the halo population. The two metal-rich outliers are found
at the bright and cool end of the sample.
%
%
%

\subsubsection{Determination of the Fe and $\alpha$-abundance}
\label{alphameasure}

An estimate of the abundances of iron and the $\alpha$-elements can be achieved
with a procedure very similar to the one above. The only difference here is the
non-uniform weighting in wavelength. Weight is put only to those wavelength
points that are influenced by line absorption by an $\alpha$-element or Fe,
respectively. Hence, the $\chi$ value to minimise was altered to

$$
\chi([\alpha,{\rm Fe}/{\rm H}]) = \frac{1}{N} \sum_{i=1}^{N} W_i \frac{\sqrt{\left(f_{{\rm obs},i} - f_{{\rm model},i}([\alpha,{\rm Fe}/{\rm H}])\right)^2}}{f_{{\rm obs},i}},
$$

where $W_i$ is the weight function. It is a binary function with value 1
for wavelength points with absorption by Fe or an $\alpha$-element of more than
2\% of the continuum flux, and 0 otherwise. The weight function was calculated
from a synthetic spectrum based on the model atmosphere with
$T_{\rm eff}=4450$\,K, $\log g = 2.1$, and $[{\rm M}/{\rm H}] = -0.50$, taking
into account only the $\alpha$-elements (O, Ne, Mg, Si, S, Ar, Ca, and Ti) or
Fe, respectively. The different $\alpha$-elements will not contribute equally to
this mean abundance. In particular, O, Ne, and Ar have no noteworthy lines in
our wavelength range. For the other elements, the relative weight on the average
$\alpha$-abundance is:
${\rm Ca} : {\rm Ti} : {\rm Si} : {\rm Mg} : {\rm S} = 1.000 : 0.751 : 0.400 : 0.027 : 0.009$.
Hence, the $\alpha$-abundance is dominated by the individual abundances of
Ca, Ti, and Si. The interpolation was done between the grid points at
$[\alpha/{\rm H}]=(-1.3,-0.8,-0.3,+0.2,+0.7)$ and
$[{\rm Fe}/{\rm H}]=(-1.5,-1.0,-0.5,0.0,+0.5)$, respectively. The fitting
routine did not converge for all stars. Eventually, $\alpha$-abundances were
determined for 368 stars, and Fe abundances for 363 stars. The determination of
the iron and $\alpha$-abundance failed in particular for many of the cool,
TiO-dominated stars. The Fe abundance is a very good tracer of the general metal
abundance: the mean of [M/H]--[Fe/H] is --0.13\,dex, with a standard deviation
of 0.12\,dex; the standard deviation decreases to only 0.07\,dex among the
sample stars fainter than $J_0=12\fm7$.
\smallskip

All measured abundances and radial velocities can be found in Table~\ref{taba1}
in the on-line Appendix.

\subsubsection{Systematic errors}\label{systematics}

A check of our relatively simple methods for the metallicity and abundance
determination was done by applying it to the spectrum of \object{Arcturus}. For
this star, our method yields a metallicity of $[{\rm M}/{\rm H}]=-0.72$, and
$[{\rm Fe}/{\rm H}]=-0.92$. If dedicated model atmospheres with the parameters
from \citet{Ryde10} are used ($T_{\rm eff}=4280$\,K, $\log g=1.7$), i.e.\ if the
grid interpolation in temperature is avoided, [M/H] increases to $-0.69$. Models
without an $\alpha$-enhancement (i.e.\ scaled solar abundances) yield
$[{\rm M}/{\rm H}]=-0.64$. A search on the Simbad
database\footnote{http://simbad.u-strasbg.fr} shows that most studies find
$[{\rm Fe}/{\rm H}\sim-0.50$ for Arcturus.

To back-up the existence of a systematic abundance offset, we made a blind
test with four stars from the sample of \citet{Gon11a}, which cover a range of
almost 1\,dex in [Fe/H]. The spectra of the stars in the FLAMES HR15 setting
were kindly provided by O.\ Gonzalez (private communication) and their stellar
parameters as determined in \citet{Zoc08} were communicated, but not their
individual abundances. The same fitting routine as for our sample stars were
applied to the spectra. The results were then compared to those of
\citet{Gon11a}. The Fe abundances derived with our method were systematically
lower by 0.19 to 0.29\,dex than those of \citet{Gon11a}.

Systematic abundance offsets of 0.2 to 0.3\,dex between equivalent width
measurements and $\chi^2$ minimisation techniques (in the sense that the
$\chi^2$ minimisation gives lower abundances) have also been reported by the
RAVE survey team \citep{Zwi08,Boe11}, in particular for giant stars. No
explanation as for the possible origin of this offset is given in the
literature, neither did we attempt to track down its cause. One explanation
could be that the classical equivalent width measurement uses only relatively
few lines, selected to be unblended, whereas the $\chi^2$ minimisation uses many
more lines, even blended and partially saturated ones. While the reasons are
only speculative at this point, we decided to add 0.2\,dex to all abundances
determined here. This does not only yield better agreement with the comparison
stars of \citet{Gon11a} and Arcturus, but, as we will see below, gives very
satisfactory agreement with other studies of abundances in GB stars. The thereby
{\it corrected abundances} will be used henceforth in the analysis of our data
set and are also reported in Table~\ref{taba1}. Despite the systematics in the
abundance scale, we are nevertheless confident that our abundance determination
is reliable in the relative sense within the sample. Furthermore, we decided to
base our analysis mainly on the [M/H] measurement because it is available for a
few more stars, in particular more of the bright stars. The results would remain
qualitatively unchanged if [Fe/H] was used instead.


\subsubsection{Uncertainty estimate}\label{errest}

The uncertainty in the determination of the metallicity was estimated for six
representative sample stars covering a range in temperature, metallicity,
brightness, and S/N ratio. This was done mainly by varying the free parameters
within their uncertainties, one at a time while keeping the others constant, and
then repeating the fitting process. Uncertainties arising from the determination
of temperature and $\log g$, as well as in the continuum placement, were taken
into account.

The uncertainty in the effective temperature was determined from the
uncertainty on $(J-K_S)_0$ by adopting the uncertainties in the $J$- and
$K_S$-band magnitudes given in the 2MASS catalogue. For the bright (cool) stars
this uncertainty is between 60 and 100\,K, while for the faint (hot) ones it is
between 130 and 200\,K. However, this does not mean that the temperature has a
lower impact on the [M/H] determination for the bright and cool stars, just on
the contrary: because their spectra are dominated by the very
temperature-sensitive TiO bands, a change in temperature has the largest impact
on their metallicity determination. Uncertainties up to 0.27\,dex from the
temperature uncertainty alone were found, but for most stars it is in the range
0.09 to 0.18\,dex.

The uncertainty coming from the $\log g$ estimate was determined by doing the
interpolation and fitting with a group of model spectra adjacent in $\log g$ to
the one that would be attributed to a given star based on its $J_0$ magnitude.
We found that the $\log g$ estimate contributes 0.06 -- 0.10\,dex of
uncertainty in the [M/H] determination. This should already include the
uncertainty coming from the fact that we interpolate model spectra to the
temperature of the stars, rather than using model spectra calculated for the
exact temperature of every star.

Finally, the uncertainty introduced by the continuum placement was estimated by
varying the normalisation factor within its uncertainty range, derived from the
scatter of the pixels in the continuum points. This uncertainty mainly depends
on the S/N of the spectrum, and contributes another 0.02 -- 0.04\,dex to the
uncertainty in [M/H]. Thus, for most stars the combined uncertainty in [M/H] is
in the range 0.11 -- 0.21\,dex, but it can be as high as 0.30\,dex for the cool
stars dominated by TiO absorption. The uncertainties in the iron and
$\alpha$-element abundances are very similar.

\subsection{The Besan\c{c}on and TRILEGAL models of the Galaxy}\label{GalMods}

\subsubsection{The Besan\c{c}on model}\label{Besancon}

The Besan\c{c}on Galaxy Model (BGM) is based on assumptions on the scenario of
formation and evolution of four main stellar populations of the Milky Way
\citep{Rob03}. It allows to simulate the stellar content in any given line of
sight, and for each simulated star the photometry, kinematics, and metallicity
are computed. For each population, the main assumtions are: a star
formation-rate history, age, and initial mass function, which allow to generate
a distribution function in absolute magnitude, effective temperature, and age of
the stars. These assumptions are confined by observations, such as the stellar
content in the solar neighbourhood, remote star counts, and photometry in the
visible and near-infrared. Density laws are assumed and controlled by dynamical
principles \citep{Bie87} for each population and are tested by means of
photometric star counts. The model has been extensively compared with 2MASS data
which allow to constrain the thin disc parameters quite well, such as the scale
length, scale height, the inner disc hole, and the warp \citep{Rey09}.
However, some model parameters are not well constrained yet, such as for the
thick disc population. When a parameter is not yet well constrained from
available data, such as metallicity gradients for example, a mean value from the
literature is assumed, or a value expected from standard scenario of formation.
For example, thick disc parameters are still controversial: its density
parameters such as the local density and the scale height are degenerated when
obtained by model fitting from high galactic latitude fields \citep{Rob12}.
However, in the direction considered here, the thick disc population remains
marginal. We conservatively use here the thick disc parameters determined from
high latitude counts by \citet{Rey01}.

Once density functions and distribution functions are obtained, the simulated
stars are generated in the whole Galaxy divided in volume elements. A number of
improvements have been made to the model since the one presented in
\citet{Rob03}. The mean metallicity and dispersion about the mean for each age
bin in the thin disc are now taken from \cite{Hay08}, rather than from
\cite{Twa80}. The model also includes a 3D extinction map \citep{Mar06} and
photometric errors are added for comparison with observations.
%

Recently, a new model of the bulge/bar region has been proposed by
\citet{Rob12}, which we will use here to interpret and compare to our observed
data. The new model includes a bar and a "thick bulge". In the following, this
latter population will be referred to as the {\em bulge} when discussing the
BGM. The bar is the most massive component, which dominates the stellar content
at low latitudes. The bulge is longer and thicker and gives a contribution at
intermediate latitudes where the bar starts to be less prominent. The main bar
population has a mean metallicity close to solar, while the bulge population is
found to be more metal-poor, with a metallicity of the order of $-0.4$ to
$-0.5$\,dex. The mean metallicties and dispersions of these two populations have
been roughly estimated from the observed MDF by \citet{Zoc08}. It should be
better constrained in the future using larger samples and more lines of sights
in the inner and outer bulge.

A large hole in the central region causes the thin disc to reach its maximum
density at about 2.5\,kpc from the Galactic centre. The kinematics of the thin
disc follow empirical estimates from Hipparcos \citep{Gomez97}, while for the
thick disc it is taken from proper motion data from \cite{Ojha96}, and for the
halo from \cite{Norris85}, as in \cite{Rob03}. In the version presented here,
the bar kinematics are taken from the dynamical model of \citet{Fux99}. The
bulge is found to have a larger velocity dispersion, close to the one of the
halo (Robin et al., in preparation).

The fact that the bar population has a smaller velocity dispersion, a higher 
metallicity, and a smaller scale height than the bulge allows to explain well
the vertical gradient of metallicity seen in bulge fields along the minor axis
\citep{Zoc08}. \citet{Bab10} already suggested that the origin of the
metallicity gradient could be a varying mix of these two populations. The bar
and bulge populations in the BGM do not have an intrinsic metallicity gradient.
This model also explains well the presence of double clumps at medium latitudes,
as seen in \cite{Nat10}, \cite{MZ10}, and \cite{Sai11}. This derives from the
fact that the bar flares from inside to outside by about 30\%. This flare can be
due to resonances in the bar that trap thin disc stars and scatter them to
higher scale height.

\subsubsection{The TRILEGAL model}

The TRILEGAL model was developed by \citet{Gir05} and is a population synthesis
code for simulating the stellar photometry of any field in the Milky Way galaxy.
The model has proven to well reproduce number counts, amongst others, in all
three pass-bands of the 2MASS catalogue, with errors smaller than $\sim30$ per
cent \citep{Gir05}. The bulge component was introduced in TRILEGAL only later by
\citet{Van09}, who also calibrated its stellar population and metallicity
distribution using photometry from the 2MASS and OGLE-II surveys of red clump
stars. The input MDF for the bulge component in TRILEGAL was obtained by
\citet{Van09} with the help of a fitting procedure to the RC star distribution,
by applying shifts to the photometric MDF of \citet{Zoc03}. The best fit was
obtained with a shift of +0.3\,dex. Note that \citet{Van09} used fields that are
considerably closer to the Galactic plane than our field, at most $5.91\degr$
south of the plane.

Their best-fit bulge model is provided as the default in the interactive web
interface to TRILEGAL\footnote{http://stev.oapd.inaf.it/trilegal}, which we
also used here to simulate the stellar content of our FLAMES field. Other
important ingredients that we adopted included: a binary fraction of 0.3 with
mass ratios between 0.7 and 1.0; an exponential extinction law with an
extinction value at infinity equal to the one used to deredden the observed
photometry ($0\fm526$ in the $V$-band); a thin and a thick disc, both with
squared hyperbolic secant density distributions; and a halo with an oblate
$r^{1/4}$ spheroid density distribution. For all these components, the default
parameters offered on the web interface were adopted.
The TRILEGAL model does not provide for the kinematics of its simulated stars,
hence a comparison with our observations was not possible in this respect.

\smallskip
A circular field of 25\arcmin\ diameter centred on the point
$(l,b)=(0\degr,-10\degr)$, equal to the area from which the observed targets
were selected, was simulated with both models. An exception here is that for the
BGM we also calculated ten simulations with identical parameters, to decrease
the Poisson noise mainly for the comparison with the observed RVs. For all
subsequent comparisons between spectroscopic observations and the Galaxy models,
the same photometric selection criteria were applied to the models as to the
observed sample (Sect.~\ref{targsel}). In Sections~\ref{results} and
\ref{compCMD}, we will confront the model predictions with our observations in
terms of kinematics, abundances, and photometric properties.

\subsubsection{Inferring metallicity biases, foreground contamination, and mass distribution with the Galaxy models}\label{selbias}


The Galaxy models can be very useful to infer a number of important properties
of the observed sample, in particular me\-tal\-li\-ci\-ty biases imposed by our
photometric selection criteria, (foreground) contamination by non-bulge stars,
and the distribution of initial masses.

The location of a bulge star in a CMD, and hence its pro\-ba\-bi\-li\-ty to fall
in our selection region, is determined by a complex interplay of metallicity,
distance, age, reddening, and photometric errors. The ratio of the input MDF for
the bulge (and bar) component to the MDF of the stars selected with our
procedure reveals any such metallicity biases. An inspection of this ratio for
the BGM simulation shows that there exists a small bias against intermediate
metallicities ($[{\rm M}/{\rm H}]\sim-0.4$). This is probably caused by the
different distance distributions of the bulge and the bar component in the BGM:
the bulge component, which contributes most stars in this intermediate to low
metallicity, is spread out between 2 and 14\,kpc distance from the sun, whereas
the bar, which is centred around solar metallicity, is much more concentrated
around 7\,kpc distance. Many of the nearby bulge stars thus do not fall in the
selection region, which causes this small bias against intermediate metallicity
stars. In fact, an inspection of the distance distribution shows that our
selection procedure is relatively efficient in picking up stars at distances
between 5 and 11\,kpc.

The TRILEGAL model includes a single bulge component, with a distance
distribution centred on 7\,kpc. Hence, the metallicity bias inferred from this
model somewhat differs from the one found from the Besan\c{c}on model. The
ratio of input MDF to the MDF in the selection region is relatively flat in a
broad range from $[{\rm M}/{\rm H}]\sim-0.4$ to $+0.5$, but gradually decreases
to lower metallicities. This is also what one would expect from a bulge
population with a relatively small distance spread: because low-metallicity
stars evolve along the RGB at relatively blue $J-K_S$ colours, in particular on
the upper RGB, they would be missed by our selection procedure (along
relatively metal-rich isochrones) with a higher probability. Consistent with
this expectation, we find in both models that the metal-poor stars in the
selection region are preferentially at larger distances than the metal-rich
ones because the blue, metal-poor stars are shifted into the selection region
when they are at large distances.

Both the Besan\c{c}on and TRILEGAL model allow to assign a simulated star to
one of the stellar components of the Milky Way, by which means we estimated the
fraction of genuine bulge stars in our sample. The selection of stars from the
Besan\c{c}on model yielded 688 stars, of which 333 (48.4\%) belong to the
thick bulge, 216 (31.4\%) to the bar (hence 79.8\% of bulge/bar stars in total),
72 (10.5\%) to the thin disc, 64 (9.3\%) to the thick disc, and 3 (0.4\%) to the
halo. On the other hand, in the TRILEGAL model 509 stars fall in the selection
region, of which 479 (94.1\%) belong to the bulge, 21 (4.1\%) to the thin disc,
9 (1.8\%) to the halo, and none to the thick disc. From this estimate we
conclude that a large fraction ($\gtrsim80\%$) of our sample stars are genuine
bulge stars. Most of the foreground contamination comes from the thin disc,
while the halo contamination should be very small. The fraction of non-bulge
stars agrees reasonably well with the fraction of foreground star candidates
selected on basis of their proper motion (12.5\%) in Sect.~\ref{foreground}.
Most notably, applying the same proper motion selection criterion of
$>20$\,mas/yr to the simulated stars in the Besan\c{c}on model retrieves 52
foreground star candidates, in excellent agreement with the number we found from
the SPM4 catalogue. Of these 52 stars, 51 are thin disc stars and one is a thick
disc star.

The distribution of initial masses of the simulated stars turns out to be
relatively sharply peaked in both models. In the Besan\c{c}on model the mean
mass is $1.05 M_{\sun}$ with a standard deviation of $0.11 M_{\sun}$, whereas in
TRILEGAL the mean is at $1.13 M_{\sun}$ with a standard deviation of
$0.14 M_{\sun}$. Genuine bulge stars have an even narrower distribution:
$1.07\pm0.05 M_{\sun}$ in the Besan\c{c}on model and $1.15\pm0.09 M_{\sun}$ in the
TRILEGAL model. The most massive stars selected from the simulated data have
1.39 and $1.59 M_{\sun}$ in the Besan\c{c}on and TRILEGAL model, respectively.
Hence, we are confident that our sample contains mostly low-mass
($M\sim1.1 M_{\sun}$) red giant stars located in the GB.

\section{Results and Discussion}\label{results}

\subsection{The observed metallicity distribution function}\label{MDF}

\subsubsection{Whole sample}\label{MDF_whole}

\begin{figure}
  \centering
  \includegraphics[width=\linewidth,bb=88 370 545 698,clip]{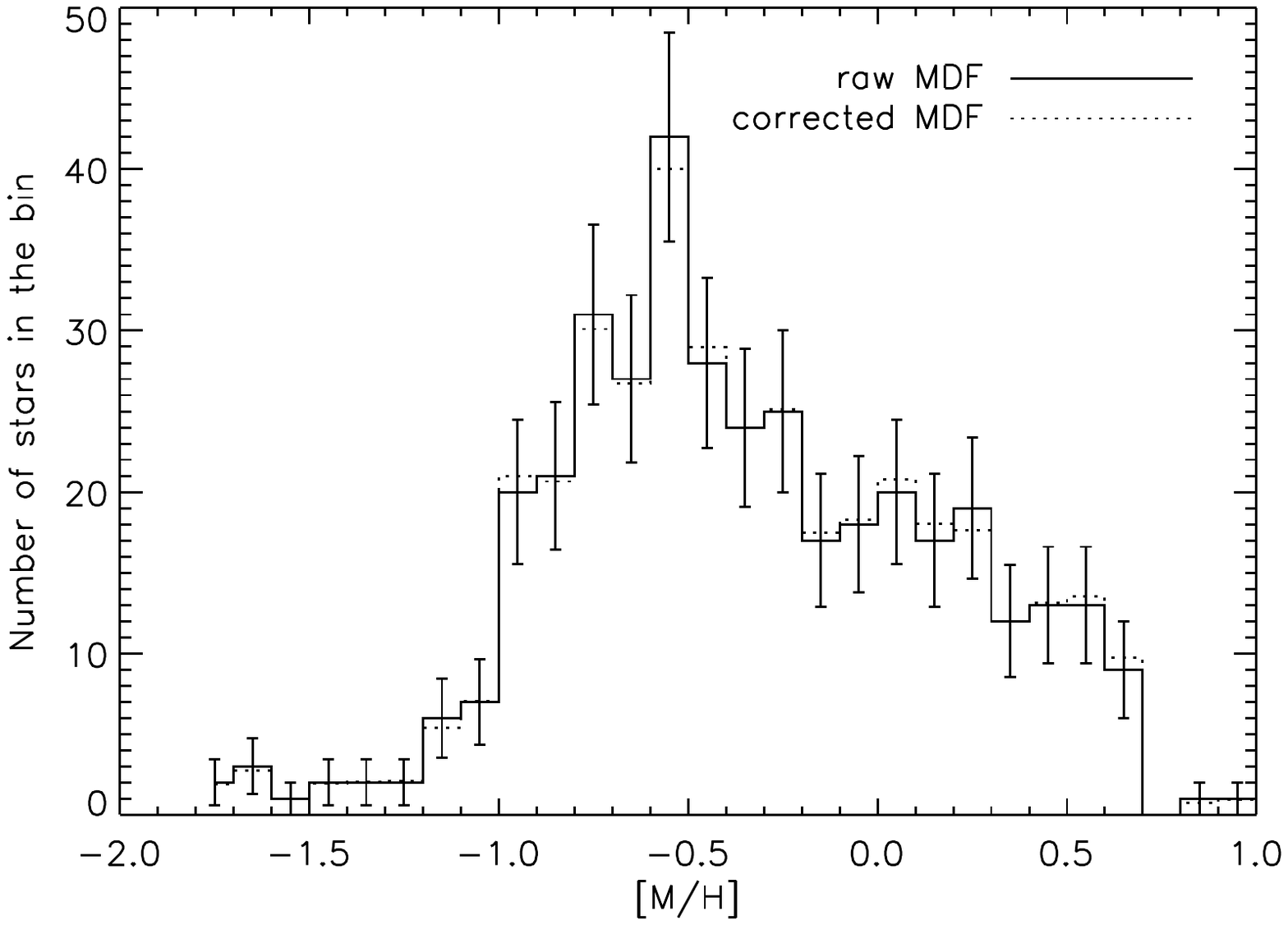}
  \caption{Metallicity distribution function of all 383 sample stars with
    determined metallicities. The solid line is the raw MDF, while the dotted
    line is an MDF corrected for the non-uniform fraction of observed vs.\
    available targets as a function of magnitude, using a bootstrapping method.
    Poissonian noise error bars are also shown for the raw MDF.}
  \label{mh_hist_obs}
\end{figure}

The MDF of our sample stars is displayed in Fig.~\ref{mh_hist_obs}.
Two versions are plotted: The raw MDF (solid line), and
a MDF corrected for the incompleteness in some brightness (and hence colour)
ranges in our sample. To correct for this, we applied a bootstrapping method
similar to method ii) used by \citet{Zoc08}. In this method, stars are drawn
randomly until a flat ratio of observed to available stars is reached.
\citet{Zoc08} considered this ratio as a function of colour. Instead, we chose
to consider it as a function of brightness because our targets were grouped for
observations according to their brightness, where one of the scheduled settings
for the intermediate brightness group was not observed. However, since our
selection region diagonally crosses the CMD, any under-sampling in colour bins
will be effectively corrected for by this method. Two-hundred such randomly
drawn MDFs were created and averaged, the result of this procedure is plotted
as dotted line in Fig.~\ref{mh_hist_obs}. The corrected MDF is only marginally
different from the raw MDF, hence we conclude that our sample well represents
the metallicities of stars in the whole selection region.

The MDF is broad with a clear peak at $[{\rm M}/{\rm H}]\sim-0.55$. The median,
mean\footnote{Doing the averaging on a {\em linear} scale, which is the
mathematically correct way, the mean is $[{\rm M}/{\rm H}]=-0.05$, converted
back to the logarithmic scale.}, and standard deviation of the MDF are $-0.41$,
$-0.34$, and 0.51\,dex. This mean metallicity agrees well with what
\citet{Zoc08} found: for fields at $b=-6\degr$ and $b=-12\degr$ they found
$\left<[{\rm Fe}/{\rm H}]\right>=-0.21$ and $-0.29$, respectively.
Interpolating between these fields to $b=-10\degr$ yields a mean of $-0.26$.
Note, however, that \citet{Zoc08} estimated a large contamination by thin and
thick disc stars in their sample at $b=-12\degr$, which may have a considerable
impact on the mean metallicity of this field. Whereas \citet{Zoc08} used a box
in an $I$ vs.\ $V-I$ colour-magnitude diagram to select their targets, we used a
selection region along isochrones, a strategy that is probably much more
discriminant against thin and thick disc contaminants. \citet{JohnC11} derived
spectroscopic iron abundances of RGB and RC stars in two fields at
$(l,b)=(0\degr,-8\degr)$ and $(l,b)=(-1\degr,-8.5\degr)$. The mean [Fe/H] found
for these fields are $-0.34$ and $-0.27$, respectively. These authors also
derive photometric metallicities for the latter field and find a median of
$[{\rm Fe}/{\rm H}]=-0.34$. Also \citet{Gon11b} derived photometric
metallicities along the GB minor axis and find
$\left<[{\rm Fe}/{\rm H}]\right>=-0.36$ in a $30\arcmin\times30\arcmin$ field
around $b=-8\degr$. These values all agree nicely with one another, and with our
result for the field at $b=-10\degr$.

The MDF in Fig.~\ref{mh_hist_obs} exhibits a relatively sharp cut-off at
$[{\rm M}/{\rm H}]\sim-1.0$. This could be related to the use of an isochrone
with $Z=0.004$ (corresponding to $[{\rm M}/{\rm H}]=-0.68$) for defining the
blue edge of the selection region. As suggested by the TRILEGAL model
(Sect.~\ref{selbias}), there probably is a bias agains metal-poor stars in the
selection, which affects the size and shape of the metal-poor peak. This
metallicity selection bias will be smoothed by the distance spread of the bulge
stars. However, it is hard to quantify if this sharp cut-off is a real feature
or an effect of the selection bias because of the lack of stars with
$[{\rm M}/{\rm H}]\lesssim-1.0$ in the input MDF of the models.

\subsubsection{Bright and faint sample stars}\label{MDF_bf}

\begin{figure}
  \centering
  \includegraphics[width=\linewidth,bb=88 370 545 699,clip]{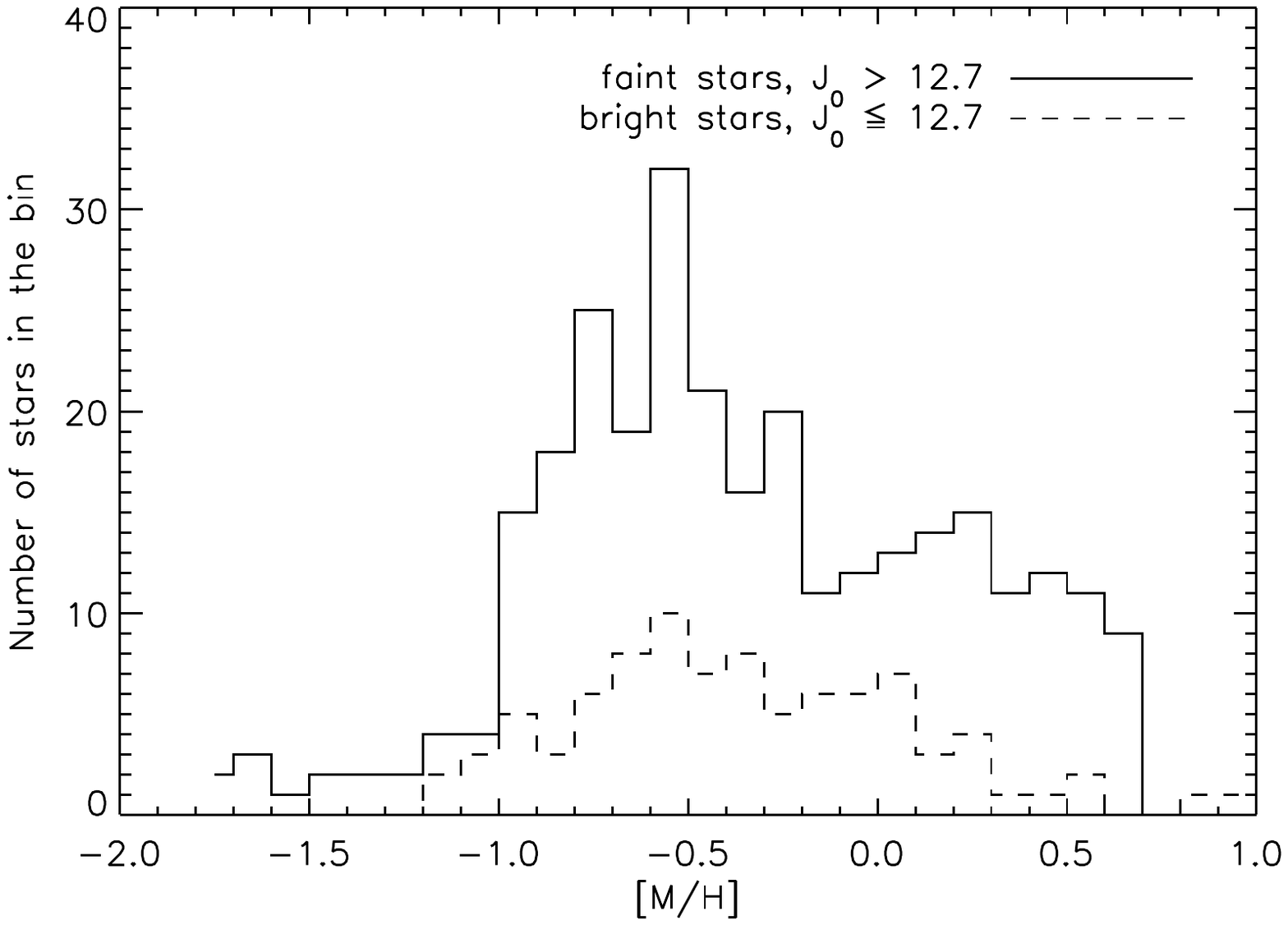}
  \caption{Metallicity distributions of bright ($J_0 \leq 12\fm7$, dashed line)
    and faint ($J_0 > 12\fm7$, solid line) sample stars. The metal-poor and
    metal-rich populations are more clearly separated among the faint sample
    stars.}
  \label{mh_hist_brightfaint}
\end{figure}

It becomes clear already from Fig.~\ref{mh_hist_obs} that there are possibly
two peaks in the MDF: one at $\sim-0.55$ and one at $\sim0.2$. This double-peak
structure becomes even more apparent if bright and faint stars are considered
separately (Fig.~\ref{mh_hist_brightfaint}). The cut between bright and faint
stars was introduced at $J_0=12\fm7$, which should safely exclude RC stars from
the bright sub-sample. The motivation to introduce a brightness cut is two-fold:
first, the uncertainty in the metallicity is larger ($\sim~0.30$\,dex) for the
coolest and brightest sample stars. Thus, with a brightness cut we exclude stars
with the most uncertain metallicity determination.
Second, a brightness cut is equivalent to a cut in evolutionary state, at least
when the distance spread within the bulge of $\sim0\fm7$ is disregarded. Hence,
with our sample we can check whether the MDF varies with evolutionary state or
not.

\begin{figure}
  \centering
  \includegraphics[width=\linewidth,bb=88 369 545 699,clip]{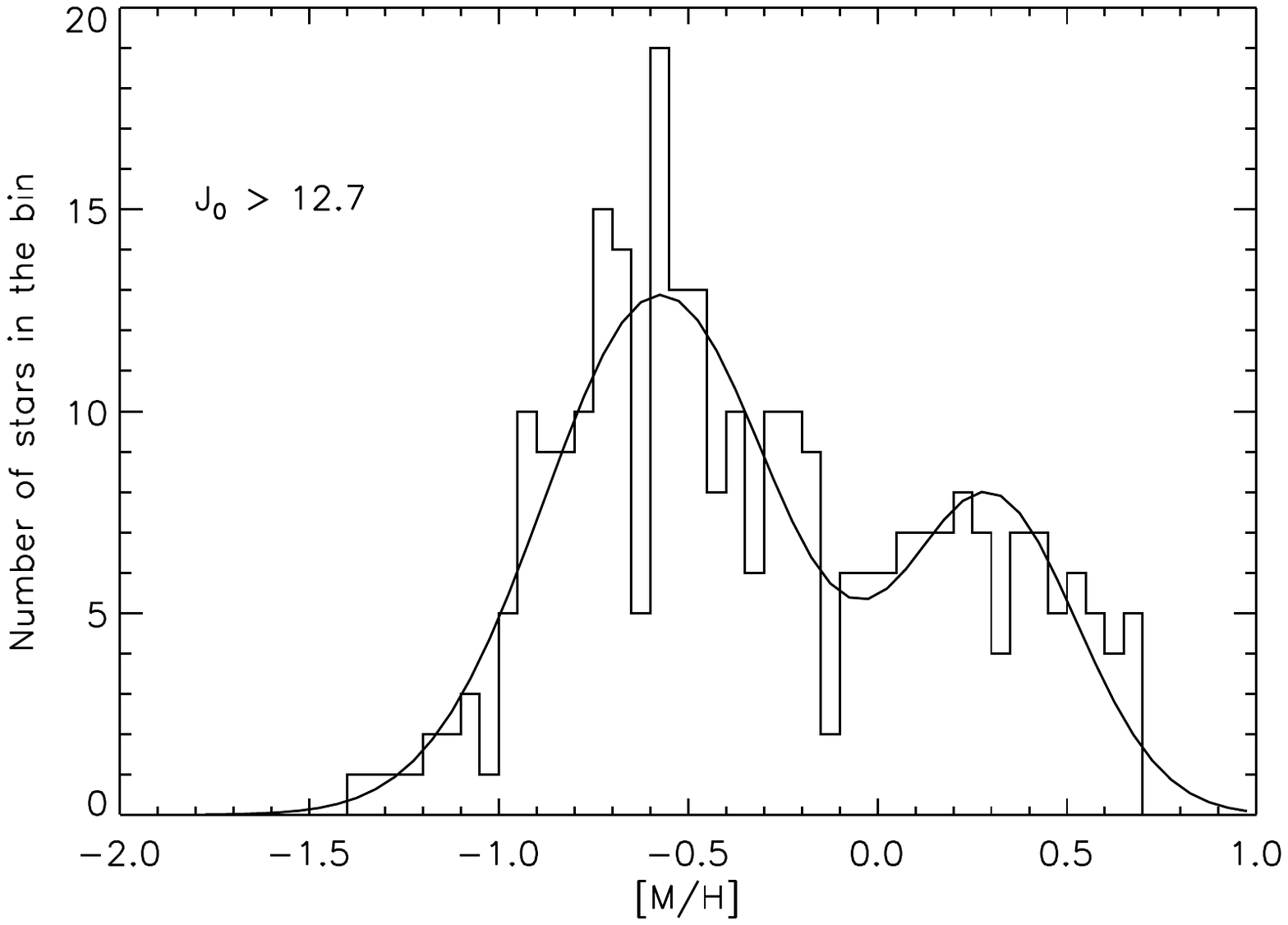}
  \caption{Metallicity distribution of the faint sample stars ($J_0 > 12\fm7$)
    together with the two best-fitting Gaussian mixture models. The metallicity
    distribution in this brightness range is clearly bimodal.}
  \label{mh_hist_faint}
\end{figure}

The most striking feature in Fig.~\ref{mh_hist_brightfaint} is that among the
faint sample stars (those with $J_0>12\fm7$) the dichotomy in metallicities
becomes even clearer than in the complete sample. To make this dichotomy more
quantitative we performed a decomposition by Gaussian mixture models. The
parameter space of two Gaussian mixture models was explored using a Markov
Chain Monte Carlo simulation and a maximum-likelihood approach to define the
best-fit parameters (for details on the code see \citet{Nat11}, for an in-depth
explanation of MCMCs see \citet{JohnJ11}). The chosen confidence level is 68\%.
Stars with $[{\rm M}/{\rm H}]<-1.4$ were omitted from the fit because we think
these stars might be halo contaminants. The results of this Gaussian
decomposition are presented in Fig.~\ref{mh_hist_faint}. We find mean
metallicities of the populations of $[{\rm M}/{\rm H}]=-0.57\pm0.03$ and
$[{\rm M}/{\rm H}]=+0.30\pm0.04$, and dispersions of $0.27\pm0.02$ and
$0.28\pm0.03$\,dex for the metal-poor and metal-rich population, respectively.
The metallicity difference between the two populations, the systematically and
statistically more robust figure, is ${\Delta}[{\rm M}/{\rm H}]=0.87\pm0.03$.
The number of stars belonging to those peaks is $198\pm12$ and $88\pm12$,
respectively, yielding a fraction of the metal-rich population of
$(30.8\pm4.2)$\%. The TRILEGAL model suggests that there could be a selection
bias against metal-poor stars (Sect.~\ref{selbias}); if true, then the fraction
of metal-poor stars at this Galactic latitude has to be regarded as a lower
limit, and that of the metal-rich ones as an upper limit.

\subsubsection{Comparison with the literature}

Recently, MDFs of bulge stars were presented by \citet{Hill11} and
\citet{Bens11}, both of which find a double-peaked MDF. \citet{Hill11} presented
[Fe/H] measurements for 219 bulge red clump stars in Baade's window
($b=-4\degr$), derived also from FLAMES/GIRAFFE spectra. Their Gaussian
decomposition yields a broader peak at $[{\rm Fe}/{\rm H}]\approx-0.30$ and a
relatively narrow peak at $[{\rm Fe}/{\rm H}]\approx+0.32$, respectively. On the
other hand, \citet{Bens11} measured from high-resolution UVES/VLT spectra the
metallicities of micro-lensed dwarf and sub-giant stars scattered throughout the
Galactic bulge at latitudes between $b=-1\degr$ and $-5\degr$, at various
longitudes. The dwarfs show also a clearly bi-modal distribution with peaks at
$[{\rm Fe}/{\rm H}]\approx-0.60$ and $\approx+0.3$. Our result is in excellent
agreement with that of \citet{Bens11}, even though our field is several degrees
(corresponding to $\sim1$\,kpc) away from the dwarfs analysed by these authors.
Only the metal-poor population identified by \citet{Hill11} has a higher
metallicity by $\sim0.3$\,dex. Unfortunately, the number of analysed bulge
dwarfs is still too low to reliably estimate the widths of the two peaks in
their MDF. In contrast to the results of \citet{Hill11}, the peaks in the MDF
found in our sample have very similar widths. In general, there are interesting
analogies between our results and those of \citet{Bens11} and \citet{Hill11},
and we conclude that there are two populations present in the GB, separated by
$0.6 - 0.9$\,dex in metallicity. The small differences might be a result of the
different selection procedures, the difference in evolutionary state of the
samples, and the different analysis methods applied in the studies.

\subsection{Evolutionary effects on the MDF}\label{MDF_evol}

Another interesting feature of Fig.~\ref{mh_hist_brightfaint} is the probable
lack of metal-rich stars among the bright sample stars ($J_0\leq12\fm7$). In the
metallicity range $+0.1 \leq [{\rm M}/{\rm H}] \leq +0.7$, eleven stars are
bright and 72 stars are faint ($13.3\pm4.3$\% bright stars, the uncertainty
being calculated from Poissonian noise), whereas in the range
$-0.5\leq[{\rm M}/{\rm H}]\leq+0.1$ the fraction is $29.5\pm5.4$\% (39 bright
and 93 faint stars). The fraction of bright stars in the whole sample is
$22.9\pm2.7$\% (92 bright and 309 faint stars). This reduced fraction of bright
stars among the metal-rich stars could be caused by a selection effect. We
therefore inspected the fraction of bright stars in the selection region of the
Galaxy models as a function of metallicity to check for differences in the
metallicity bias between faint and bright RGB stars (i.e.\ less and more evolved
stars). However, no such trend is found, which suggests that the metallicity
biases equally apply to faint and bright stars.

Apparently, the shape of the MDF depends on the evolutionary state of the sample
stars one is considering. The simplest explanation for this finding is that the
metal-rich stars do not evolve all the way from the main sequence up to the RGB
tip or the early AGB, but get ``lost'' somewhere on the way
\citep[e.g.\ due to enhanced mass loss;][]{Cas93}. A similar scenario was put
forward by \citet{Chi09} to explain the shift of 0.3\,dex to lower values in the
planetary nebulae oxygen abundance distribution with respect to the distribution
of giant stars. Because of the still relatively small number of stars in the
present sample this conclusion has to be taken with some caution. Also, some
(presumably metal-rich) stars on the red side of the selection region have not
been observed in the present programme (see Fig.~\ref{selcmd}). However, in
Sect.~\ref{sigma_evol} we will present another piece of evidence from the radial
velocity dispersion of evolved stars that strongly supports the scenario of
metallicity-dependent mass loss.

 \subsection{The metallicity of the red clump stars: Is the double RC only connected to the metal-rich population?}\label{MH_RC}

\begin{figure}
  \centering
  \includegraphics[width=\linewidth,bb=88 369 537 699,clip]{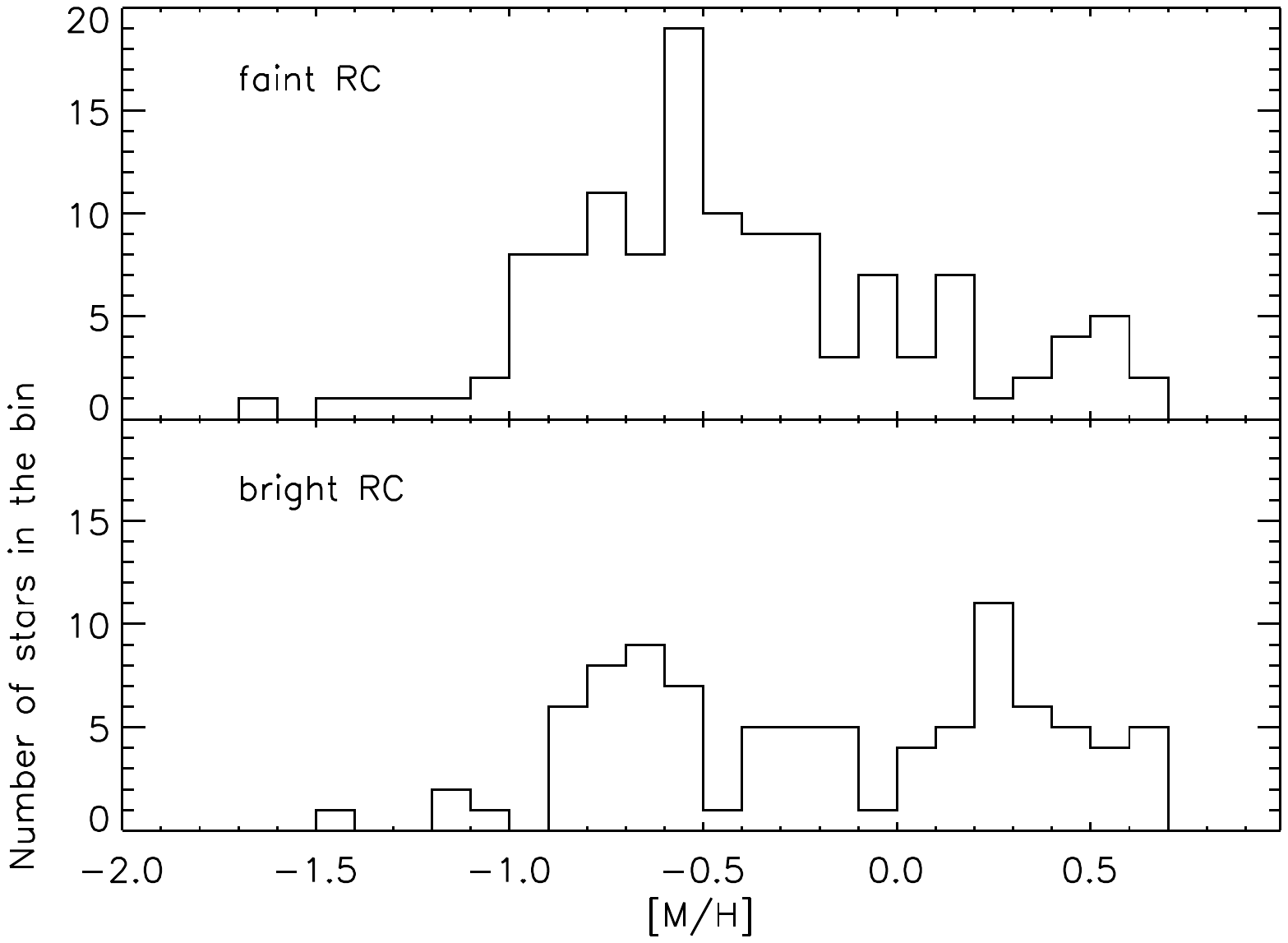}
  \caption{Metallicity distribution of stars belonging to the faint
    ({\em upper panel}) and bright ({\em lower panel}) RCs, respectively.}
  \label{mh_hist_RCs}
\end{figure}

Figure~\ref{mh_hist_RCs} shows the MDF of the bright and faint RC, with the
definition of the two RCs as introduced in Sect.~\ref{maghist2rcs}. It is clear
from that figure that the faint RC has a much more prominent low-metallicity
peak than the bright RC, which in turn shows both peaks. Their mean
metallicities are $[{\rm M}/{\rm H}]=-0.18$ and $-0.41$, respectively. This
would suggest that the two RCs have intrinsically different metallicities, in
contradiction to findings by \citet{DePro10}, and would therefore not represent
the same underlying population of stars. However, we believe that this seemingly
different mean metallicity is actually caused by our selection procedure.
Because of the way how we selected the stars along two isochrones, which become
redder at brighter magnitudes, there are preferentially redder stars selected
from the bright RC, and bluer stars from the faint RC (see also
Fig.~\ref{CMD_MH}). At a given evolutionary state, redder stars are more
metal-rich than bluer ones. Hence, we selected preferentially more metal-rich
stars in the bright RC than in the faint RC. This interpretation is also
favoured by the RV distributions of the two RCs, which are practically
indistinguishable (cf.\ Sect.~\ref{res_rv}). Introducing an additional colour
criterion does not strengthen this conclusion because of the small number of
stars that would be compared then.

\begin{figure}
  \centering
  \includegraphics[width=\linewidth,bb=88 367 537 698,clip]{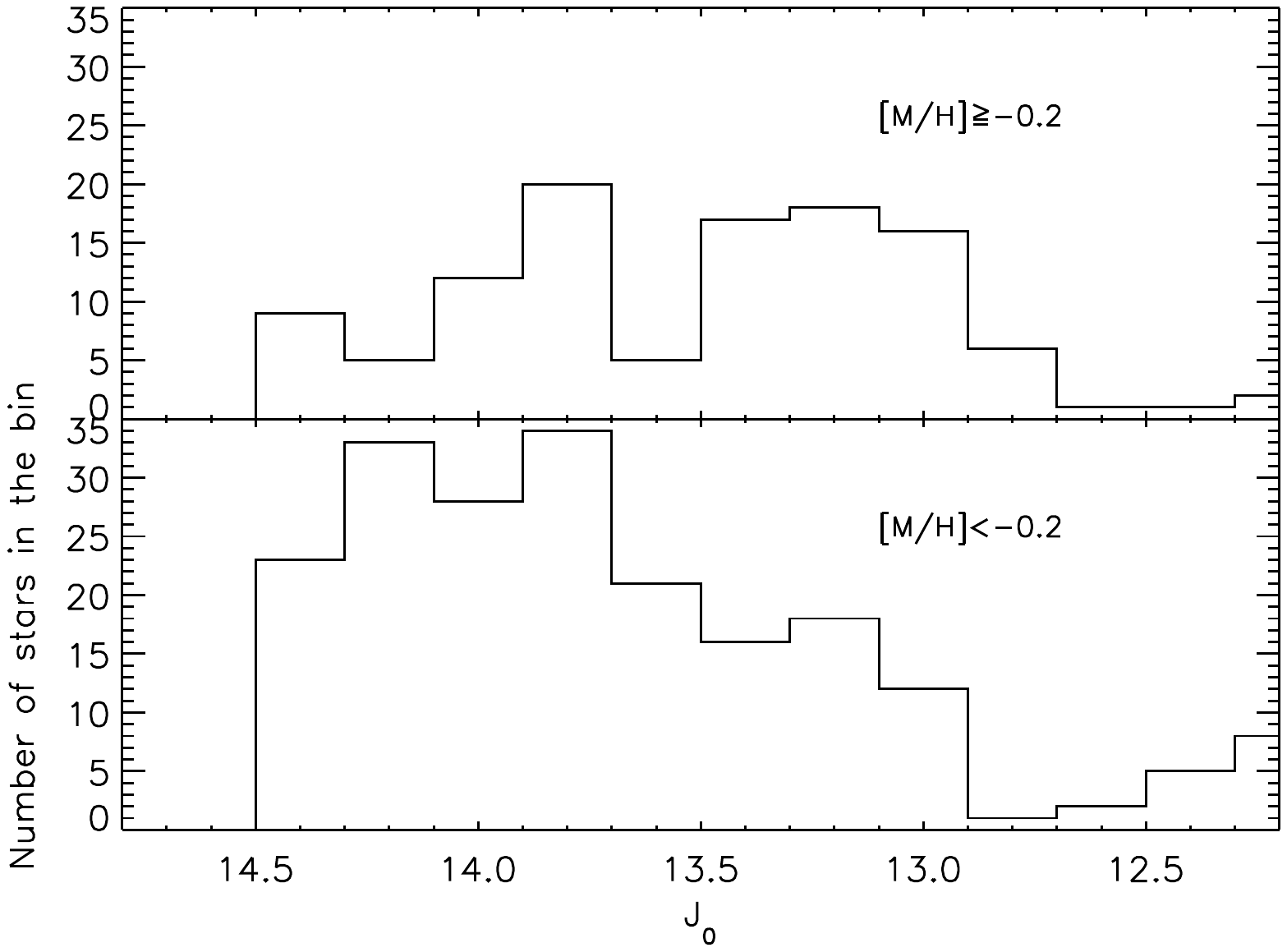}
  \caption{Luminosity functions of the metal-rich ($[{\rm M}/{\rm H}]\geq-0.2$,
    {\em upper panel}) and the metal-poor ($[{\rm M}/{\rm H}]<-0.2$,
    {\em lower panel}) populations around the red clump brightness.}
  \label{mag_hist_RCs}
\end{figure}

In fact, we suggest that the double RC feature is connected only to the
metal-rich population. A direct hint may come from the luminosity functions of
the metal-poor and metal-rich populations around the RC brightness, see
Fig.~\ref{mag_hist_RCs}. The metallicity cut was introduced at
$[{\rm M}/{\rm H}]=-0.2$ for this diagram. The metal-rich population has two
peaks in the $J_0$ histogram, one at $J_0\sim13\fm2$ and one at $J_0\sim13\fm9$.
It stands to reason that these two peaks are the manifestation of the front and
back arms of the X-shaped bulge. The fainter peak contains fewer stars
because its stars are at larger distance from the Galactic plane where the
density of stars is lower \citep{Sai11}. The metal-poor population has one very
broad, symmetric peak at $J_0\sim14\fm0$ with a decline towards brighter
magnitudes. Because the RC and the RGB bump have the same magnitude in $J_0$ at
a metallicity of $[{\rm M}/{\rm H}]\approx-0.7$ \citep{Nat11}, this peak may
also contain some RGB bump stars of the metal-poor population. A statistical
t-test shows that the probability that the two luminosity functions are drawn
from the same parent distribution is only 5.1\%. Also \citet{Ness12}, from an
analysis of spectra of clump stars along the minor bulge axis, find that a split
in the luminosity function appears only for stars with
$[{\rm Fe}/{\rm H}]>-0.5$, in good agreement with our finding.

\begin{figure}
  \centering
  \includegraphics[width=8.0cm,bb=21 202 570 682,clip]{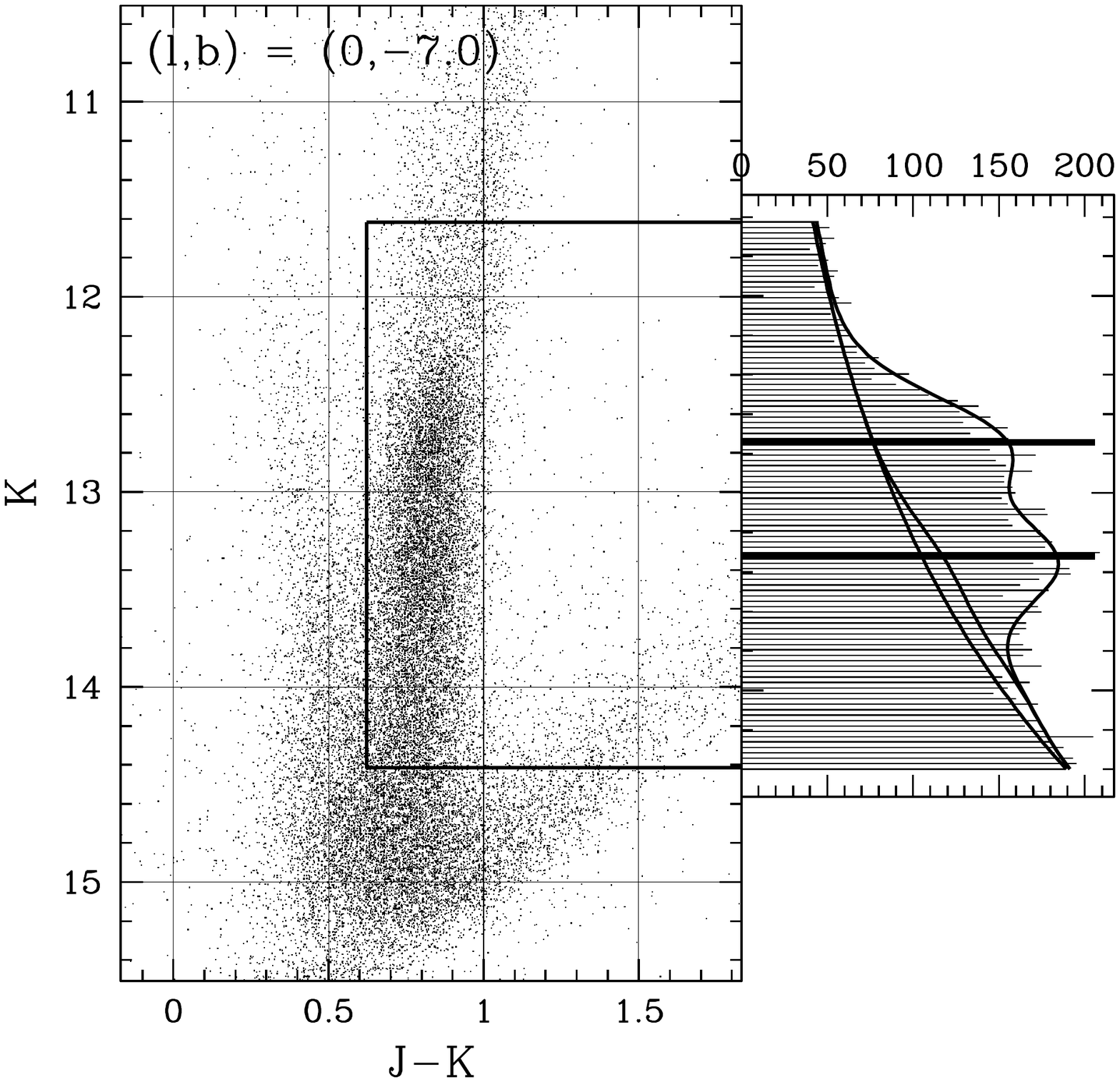}
  \includegraphics[width=8.0cm,bb=21 202 570 682,clip]{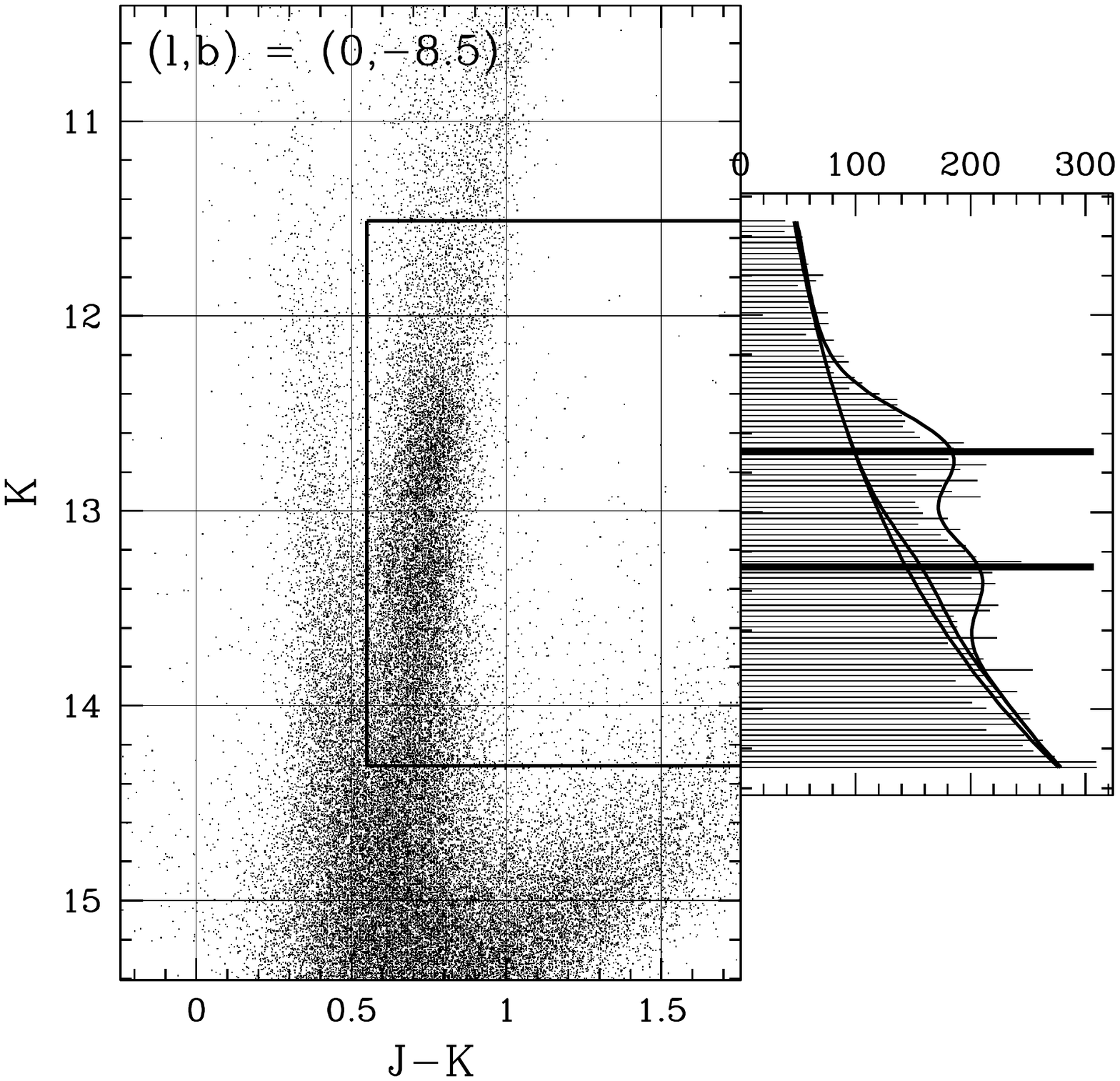}
  \includegraphics[width=8.0cm,bb=21 156 570 682,clip]{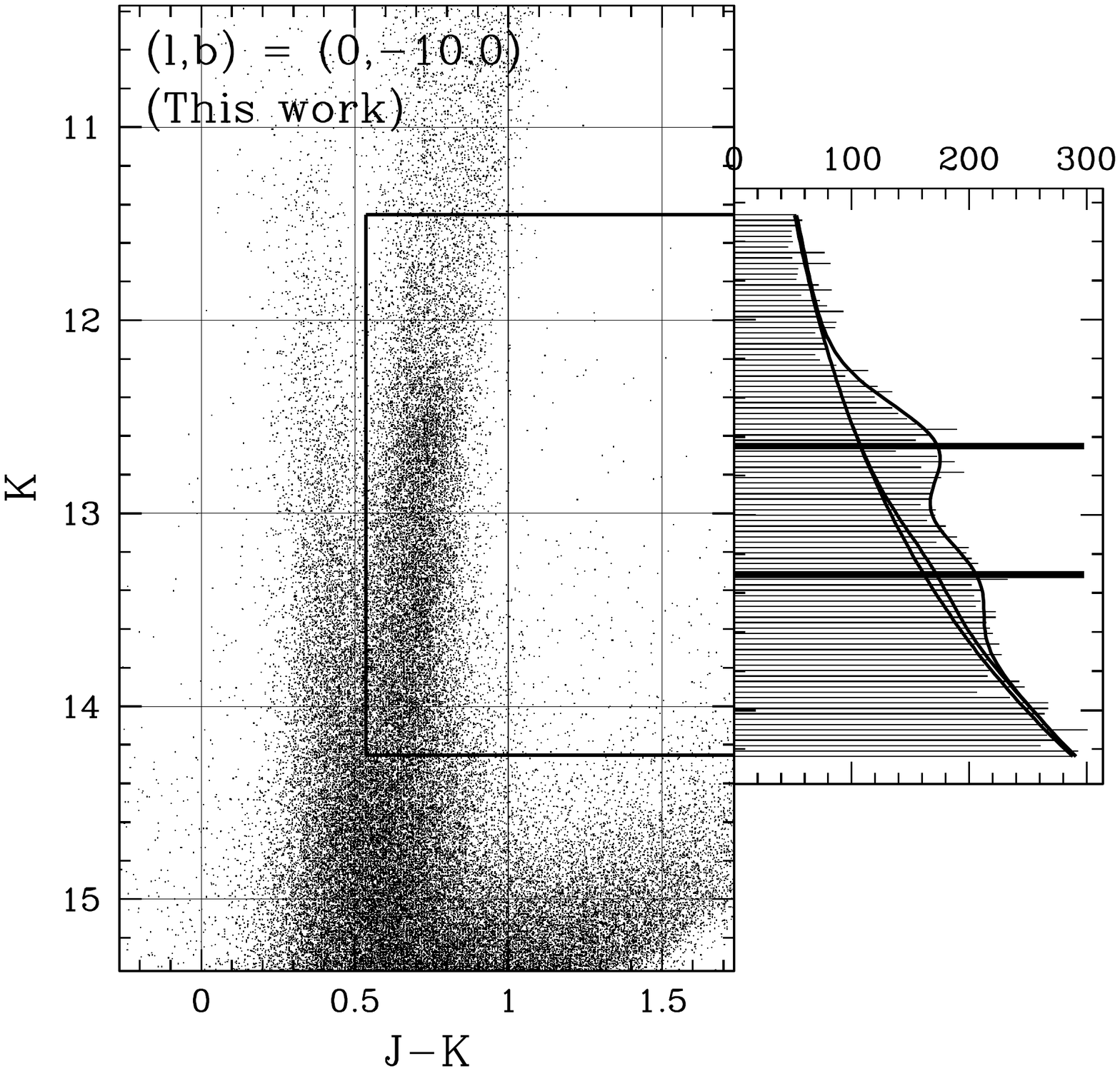}
  \caption{Colour-magnitude diagrams of fields at $(l,b)=(0\degr,-7.0\degr)$
  ({\em top panel}), $(l,b)=(0\degr,-8.5\degr)$ ({\em middle panel}), and
  $(l,b)=(0\degr,-10.0\degr)$ ({\em bottom panel}). For each field, a magnitude
  histogram of the stars in the thick box in the CMD is shown in the right hand
  panel, including a fit to the double RC, the RGB bump, and the underlying
  distribution of RGB stars. The thick horizontal bars in the magnitude
  histograms mark the peak magnitudes of the two RCs.}
  \label{CMDs_David}
\end{figure}

Another argument for the suggestion that the double RC feature is connected
only to the metal-rich population may come from the analogy of how the double
RC feature diminishes with increasing distance from the Galactic plane and the
presence of a metallicity gradient in the bulge. CMDs based on 2MASS photometry
in three different fields along the minor axis of the bulge (at $b=-7.0\degr$,
$-8.5\degr$, and $-10.0\degr$, from top to bottom) are shown in
Fig.~\ref{CMDs_David}. Fields with radii of 20\arcmin, 30\arcmin, and 40\arcmin,
respectively, were used to keep the total number of stars approximately
constant. On the right hand side of each panel, a magnitude histogram of the
stars inside the box marked by thick lines is shown. The luminosity function
was fitted with Gaussian and exponential functions to reproduce the RCs, RGB
bump, and the underlying distribution of RGB stars. It is clear from that
figure that not only the peak brightness positions of the RCs are moving apart
when going away from the plane, but also the relative number of stars making up
the double RC feature is decreasing with distance from the plane.

Together with the presence of a metallicity gradient in the bulge \citep{Zoc08},
this suggests that the double RC feature is connected only to the metal-rich
population, which is much more concentrated towards the Galactic plane. In both
\citet{Hill11} and \citet{Bens11} the stars evenly distribute among the two
sub-populations. In our sample, the metal-rich population makes up only
$\sim30$\% of the stars (upper limit), which would suggest that the fraction of
the populations varies with Galactic latitude. Hence, the metal-rich population
seems to be more concentrated towards the plane than the metal-poor one. This
would also explain the metallicity gradient found by \citet{Zoc08}: If the
fraction of metal-rich stars decreases with increasing distance from the plane,
also the mean metallicity would shift to lower values. Hence, rather than a
decreasing mean metallicity within one monolithic bulge population, probably it
is the varying percentage of two sub-populations that {\em mimics} a metallicity
gradient in the bulge.

\subsection{Comparison with simulated MDFs}\label{compMDF}

\begin{figure}
  \centering
  \includegraphics[width=\linewidth,bb=76 369 536 699,clip]{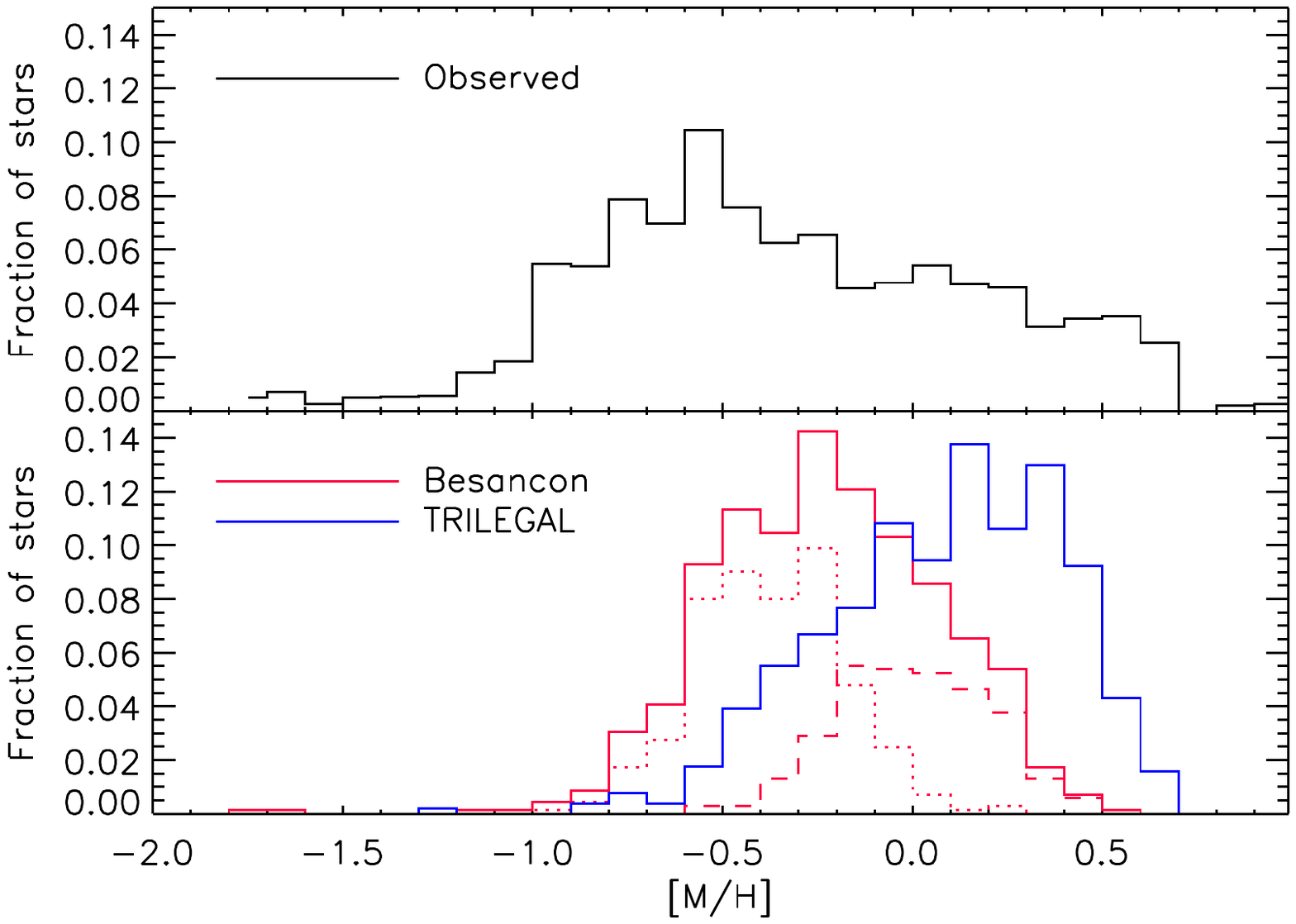}
  \caption{Comparison of the observed MDF corrected for sampling effects
    ({\em upper panel}), with the ones predicted by the Besan\c{c}on and
    TRILEGAL models in the selection region ({\em lower panel}). The
    histograms are normalised to the same area. In the lower panel, the red
    dotted line shows the MDF of ``thick bulge'' stars in the Besan\c{c}on
    model, whereas the dashed red line is the one for the bar component.}
  \label{mh_hist_model}
\end{figure}

Our observed MDF, corrected for sampling effects (Sect.~\ref{MDF_whole}), is
compared to the simulated MDFs from the Besan\c{c}on and TRILEGAL models in
Fig.~\ref{mh_hist_model}. The simulated MDFs do not reproduce the observed MDF
in all aspects. The MDF of the TRILEGAL model has clearly a much higher mean
metallicity than the observed mean ($[{\rm M}/{\rm H}]=+0.09$, compared to
$-0.34$). The number of metal-poor stars is underestimated, whereas the
super-solar metallicity stars are overrepresented in the TRILEGAL model.
Clearly, the shift of $+0.3$\,dex applied by \citet{Van09} to the MDF of
\citet{Zoc03} does not reproduce the MDF of our sample (which is at $-10\degr$
from the Galactic plane). This probably means that an MDF calibrated on fields
at low Galactic latitude cannot be simply applied to higher latitudes because of
the metallicity gradient. The mean metallicity in the Besan\c{c}on model, on
the other hand, is close to the observed one ($[{\rm M}/{\rm H}]=-0.23$). For
this model, the distributions of the bar (red dashed line) and the thick bulge
(red dotted line) are shown separately in the lower panel of
Fig.~\ref{mh_hist_model}. The bar stars reproduce the metal-rich peak of the
observed MDF relatively well. However, the two components are not separated in
metallicity as much as they are in the observations. If the bulge component in
the Besan\c{c}on model was shifted to lower metallicities by 0.2 to 0.3\,dex, it
would very nicely reproduce the metal-poor peak in the observed MDF (cf.\
Sect.~\ref{RV_BGM}). We remind that these stellar population synthesis models
assume certain metallicity distributions for each Galactic component.
Self-consistent chemical evolution models such as \citet{Bal07} are needed to do
a detailed comparison.

\subsection{$\alpha$-element abundances}

\begin{figure}
  \centering
  \includegraphics[width=\linewidth,bb=72 369 536 698,clip]{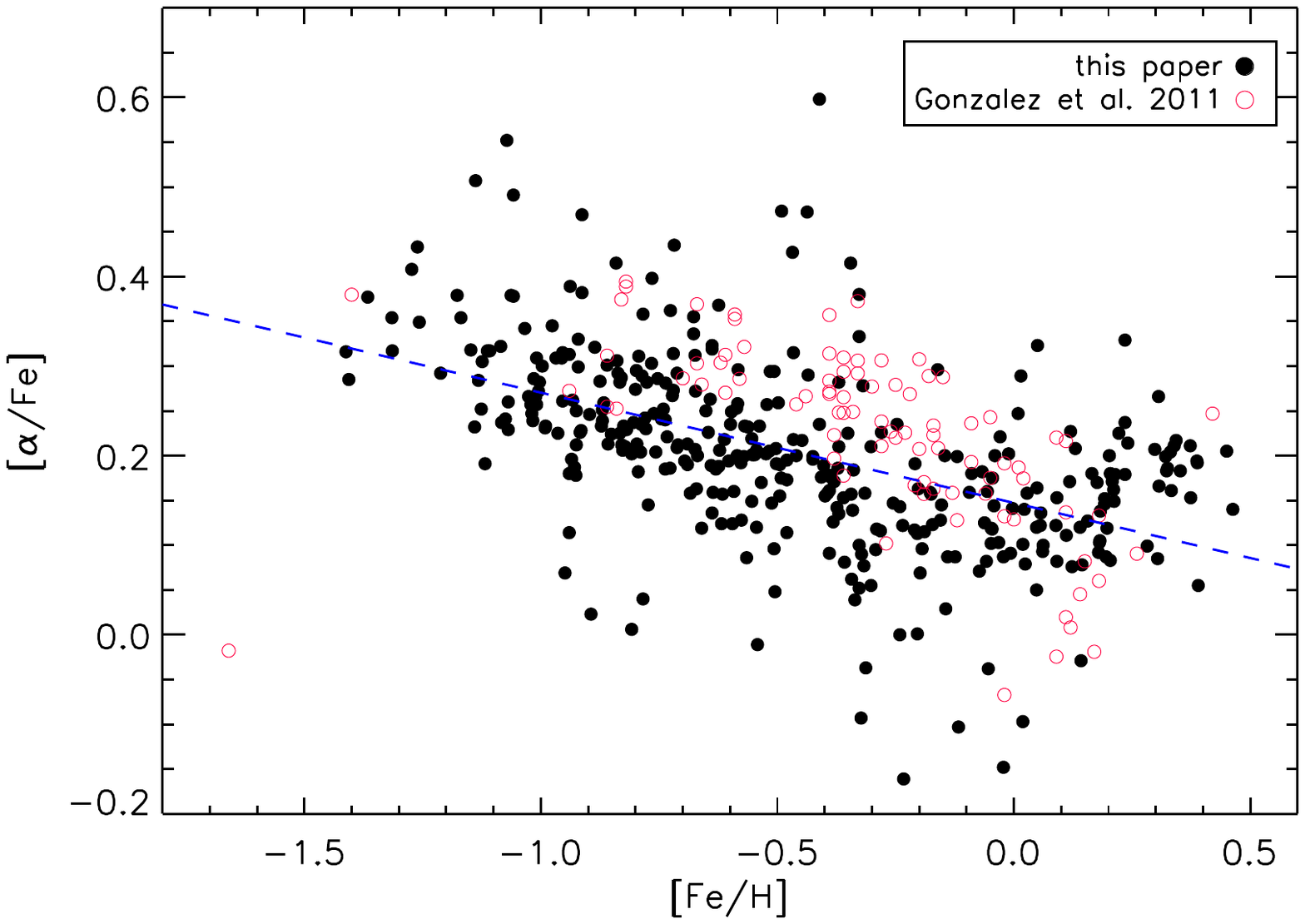}
  \caption{Abundance of $[\alpha/{\rm Fe}]$ as a function of [Fe/H]. Filled
      black circles: data from this study; open red circles: data of the field
      at $b=-12\degr$ from \citet{Gon11a}. The dashed line is a linear fit
      trough the data of this paper.}
  \label{alphaFe_FeH}
\end{figure}

An important piece of evidence for the understanding of the GB population is the
abundance of the $\alpha$-elements because they convey information of the star
formation history. In Fig.~\ref{alphaFe_FeH} we present the abundance of
[$\alpha$/Fe] as a function of the iron abundance [Fe/H].
The metal-rich and metal-poor stars also differ in this diagram: The metal-rich
stars have clearly a lower over-abundance in the $\alpha$-elements than the
metal-poor ones. Among the metal-rich stars, most stars fall in the range
$[\alpha/{\rm Fe}]=+0.05$ to $+0.25$, while among the metal-poor stars many
stars can be found at over-abundances of $+0.30$\,dex and more.

Because of proximity to our field, we over-plot the abundances measured by
\citet{Gon11a} in the field at $b=-12\degr$. To convert the individual
abundances to a combined $[\alpha/{\rm Fe}]$ value, we computed the weighted
mean of their Ca, Ti, Si, and Mg abundances, using as weight the contribution to
the absorption in the synthetic spectrum that we applied in the determination of
the $\alpha$-abundances (Sect.~\ref{alphameasure}). The match between our data
and those of \citet{Gon11a} is not perfect. The over-abundances measured by
\citet{Gon11a} are on average slightly higher than in our data, even though a
shift was applied to our data to take into account a systematic offset
(Sect.~\ref{systematics}). It is also clear that there are many more metal-poor
stars ($-1.5\le[{\rm Fe}/{\rm H}]\le-0.5$) in our sample than in that of
\citet{Gon11a}, and also at the metal-rich end we find somewhat more stars than
they do.

We also confirm the trend of decreasing $\alpha$-element over-abundance
with increasing metallicity, as was found already by a number of previous
studies \citep[e.g.][]{Ful07,Ryde10,Alv10,Hill11,Gon11a,JohnC11,Bens10a}. A
linear regression including all sample stars yields a slope of $-0.12$ (dashed
line in Fig.~\ref{alphaFe_FeH}). The $1\sigma$ scatter around this slope is only
$\sim0.08$\,dex, which suggests that it originates only from measurement error,
not from real abundance scatter within the sample, at a given metallicity.
At the high metallicity end, the trend of decreasing $\alpha$-element
over-abundance might be shallower or even reversed. This seems to disagree with
the data of \citet{Gon11a} overplotted in Fig.~\ref{alphaFe_FeH}, however. On
the other hand, a flat trend with [Fe/H] is also found in metal-rich,
microlensed bulge dwarf stars for some $\alpha$-elements, e.g.\ Ca, Ti, Si, and
Mg \citep[Fig.~10 of ][]{Bens11}. Only oxygen is clearly found to continue its
decrease at high metallicity. If this is a real feature or the result of
uncertainties needs to be investigated with more precise line-by-line studies
and larger samples, respectively.

\subsection{Radial velocities}\label{res_rv}

\begin{figure}
\centering
\includegraphics[width=\linewidth,bb=87 370 538 700,clip]{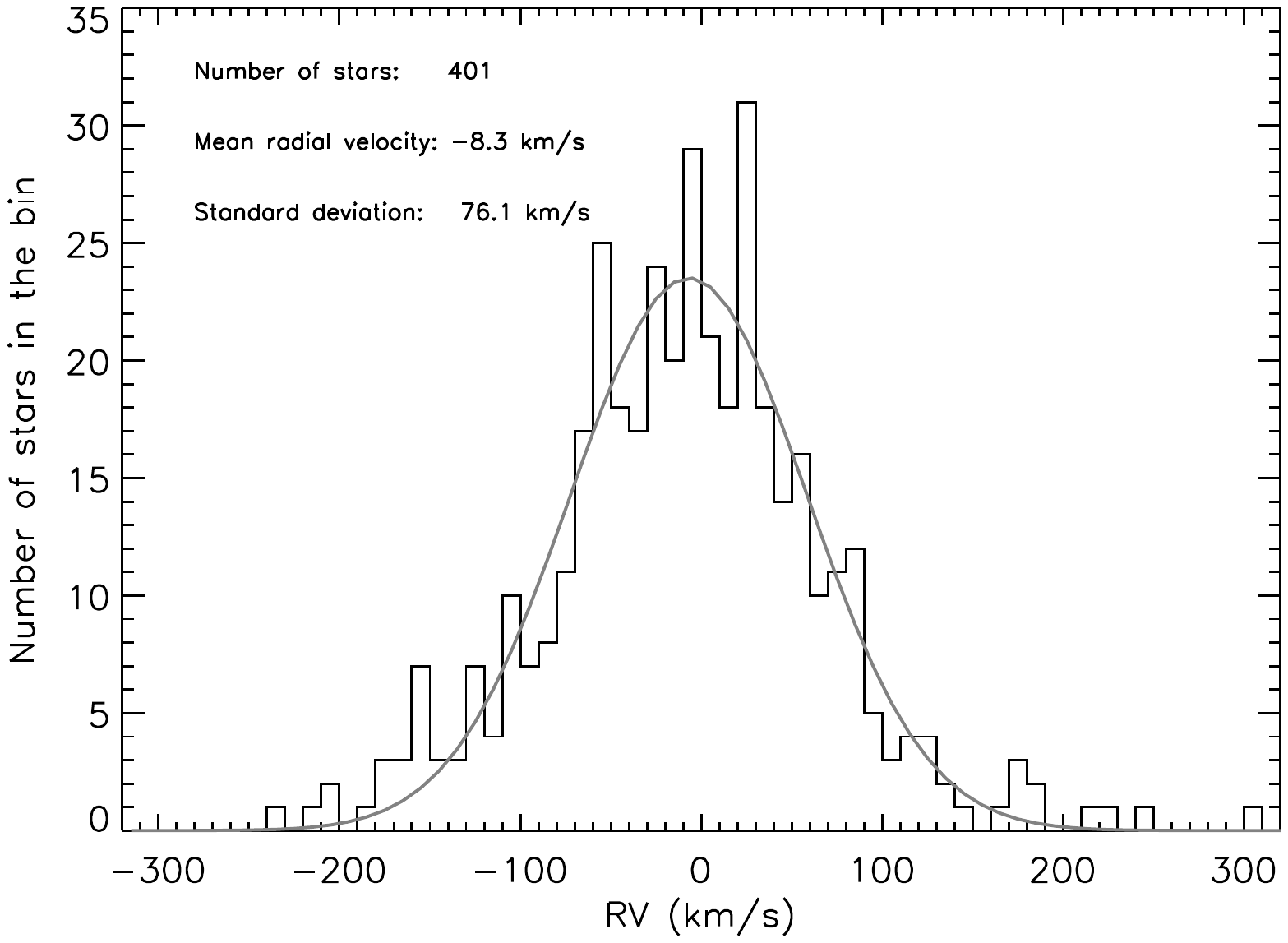}
\caption{Radial velocity distribution of our sample stars. The grey graph is a
  Gaussian fit to the data with mean and standard deviation as indicated in the
  figure.}
\label{rv_all}
\end{figure}

The RV distribution is an invaluable constraint on the nature of the GB, i.e.\
to decide whether it is a classical or a pseudo-bulge.
The mean RV of our whole sample is $-8.3\pm3.8$\,km\,s$^{-1}$, with a velocity
dispersion $\sigma_{\rm RV}=76.1\pm2.7$\,km\,s$^{-1}$ (Note: In the following, the
quoted uncertainties in the mean velocity and the standard deviation of the
distribution are always the purely statistical uncertainties.) The distribution
of RVs of our sample stars is displayed in Fig.~\ref{rv_all}. No cold streams
are obvious in our sample. Excluding the 50 foreground candidates identified in
Sect.~\ref{foreground} by their high proper motion (and also the 10 stars for
which no proper motion measurement was found), we get
$\left<RV\right>=-8.9\pm4.2$\,km\,s$^{-1}$ and
$\sigma_{\rm RV}=77.1\pm3.0$\,km\,s$^{-1}$, hence no significant change.

\begin{figure}
  \centering
  \includegraphics[width=\linewidth,bb=78 370 539 698,clip]{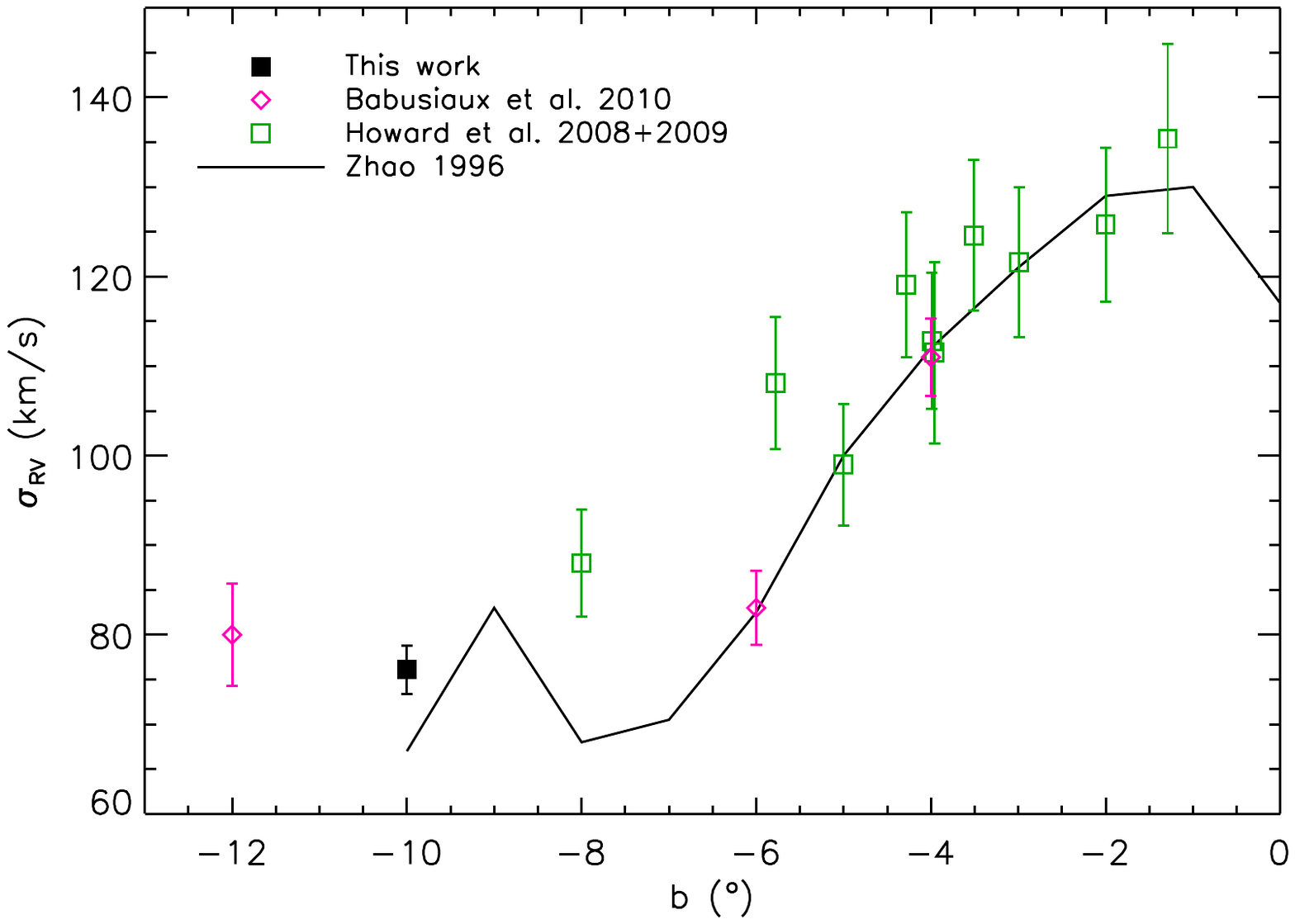}
  \caption{Comparison of observed velocity dispersions as a function of
    Galactic latitude $b$ along the bulge minor axis ($l=0\degr$) with the model
    of \citet{Zhao96}. Cf.\ Fig.~7 of \citet{Bab10}.}
  \label{Zhao_model}
\end{figure}

We can compare our observed $\sigma_{\rm RV}$ to those of \citet{Bab10} and
the BRAVA team \citep{How08,How09}. The comparison is done in
Fig.~\ref{Zhao_model}, which is a reproduction of Fig.~7 of \citet{Bab10}. We
emphasise that the sample selection for the BRAVA survey, \citet{Bab10}, and for
the present study are quite different. From the BRAVA survey, only the results
for fields along the bulge minor axis ($l \approx 0\degr$) are shown. Our
measurement of $\sigma_{\rm RV}$ is the lowest one in this figure, but within the
error bar of the data point at $b=-12\degr$ from \citet{Bab10}.
Figure~\ref{Zhao_model} also includes the prediction from the model by
\citet{Zhao96}, which is a ``3D steady-state stellar dynamic model for the
Galactic bar'', hence a pseudo-bulge model. The observed $\sigma_{\rm RV}$ as a
function of $b$ is well described by the model of \citet{Zhao96}, except for the
two BRAVA data points farthest from the Galactic plane. The $\sigma_{\rm RV}$ of
our sample is only slightly higher than what is predicted by this model.
Furthermore, our measurement is also in excellent agreement with the predictions
of the pseudo-bulge model developed by
\citet[][their Fig.~2, lower right panel]{Shen10}.

\begin{figure}
\centering
\includegraphics[width=\linewidth,bb=88 370 536 699,clip]{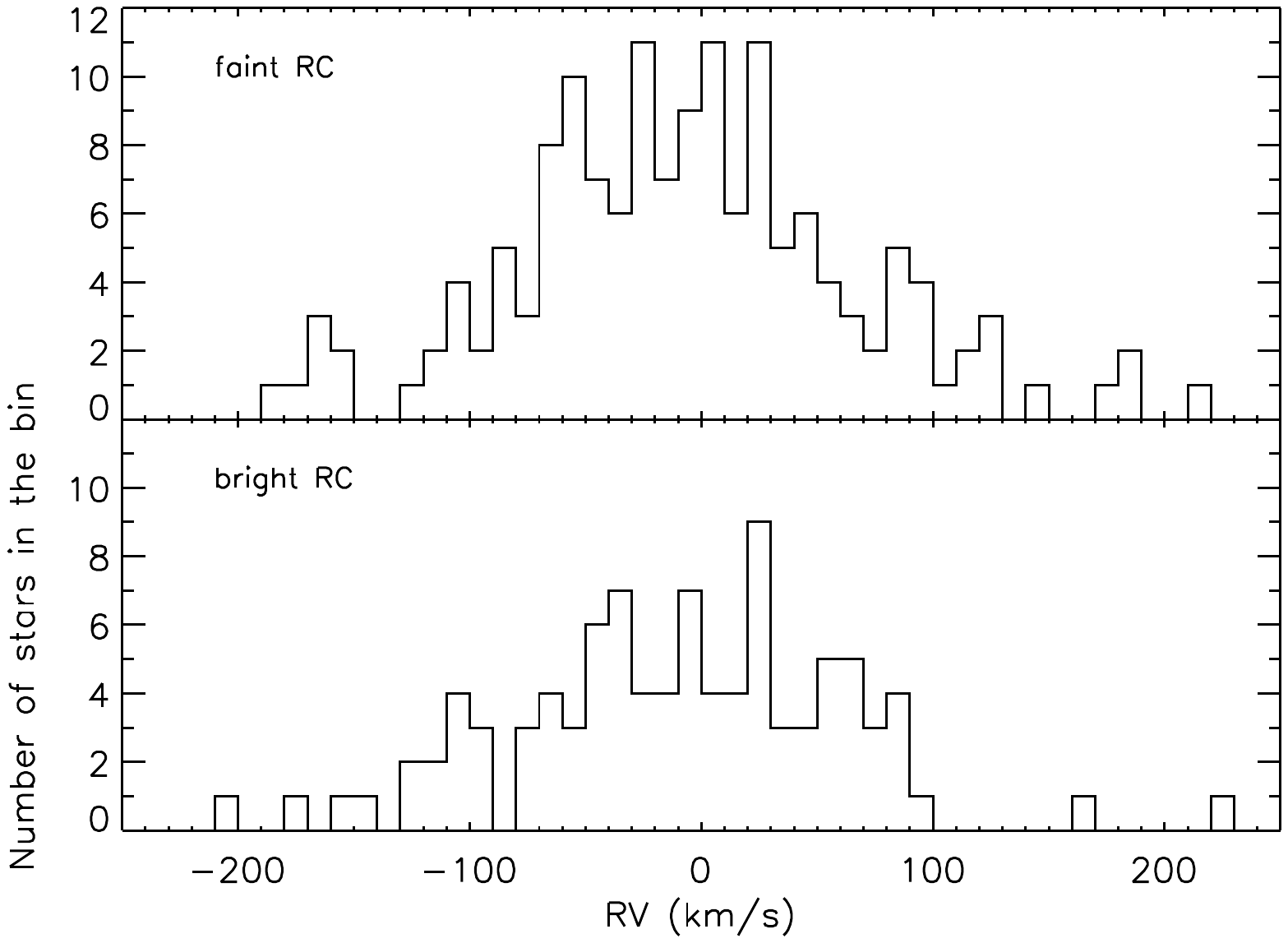}
\caption{Radial velocity distribution of the stars in the bright and faint RC,
  respectively.}
\label{rv_rcs}
\end{figure}

The kinematics of the RC stars is an important test for the understanding of the
nature of the double RC feature. Figure~\ref{rv_rcs} displays the RV
distribution of the stars in the two RCs as defined in Sect.~\ref{maghist2rcs}.
The mean RVs are $-10.4\pm7.3$ and $-6.0\pm6.1$\,km\,s$^{-1}$, and the standard
deviations are $ 71.6\pm5.2$ and $74.1\pm4.3$\,km\,s$^{-1}$, respectively.
Although the velocity dispersion of the faint RC is slightly larger than that of
the bright RC, the two distributions are indistinguishable, within the error
bars. To make this result more significant, we derived probabilities that a
particular star belongs to either the front or back over-density (bright or
faint RC). For this end we fitted two Gaussians (for the two RCs) and an
exponential function (for the underlying RGB stars) to the magnitude histogram
to derive these probabilities. This procedure was applied to a 40\arcmin\
diameter field to increase the statistics for a Gaussian fitting of the two RCs.
Using these membership probabilities as weights, we get
$\left<RV\right>=-10.3\pm7.3$\,km\,s$^{-1}$ and
$\left<RV\right>= -5.1\pm6.0$\,km\,s$^{-1}$, and
$\sigma_{\rm RV}=71.3\pm5.2$\,km\,s$^{-1}$ and
$\sigma_{\rm RV}=73.5\pm4.1$\,km\,s$^{-1}$ for the bright and faint RC,
respectively. Even in the weighted mean, the kinematics of the two RCs remain
indistinguishable. Furthermore, a statistical t-test yields a probability of
64.2\% that the two distributions are drawn from the same parent population.

The above discussed mean RVs and velocity dispersions of the different
sub-samples are summarised in Table~\ref{rv_subsam}.

\begin{table}
\caption{Radial velocity distributions of different sub-samples. The top group
  are the values measured from the observed sample, the bottom group are the
  values derived from ten simulation runs of the Besan\c{c}on model
  \citep{Rob12}.}
\label{rv_subsam}          
\centering                 
\begin{tabular}{lrrc}      
\hline\hline               
Selection & $N$ & $\left<RV\right>$ (km\,s$^{-1}$) & $\sigma_{\rm RV}$ (km\,s$^{-1}$)\\
\hline                       
{\bf Observations}      & ~   & ~            & ~            \\
whole sample            & 401 & $-8.3\pm3.8$ & $76.1\pm2.7$ \\
$\mu<20$\,mas           & 341 & $-8.9\pm4.2$ & $77.1\pm3.0$ \\
bright RC               &  96 &$-10.4\pm7.3$ & $71.6\pm5.2$ \\
bright RC, weighted     &  96 &$-10.3\pm7.3$ & $71.3\pm5.2$ \\
faint RC                & 150 & $-6.0\pm6.1$ & $74.1\pm4.3$ \\
faint RC, weighted      & 150 & $-5.1\pm6.1$ & $73.5\pm4.1$ \\
metal-rich third        & 128 & $-7.1\pm4.6$ & $52.3\pm3.3$ \\
metal-poor third        & 128 & $-5.4\pm8.0$ & $90.6\pm5.7$ \\
metal-rich pop.$^{\rm 1}$ &  88 & $-6.1\pm5.3$ & $47.7\pm5.6$ \\
metal-poor pop.$^{\rm 1}$ & 198 & $-9.1\pm6.4$ & $90.3\pm6.5$ \\
\hline
{\bf Besan\c{c}on model}& ~    & ~             & ~            \\
whole sample            & 6933 & $-11.1\pm0.9$ & $77.8\pm0.7$ \\
$\mu<20$\,mas           & 6555 & $-11.0\pm1.0$ & $79.6\pm0.7$ \\
bright RC               & 1408 & $-10.0\pm2.0$ & $74.1\pm1.4$ \\
faint RC                & 2802 & $-11.1\pm1.5$ & $80.9\pm1.1$ \\
metal-rich third        & 2311 & $-13.1\pm1.4$ & $65.9\pm1.0$ \\
metal-poor third        & 2312 &  $-9.2\pm1.8$ & $87.2\pm1.3$ \\
\hline                       
\end{tabular}
\tablefoot{(1): Metal-rich and metal-poor population according to the Gaussian
  decomposition in Sect.~\ref{MDF_bf}.}
\end{table}

\subsection{Combining kinematics with abundances}\label{RV_abu}

\begin{figure}
  \centering
  \includegraphics[width=\linewidth,bb=69 406 545 699,clip]{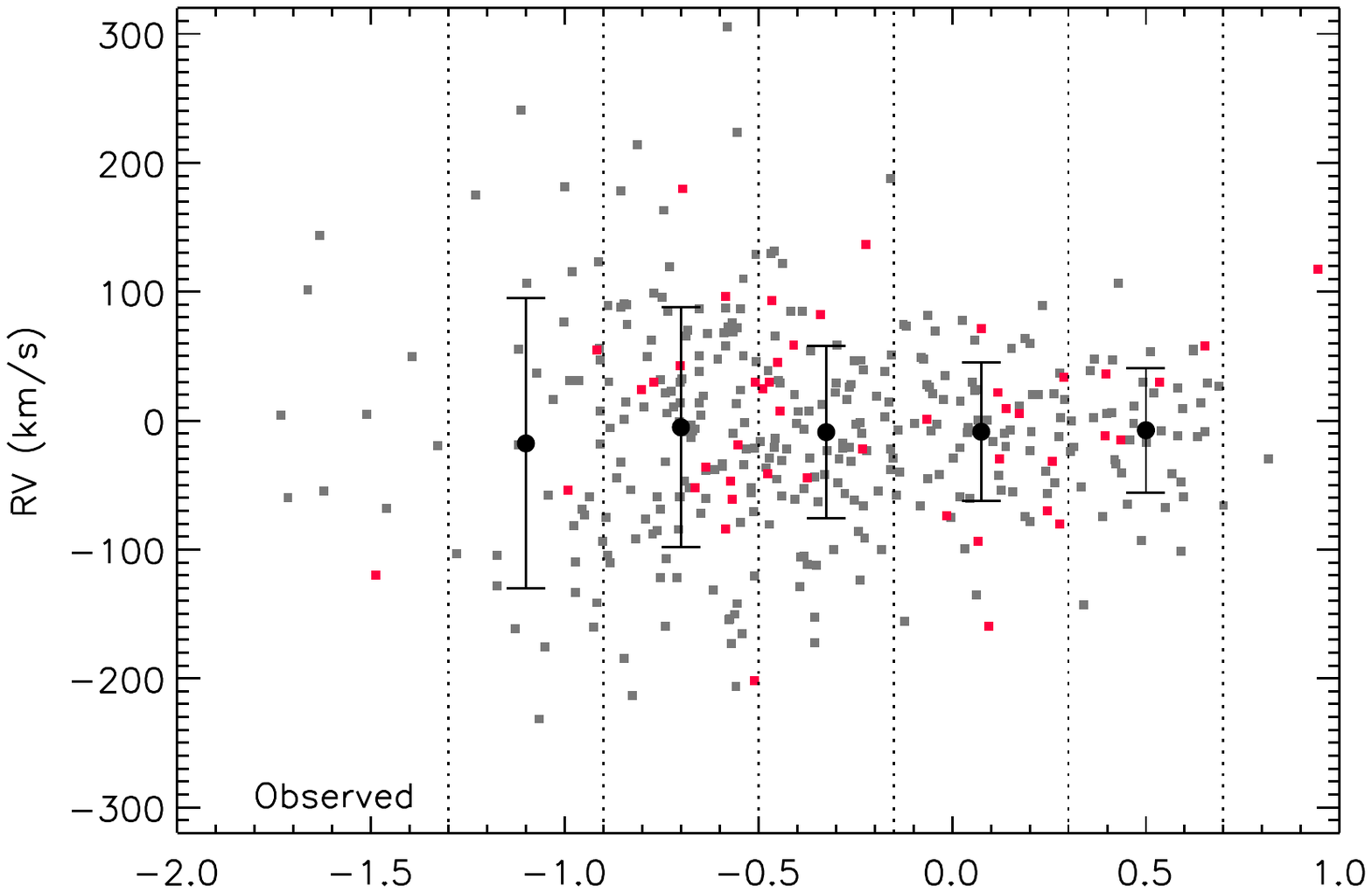}
  \includegraphics[width=\linewidth,bb=69 370 545 699,clip]{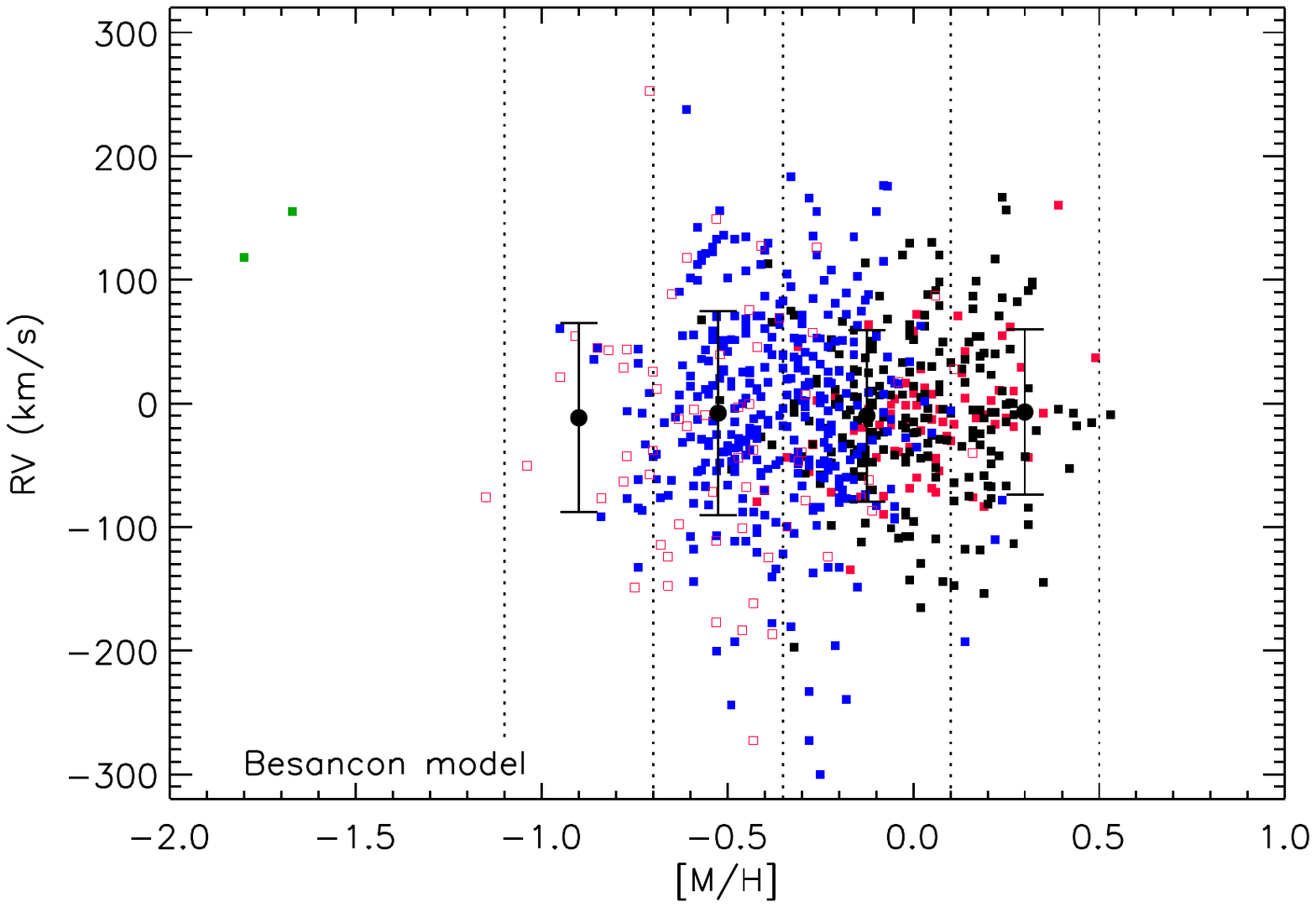}
  \caption{Radial velocity versus metallicity. {\em Top panel:} observed
    sample stars. Foreground star candidates are represented by red squares, all
    other stars by grey squares. {\em Bottom panel:} simulated stars in the
    Besan\c{c}on model \citep{Rob12}. Blue squares represent the thick bulge,
    black squares the bar, filled red squares the thin disc, open red squares
    the thick disc, and green symbols represent the halo population. The
    over-plotted error-bar symbols in both panels show the mean and standard
    deviation of the RV over the metallicity range between the vertical dotted
    lines.}
  \label{RV_vs_MH}
\end{figure}

The upper panel of Fig.~\ref{RV_vs_MH} shows how the sample stars distribute in
the RV vs.\ [M/H] plane. It it immediately clear from this diagram that the RV
distribution becomes broader with decreasing metallicity and that at low
metallicity more extreme RV values are found than at high metallicity. This is
also illustrated by the large over-plotted ``error-bar'' symbols, which show the
mean RV and velocity dispersion over the metallicity range between the vertical
dotted lines. The mean RV stays more or less constant, whereas $\sigma_{\rm RV}$
increases monotonically with decreasing metallicity. This is further evidence
that there are two different populations present in our sample, each with its
own velocity distribution. Whereas the increase in the velocity dispersion
agrees with the results of \citet{DePro10}, the lack of a trend of the mean RV
with metallicity is in clear contrast to their results. This also holds if only
the RC stars in our sample are considered. This is somewhat surprising because
the field studied by \citet{DePro10} is only two degrees closer to the Galactic
plane than our field, along the bulge minor axis. At this point we do not have
an explanation for this discrepancy.

The metal-rich stars on the near side of the X (stars in the bright peak in
the upper panel of Fig.~\ref{mag_hist_RCs}) may be expected to differ in their
mean radial velocity from those on the far side (faint peak in
Fig.~\ref{mag_hist_RCs}). We find $\left<RV\right>=+1.5\pm7.5$\,km\,s$^{-1}$ and
$\sigma_{\rm RV}=53.4\pm5.3$\,km\,s$^{-1}$ for the 51 bright, metal-rich RC stars
and $\left<RV\right>=-10.7\pm8.2$\,km\,s$^{-1}$ and
$\sigma_{\rm RV}=51.5\pm5.9$\,km\,s$^{-1}$ for the 39 faint, metal-rich RC stars,
respectively (using a metallicity cut at $[{\rm M}/{\rm H}]\ge-0.2$ and the
brightness criteria for the RCs of Sect.~\ref{maghist2rcs}). The error bars on
the mean RVs overlap, hence the difference is not significant. Note, however,
that RGB/early AGB stars brighter than the RC located on the far side and stars
fainter than the RC located on the near side will dilute any difference between
these two groups. A still larger sample would be needed to confirm this
difference in mean RV.

\subsubsection{Comparison with the Besan\c{c}on model}\label{RV_BGM}

It is interesting to compare this result with the kinematics as provided by the
simulations to see whether the BGM, based on a decomposition in several
populations, mainly disc, bulge and bar in the central region, is in reasonable
agreement with our data. The lower panel of Fig.~\ref{RV_vs_MH} shows for
comparison the distribution of simulated stars in the BGM in the RV vs.\
metallicity plane. The width of the RV distribution as well as the extreme RV
values agree well with those in the observed sample. However, the trend of
increasing velocity dispersion with decreasing metallicity is much smaller in
the simulation than in the observations. The main reason for this is that the
bulge population (blue symbols in the lower panel of Fig.~\ref{RV_vs_MH}), whose
velocity dispersion is relatively large, is close in metallicity to the bar
population (black symbols), which has a smaller velocity dispersion. If the
bulge population in the Besan\c{c}on model was shifted to lower metallicities by
0.2 to 0.3\,dex, it would nicely reproduce this trend, and also bring the range
of metallicities in the model closer to the observed range. This is in line with
the conclusions drawn from the comparison of the MDFs (Sect.~\ref{compMDF}).

\begin{figure}
\centering
\includegraphics[width=\linewidth,bb=79 372 538 700,clip]{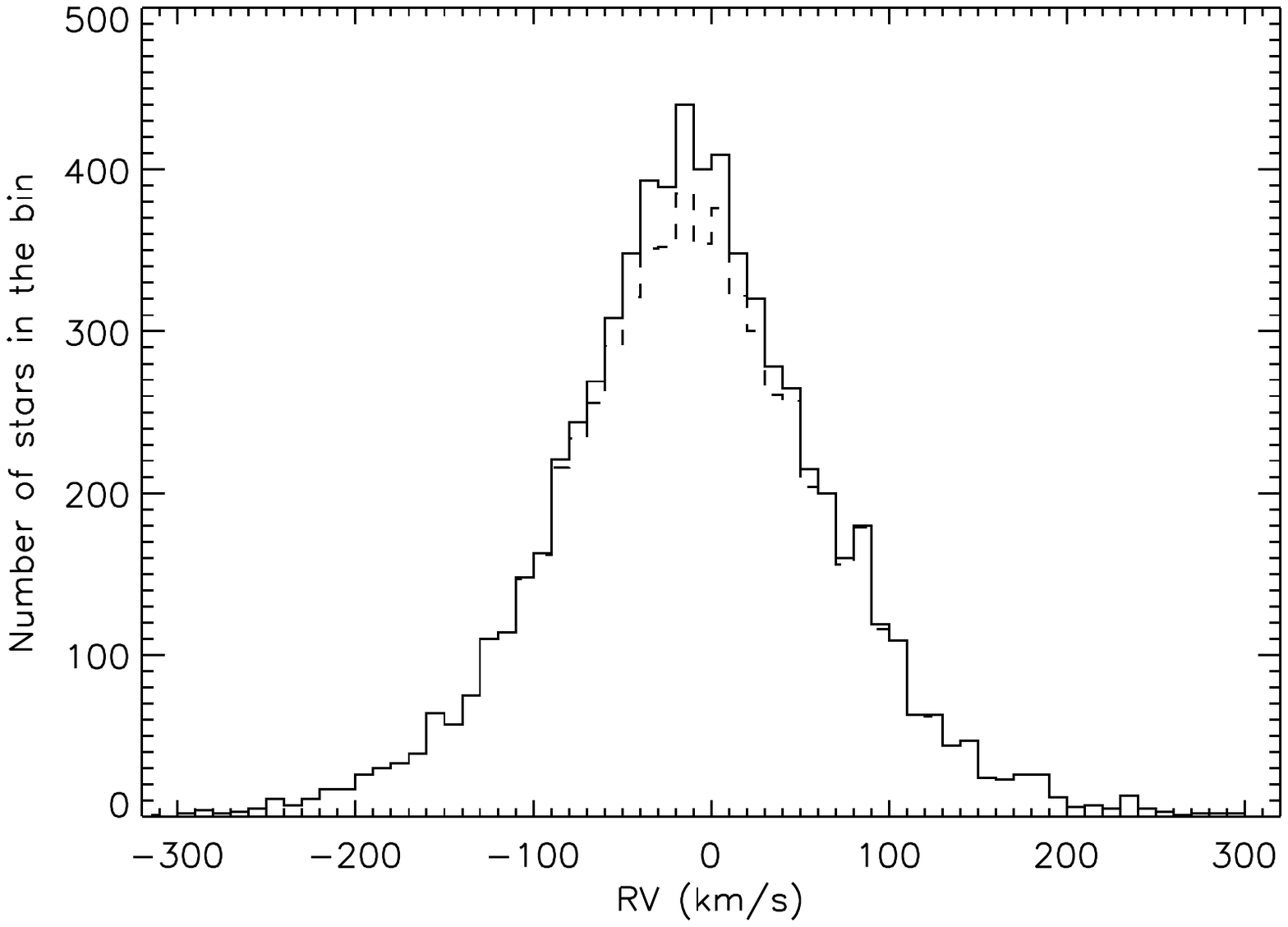}
\caption{Radial velocity distribution of stars in ten simulation runs of the
  Besan\c{c}on model \citep{Rob12} for our field. The full line is for all stars
  in the selection region, the dashed line shows the distribution for stars with
  total proper motion $\mu<20$\,mas/yr.}
\label{RV_hist_BGM}
\end{figure}

To compare the observed with the simulated velocity dispersions, and to decrease
the Poisson noise in the model, we calculated ten simulation runs of the
Besan\c{c}on model for our field. From these ten simulations, we obtain 6933
stars in the selection area. The results of this exercise are summarised in the
lower part of Table~\ref{rv_subsam}. The mean RV as well as the velocity
dispersion of the simulation agree well with the observed values for the whole
sample. Excluding stars with a total proper motion $\mu>20$\,mas/yr, the
velocity dispersion slightly increases in both observations and simulations. As
discussed in Sect.~\ref{selbias}, this selection by proper motion is very
efficient in removing the foreground disc. This increase in velocity dispersion
is not significant in the observed sample, but it is significant in the
simulation. In Fig.~\ref{RV_hist_BGM} we show the histogram of the distribution
of radial velocity from the simulations, with (dashed line) and without (solid
line) the proper motion selection. As can be seen from this figure, the proper
motion selection removes a cold (disc) component.

Table~\ref{rv_subsam} also lists the mean RV and dispersions of the simulated
bright and faint red clump sub-samples, as well as for the metal-rich and
metal-poor 33.3\% quantile. In both the observations and in the simulations the
bright RC has a slightly smaller velocity dispersion; while this increase is
again not significant in the observed sample, it is significant in the
simulations. The reason for this probably is that, as already discussed in
Sect.~\ref{MH_RC}, the faint RC is more contaminated by stars from the
metal-poor population, which has a higher velocity dispersion than the bar
population, the probable carrier of the double clump feature.

The metal-rich sample has a smaller velocity dispersion than the metal-poor one.
This is due to the fact that the metal-rich one is dominated by the bar
population, while the more metal-poor one is dominated by the bulge. The
velocity dispersion of the metal-poor third of the simulation agrees nicely with
the observed one (87.2\,km\,s$^{-1}$ to be compared with 90.6\,km\,s$^{-1}$, see
below). However, the metal-rich third has a larger dispersion in the simulations
(65.9\,km\,s$^{-1}$) than in the observed sample (52.3\,km\,s$^{-1}$). As seen in
the lower panel of Fig.~\ref{RV_vs_MH}, the metal-rich part of the sample is
dominated by the bar, which follows \citet{Fux99} dynamics. But there is also a
contribution from the disc, which is mainly removed using the proper motion
selection. In the very low metallicity tail, there are a few halo stars and a
noticeable contribution from the thick disc. It has, in our model, a smaller
dispersion than the bulge, which has about the same metallicity. The velocity
dispersion of the thick disc in the inner galaxy is not known and the simulation
we are proposing here is just an attempt to extrapolate the thick disc as seen
in the solar neighbourhood to the central region. Though, the dispersion we
obtain for this intermediate and low metallicity bin is in very good agreement
with the data.

We conclude that this new model, having both a flaring bar and a bulge component
in the central region, explains very well the observed relation between the
metallicity and the kinematics in this bulge field.


\subsubsection{Comparison with the Fux model}\label{Fux}

\begin{figure}
  \centering
  \includegraphics[width=\linewidth,bb=78 370 539 698,clip]{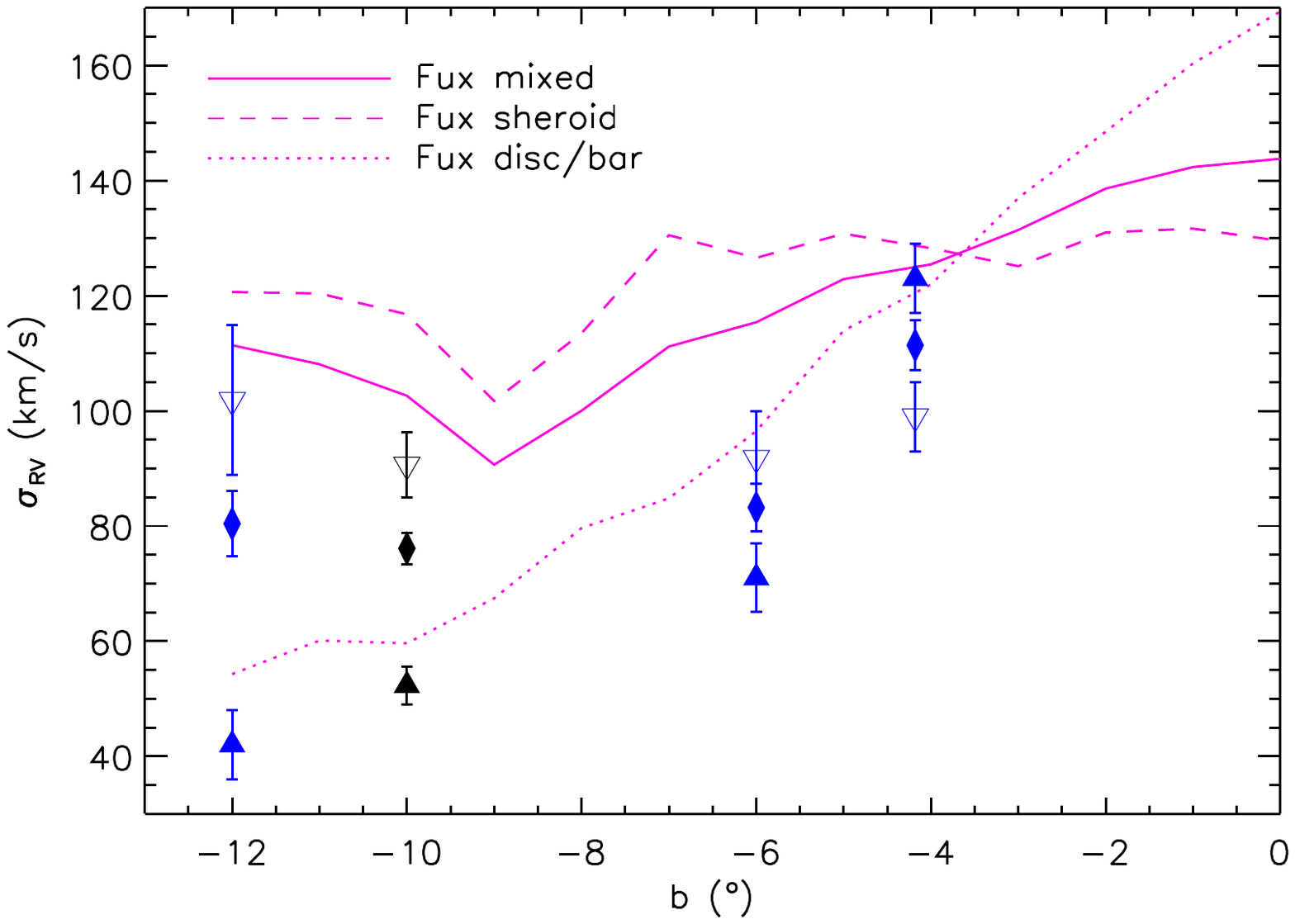}
  \caption{Observed RV dispersions as a function of Galactic latitude $b$ along
    the bulge minor axis, compared to the models of \citet{Fux99}, cf.\ Fig.~8
    of \citet{Bab10}. The blue symbols are the data points from \citet{Bab10},
    the black symbols are from this work. The diamond symbol represents the
    velocity dispersion of the complete sample, while the open
    downward-pointing triangle represents the 33.3\% metal-poor quantile and the
    filled upward-pointing triangle represents the 33.3\% metal-rich quantile,
    respectively.}
  \label{Fux_model}
\end{figure}

The increase of $\sigma_{\rm RV}$ with decreasing metallicity is reminiscent of
the results of \citet{Bab10}, who also found a clear distinction in kinematics
between the metal-poor and the metal-rich stars in bulge fields at different
Galactic latitudes. In a reproduction of their Fig.~8, we show in
Fig.~\ref{Fux_model} a comparison with the bulge dynamics as predicted by the
3D self-consistent N-body barred models of the Milky Way by \citet{Fux99}. As
in \citet{Bab10}, $\sigma_{\rm RV}$ is plotted in different symbols for the
complete sample (diamonds), the 33.3\% metal-poor quantile (open
downward-pointing triangles), and the 33.3\% metal-rich quantile (filled
upward-pointing triangles). We find $\sigma_{\rm RV}=52.3\pm3.3$\,km\,s$^{-1}$ for
the metal-rich third of the sample and $\sigma_{\rm RV}=90.6\pm5.7$\,km\,s$^{-1}$
for the metal-poor third (Table~\ref{rv_subsam}). By adding only one more data
point at $b=-10\degr$ we can put much more weight on the conclusions of
\citet{Bab10}. Within the error bars, $\sigma_{\rm RV}$ of the metal-poor
quantile is constant throughout the bulge, from $b=-4\degr$ to $-12\degr$. The
spheroid model of \citet{Fux99} also predicts $\sigma_{\rm RV}$ to be fairly
constant as a function of $b$, except for a dip around $b=-9\degr$. However,
the observed $\sigma_{\rm RV}$ is systematically lower than the $\sigma_{\rm RV}$
predicted by the spheroid model. The velocity dispersion of the 33.3\%
metal-rich quantile, on the other hand, decreases monotonically with increasing
distance from the Galactic plane. This follows closely the prediction by the
disc/bar model of \citet{Fux99}, but except at $b=-4\degr$ it is also always
below the predicted value.

We must caution at this point that, while the definition of quantiles is a
robust measure, the metallicity cuts will be at different absolute values,
depending on the metallicity scale used in a given study and on the metallicity
distribution in the different fields. We therefore also derived the RV
dispersions of the two populations from the Gaussian decomposition done in
Sect.~\ref{MDF_bf}. Assuming that both have Gaussian RV distributions, we find
that the metal-poor population has $\left<RV\right>=-9.1\pm6.4$\,km\,s$^{-1}$
with $\sigma_{\rm RV}=90.3\pm6.5$\,km\,s$^{-1}$, while the metal-rich population
has $\left<RV\right>=-6.1\pm5.3$\,km\,s$^{-1}$ and
$\sigma_{\rm RV}=47.7\pm5.6$\,km\,s$^{-1}$ (Table~\ref{rv_subsam}).

\subsubsection{Evidence for metallicity-dependent mass loss from the velocity
  dispersion of evolved stars}\label{sigma_evol}

\begin{figure}
  \centering
  \includegraphics[width=\linewidth,bb=78 406 536 699,clip]{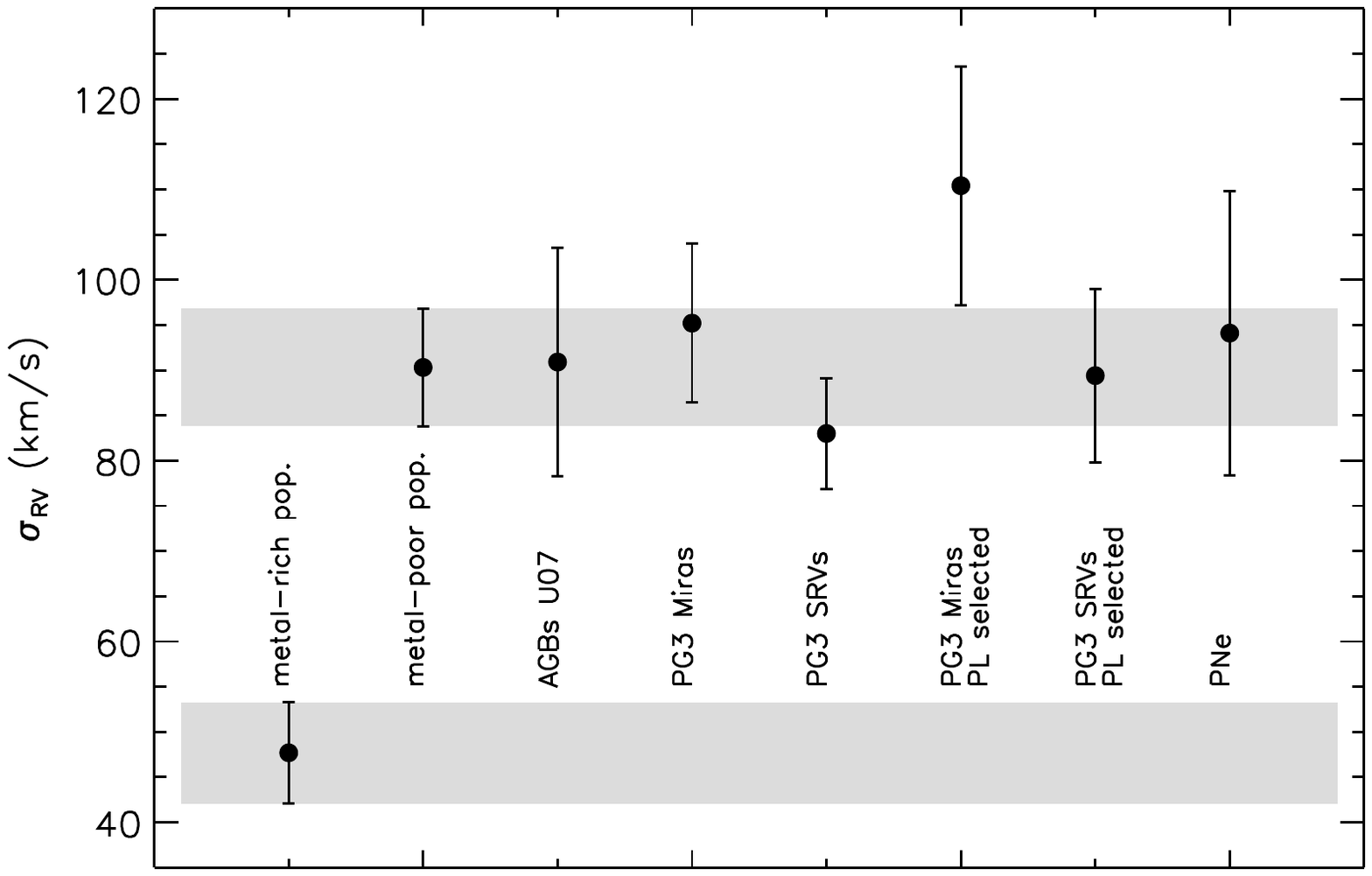}
  \caption{Velocity dispersion of the metal-rich and metal-poor populations
    (Table \ref{rv_subsam}), as well as of several samples of evolved stars
    towards the PG3 field: AGB stars from \citet[][U07]{Utt07}, Miras and SRVs
    from \citet{Schulthe98} without and with selection using a
    period - $K$-magnitude relation, and PNe from \citet{Dur98}. The grey shaded
    areas mark the range of uncertainty of the metal-rich and metal-poor
    population, respectively.}
  \label{sigma_evol_fig}
\end{figure}

In Sect.~\ref{MDF_evol} we speculated that the lack of metal-rich stars
among the bright sample stars might be caused by strong, metallicity-dependent
mass loss that preferentially removes metal-rich stars from the evolutionary
pathes (the so-called AGB manqu\'e). There is another, even stronger piece of
evidence coming from the RV distributions of AGB stars and planetary nebulae
(PNe) in the PG3 field that supports this explanation. RVs of Mira and
semi-regular variables (SRVs) in the PG3 have been presented by
\citet{Schulthe98} and \citet{Utt07}. The 27 AGB stars studied by \citet{Utt07}
by means of high-resolution UVES/VLT spectra have a velocity dispersion of
$90.9\pm12.6$\,km\,s$^{-1}$. \citet{Schulthe98} found
$\sigma_{\rm RV}=95.2\pm8.8$\,km\,s$^{-1}$ for 59 PG3 Miras and
$\sigma_{\rm RV}=83.0\pm6.1$\,km\,s$^{-1}$ for 94 SRVs. If additional criteria
based on the period - $K$-magnitude relation are applied to select only
probable bulge members, these numbers change to
$\sigma_{\rm RV}=110.4\pm13.2$\,km\,s$^{-1}$ (Miras, $N=36$) and
$\sigma_{\rm RV}=89.4\pm9.6$\,km\,s$^{-1}$ (SRVs, $N=43$). Furthermore, we
retrieved RV measurements of PNe from the compilation of \citet{Dur98}.
Nineteen PNe were found within a search radius of $4\degr$ from the centre of
our FLAMES field. Those 19 PNe have a velocity dispersion of
$94.1\pm15.7$\,km\,s$^{-1}$. Although this last number is not highly significant
owing to the small number of objects, it is clear that these velocity
dispersions agree very well with the one determined for the metal-poor
sub-population, but disagree with that of the metal-rich one. This statement is 
summarised in Fig.~\ref{sigma_evol_fig}. We conclude from this that the AGB
stars and PNe in the PG3 field descend mainly from the metal-poor
sub-population. Or, to say it the other way round, we conclude that the
metal-rich stars probably never evolve up to the AGB stage and beyond because
they terminate their evolution before. This conclusion has also important
consequences on the interpretation of the dual chemistry found in many bulge PNe
\citep{Guz11}, a phenomenon which was previously assumed to be a consequence of
high-metallicity PNe precursor stars \citep{PC09}.

Apparently, there is a significant metal-rich population present out to
$b\geq-10\degr$, however it seems to be unable to produce highly evolved stars.
Also the dearth of super-solar metallicity stars among M giants in fields close
to the Galactic plane \citep{Rich07,ROV12} may be caused by enhanced mass loss
off metal-rich stars, in line with our result and interpretation. Finally,
enhanced mass loss might also be required to understand the binary fraction of
low-mass white dwarfs \citep{Brown11}. A puzzle in this context is the presence
of technetium-rich, long-period Mira stars towards the PG3 field \citep{Utt07},
which suggest the presence of a younger and more massive ($M\approx1.5M_{\sun}$)
population in the GB, but at the same time carbon stars are extremely rare or
even absent in the bulge \citep{BT89,Ng97,Schulthe98}. This seems to be
incompatible with the hypothesis of bulge AGB stars descending only from the
old, metal-poor population. A solution to this problem could be that the Tc-rich
AGB stars are the result of blue straggler evolution. Also problematic in this
respect are the various selection effects that are certainly present on the AGB
and PN samples in the GB, and only more detailed investigations, e.g.\ on the
metal content of bulge AGB stars, can give more reliable answers.

A natural consequence of the loss of metal-rich stars by enhanced stellar
winds would be that samples of bright M-type giants, as used by the BRAVA survey
to probe the bulge velocity dispersion, could be biased towards {\em metal-poor}
stars. In Sect.~\ref{res_rv} we found that the observed $\sigma_{\rm RV}$ as a
function of $b$ is well described by the model of \citet{Zhao96}, except for the
two BRAVA data points farthest from the Galactic plane (Fig.~\ref{Zhao_model}).
A bias towards low metallicity could be the reason why the $\sigma_{\rm RV}$
measured by BRAVA at large distances from the plane are systematically higher
than the predictions by the \citet{Zhao96} model: at this Galactic latitude, the
velocity dispersions of the metal-rich and the metal-poor populations differ
considerably, thus a bias towards more metal-poor stars will yield an enhanced
velocity dispersion.

\section{Comparison of observed and simulated CMDs}\label{compCMD}

\begin{figure*}
  \centering
  \includegraphics[width=\linewidth]{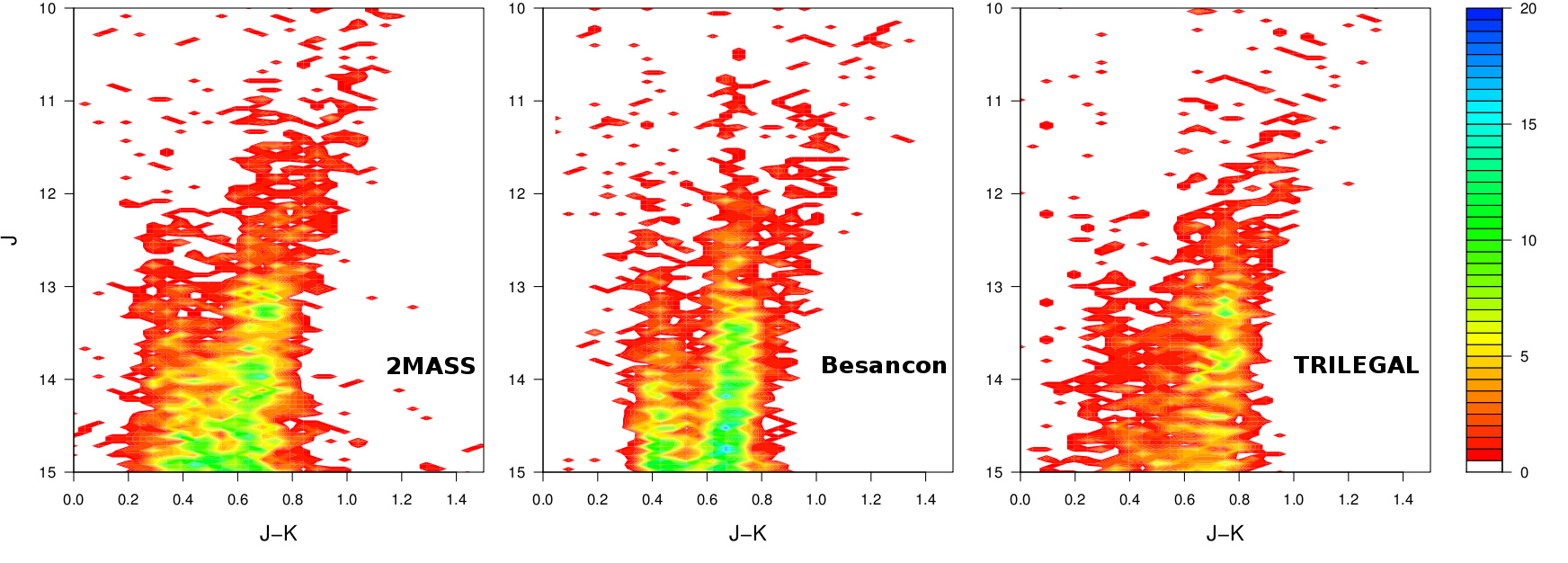}
  \caption{Observed and simulated CMDs plotted as density maps.
    {\em Left panel:} observed CMD from 2MASS. {\em Middle panel:} simulated
    CMD from the Besan\c{c}on model. {\em Right panel:} simulated CMD from the
    TRILEGAL model. The number of stars is not normalised. The scale on the
    right hand side illustrates to what colours the density of stars are
    connected in the diagrams.}
  \label{CMDs}
\end{figure*}

Comparisons with observed CMDs are an important test of Galaxy models. The 2MASS
all-sky catalogue is an ideal source for such constraints
\citep[see e.g.][]{Rob12}. In Fig.~\ref{CMDs} we provide a comparison for our
field. No extinction correction was applied to the observed CMD. Rather, we
applied the  \citet{Mar06} 3D extinction model to both the TRILEGAL and BGM
simulations.
In addition, photometric errors in each filter of the order of those in the
2MASS catalogue were added to both the Besan\c{c}on and the TRILEGAL simulation.

The CMD in this area of the sky is dominated by three sequences: a
sequence on the blue side formed by main-sequence disc stars in the foreground,
a vertical sequence formed by red clump stars distributing at various distances
from the sun, and a red sequence formed by bulge stars, along which our
selection of spectroscopic targets was made. These three sequences are most
clearly resolved in the Besancon model (middle panel of Fig.~\ref{CMDs}), even
though photometric errors of the same order of magnitude as in the 2MASS
catalogue have been added to the simulated stars.

The two red clumps in the observed CMD at $J\sim13\fm3$ and $J\sim14\fm0$ are
not well reproduced in the simulated CMDs for this field. This is not surprising
for the TRILEGAL model, which uses a triaxial structure to describe the bulge.
On the other hand, it has been shown that the flaring bar model developed by
\citet{Rob12} can reproduce the double clump feature, at least for a field at
$(l,b)=(0\degr,-7\degr)$. The double clump feature is not obvious in this single
simulation plotted here, but an inspection of the ten combined simulations
(Sect.~\ref{RV_BGM}), which were calculated to reduce the Poisson noise, the
two clumps are well discernible at $J$ magnitudes of $\sim13\fm6$ and
$\sim14\fm1$. Hence, while the faint RC is well matched by the Besan\c{c}on
model, the bright one is slightly underestimated in its brightness (e.g.\ due to
an overestimation of its distance).

In the TRILEGAL CMD a vertical sequence of RC stars is dominating in the bulge,
which is probably a result of the triaxial bulge structure seen almost end-on,
such that the bulge RC stars distribute over a relatively large range of
distances from the sun. Furthermore, the number of stars in the disc sequence
seems to be lower in the TRILEGAL simulation than it is in both the observed
and the Besan\c{c}on CMD. The number of stars in the CMDs of Fig.~\ref{CMDs} is
not normalised; star counts constitute an important test of population synthesis
models \citep{Gir05}. In the colour and magnitude range of Fig.~\ref{CMDs}, the
observed CMD contains 2148 stars, while the Besan\c{c}on and TRILEGAL models
contain 2456 and 1476 stars, respectively.

While some details in the simulated CMDs somewhat deviate from observed CMDs, it
is clear that the Galaxy models have reached already a very high level of
accuracy and can reproduce most of the stellar populations at least
qualitatively.

\section{Conclusions}\label{conclusio}

We presented an analysis of spectra of $\sim400$ red giant stars towards a field
at $(l,b)=(0\degr,-10\degr)$. From these spectra, we derive metallicities, iron
and $\alpha$-element abundances, and radial velocities. It is the first study
that presents a homogeneous analysis of stars on the upper and lower RGB of the
Galactic bulge. These data, as well as photometric data from the 2MASS survey,
are compared to predictions by the TRILEGAL \citep{Gir05} and Besan\c{c}on
\citep{Rob12} models of the Galaxy. These models are also used extensively to
interpret our observations.

The mean metallicity of the whole sample is $[{\rm M}/{\rm H}]=-0.34$, and the
radial velocity dispersion is $\sigma_{\rm RV}\sim76$\,km\,s$^{-1}$. In this study
we confirm the presence of two sub-populations in the bulge at peak
metallicities that differ by $\sim0.9$\,dex. The peak metallicities are found at
$[{\rm M}/{\rm H}]\sim-0.6$ and $\sim+0.3$, with roughly equal dispersions. The
sub-populations have significantly different kinematics and $\alpha$-element
abundances: the metal-rich population has a narrow velocity distribution
($\sigma_{\rm RV}\sim50$\,km\,s$^{-1}$) and low $\alpha$-abundances, whereas the
metal-poor one has a broad distribution in radial velocities
($\sigma_{\rm RV}\sim90$\,km\,s$^{-1}$) and high $\alpha$-abundances. The
metal-rich population makes up $\sim30$\% of our sample. This confirms and
fosters recent results by \citet{Bab10}, \citet{Bens11}, and \citet{Hill11} in
most aspects. Also the kinematic models of \citet{Zhao96} and \citet{Fux99} are
confronted with our data, and good agreement is found.

Furthermore, we find support for the suggestion by \citet{Bab10} that the ratio
in number of these two sub-populations is a function of angular distance from
the Galactic plane, which explains the observed metallicity gradient. The new
scheme in the Besan\c{c}on Galaxy model explains well the features seen in this
sample, with two populations, a flaring metal-rich bar and a more metal-poor
classical or thick bulge, even though the mean metallicity of this thick bulge
should be adjusted downward by 0.2 to 0.3\,dex. The metallicity distribution
simulated with the TRILEGAL model yields a clearly too high mean metallicity to
fit our observations.

This study also presents the first medium-resolution, high-S/N spectra of stars
belonging to the two red clumps of the bulge. A small difference in mean
metallicity between the two RCs is attributed to selection effects. We find that
the double RC might be entirely due to the metal-rich sub-population. The
kinematic difference between the two RCs when measured with radial velocities
alone is at most very small. Some difference in the mean radial velocity of
bright and faint metal-rich RC stars is found, although a still larger sample
would be needed to confirm that stars on the far side preferentially move faster
towards us than the ones on the near side.

We also find indications that the metallicity distribution function of the bulge
might depend on the evolutionary state of the considered sample stars. In
particular, there are fewer metal-rich stars present among the brighter, more
evolved stars in our sample. Although this result could be a consequence of
selection bias or the still limited number of stars in our sample, we interpret
this as the result of strong mass loss that causes the most metal-rich stars to
terminate their evolution before the most advanced states. A very robust
indication for that comes from the fact that AGB stars and PN in the outer bulge
have a velocity dispersion that agrees with that of the metal-poor population
identified here, but clearly disagrees with that of the metal-rich population.
We conclude, hence, that metal-rich stars might lose so much mass that they skip
phases of post-main-sequence stellar evolution. This means that the selection of
M giants in the BRAVA survey could be biased towards metal-poor stars. We also
confirm the trend of decreasing $\alpha$-element over-abundance with increasing
iron abundance, as found in previous studies \citep[e.g.][]{Gon11a}.

Our study also provides for an explanation how the existence of a dual bulge
identified here and elsewhere can be reconciled with kinematic studies that
conclude that the Milky Way bulge might be a pure pseudo-bulge
\citep[e.g.][]{Shen10}. In our version of the Besan\c{c}on model, the
(thick or classical) bulge population has only 4\% the mass of the bar, even
when its contribution is significant when one moves away from the plane. Thus it
is compatible with the result of \citet{Shen10}, which finds that the fraction
of the classical bulge cannot be more than 8\% of the disk mass. The
pseudo-bulge dominates, but there is some space for a classical bulge, which we
see in the data.

\begin{acknowledgements}
  We thank Marica Valentini for pointing us to the abundance measurements of
  the RAVE survey and for helpful discussion. SU acknowledges support from the
  Austrian Science Fund (FWF) under project P~22911-N16 and from the Fund for
  Scientific Research of Flanders (FWO) under grant number G.0470.07. DMN was
  primarily supported by the NSERC grant PGSD3-403304-2011, and partially
  supported by the NSF grant AST-1103471. TL acknowledges support from the FWF
  under projects P~23737-N16 and P~21988-N16. The Besan\c{c}on model simulations
  were executed on computers from the Utinam Institute of the Universit\'{e} de
  Franche-Comt\'{e}, supported by the R\'{e}gion de Franche-Comt\'{e} and
  Institut des Sciences de l'Univers (INSU). We acknowledge the support of the
  French Agence Nationale de la Recherche under contract
  ANR-2010-BLAN-0508-01OTP. This publication makes use of data products from the
  Two Micron All Sky Survey, which is a joint project of the University of
  Massachusetts and the Infrared Processing and Analysis Center/California
  Institute of Technology, funded by the National Aeronautics and Space
  Administration and the National Science Foundation.
\end{acknowledgements}

\Online
\begin{appendix}
\onllongtab{3}{
\begin{longtable}{lccccccrcccc}
\caption{Parameters and abundances of the sample stars.}\label{taba1}\\
\hline
\hline
No.\ & RA(J2000) & Dec(J2000) & $J$ & $H$ & $K$ & $T_{\rm eff}$ & RV          & [M/H]  & [Fe/H] & [$\alpha$/Fe] & FG  \\
     &  deg.\    & deg.\      & mag & mag & mag & K           & km\,s$^{-1}$ &        &        &               &     \\
 (1) & (2)       & (3)        & (4) & (5) & (6) & (7)         & (8)         & (9)    & (10)   & (11)          & (12)\\
\hline
\endfirsthead
\caption{Continued.}\\
\hline
\hline
No.\ & RA(J2000) & Dec(J2000) & $J$ & $H$ & $K$ & $T_{\rm eff}$ & RV          & [M/H]  & [Fe/H] & [$\alpha$/Fe] & FG  \\
     &  deg.\    & deg.\      & mag & mag & mag & K           & km\,s$^{-1}$ &        &        &               &     \\
 (1) & (2)       & (3)        & (4) & (5) & (6) & (7)         & (8)         & (9)    & (10)   & (11)          & (12)\\
\hline
\endhead
\hline
\endfoot
\hline
\endlastfoot
001 & 276.814513 & $-33.589966$ &  9.495 &  8.524 &  8.256 & 3465 & $+117.328$ & $+0.94$ &         &         & * \\
002 & 276.544710 & $-33.787083$ &  9.512 &  8.624 &  8.369 & 3697 & $ +20.277$ & $+0.21$ &         &         &   \\
003 & 276.608107 & $-33.667065$ &  9.516 &  8.615 &  8.298 & 3520 & $ +76.484$ & $-1.00$ &         &         &   \\
005 & 276.918684 & $-33.751503$ &  9.696 &  8.820 &  8.527 & 3629 & $ +55.090$ & $-1.12$ & $-0.95$ & $+0.07$ &   \\
006 & 276.432104 & $-33.784958$ &  9.926 &  9.061 &  8.704 & 3514 & $ -29.587$ & $+0.82$ &         &         &   \\
007 & 276.714893 & $-33.891556$ &  9.978 &  9.096 &  8.838 & 3698 & $ +12.913$ & $-0.56$ & $-0.48$ & $+0.11$ &   \\
008 & 276.600416 & $-33.681347$ & 10.007 &  9.121 &  8.891 & 3756 & $-135.108$ & $+0.06$ & $+0.05$ & $+0.32$ &   \\
009 & 276.807769 & $-33.757442$ & 10.057 &  9.178 &  8.941 & 3749 & $+106.494$ & $-1.10$ & $-0.81$ & $+0.01$ &   \\
010 & 276.873765 & $-33.726830$ & 10.098 &  9.183 &  8.932 & 3637 & $-128.768$ & $-0.39$ & $-0.41$ & $+0.60$ &   \\
011 & 276.817925 & $-33.649303$ & 10.115 &  9.191 &  8.923 & 3578 & $ -59.147$ & $-0.76$ &         &         &   \\
012 & 276.611791 & $-33.631435$ & 10.170 &  9.330 &  9.094 & 3843 & $ +74.703$ & $-0.13$ & $+0.06$ & $+0.09$ &   \\
013 & 276.506457 & $-33.723846$ & 10.213 &  9.322 &  9.062 & 3681 & $ +27.664$ & $-0.06$ & $+0.02$ & $+0.29$ &   \\
015 & 276.488172 & $-33.844830$ & 10.232 &  9.400 &  9.158 & 3847 & $ +85.083$ & $-0.39$ & $-0.30$ & $+0.05$ &   \\
016 & 276.682337 & $-33.955391$ & 10.266 &  9.421 &  9.194 & 3845 & $ +25.704$ & $-0.27$ & $-0.06$ & $+0.08$ &   \\
017 & 276.595604 & $-33.694199$ & 10.406 &  9.574 &  9.312 & 3803 & $ +50.134$ & $-0.54$ & $-0.32$ & $+0.09$ &   \\
018 & 276.592533 & $-33.625336$ & 10.409 &  9.579 &  9.366 & 3914 & $ -59.013$ & $-0.70$ & $-1.07$ & $+0.55$ &   \\
019 & 276.586978 & $-33.867764$ & 10.470 &  9.639 &  9.438 & 3934 & $+178.014$ & $-0.85$ & $-1.15$ & $+0.32$ &   \\
020 & 276.704314 & $-33.794868$ & 10.542 &  9.678 &  9.446 & 3794 & $ +33.422$ & $+0.29$ &         &         & * \\
021 & 276.653399 & $-33.955818$ & 10.642 &  9.747 &  9.556 & 3815 & $ +72.019$ & $-0.56$ & $-0.31$ & $-0.04$ &   \\
022 & 276.471245 & $-33.828197$ & 10.657 &  9.769 &  9.520 & 3711 & $ -80.548$ & $+0.28$ &         &         & * \\
023 & 276.565931 & $-33.660713$ & 10.690 &  9.821 &  9.542 & 3687 & $ -42.688$ & $+0.12$ &         &         &   \\
024 & 276.667873 & $-33.869152$ & 10.726 &  9.949 &  9.659 & 3857 & $  -7.078$ & $-0.14$ & $+0.12$ & $+0.08$ &   \\
025 & 276.771326 & $-33.748791$ & 10.735 &  9.924 &  9.721 & 3969 & $-131.548$ & $-0.62$ & $-0.29$ & $+0.12$ &   \\
026 & 276.566599 & $-33.576355$ & 10.747 &  9.916 &  9.673 & 3849 & $ +15.044$ & $+0.02$ & $+0.12$ & $+0.17$ &   \\
027 & 276.621730 & $-33.925117$ & 10.779 &  9.943 &  9.708 & 3849 & $  -0.464$ & $+0.31$ & $+0.38$ &         &   \\
028 & 276.566492 & $-33.894100$ & 10.807 & 10.005 &  9.774 & 3932 & $ -78.865$ & $-0.55$ & $-0.19$ & $+0.12$ &   \\
029 & 276.632910 & $-33.850189$ & 10.843 & 10.032 &  9.847 & 4008 & $+305.574$ & $-0.58$ & $-0.87$ & $+0.23$ &   \\
030 & 276.876718 & $-33.749680$ & 10.911 & 10.121 &  9.909 & 3991 & $ +13.784$ & $-0.70$ & $-0.93$ & $+0.18$ &   \\
031 & 276.833702 & $-33.646965$ & 10.933 & 10.095 &  9.844 & 3808 & $ -12.983$ & $-0.68$ & $-0.33$ & $+0.05$ &   \\
032 & 276.664170 & $-33.889168$ & 11.002 & 10.132 &  9.895 & 3770 & $ +59.999$ & $+0.20$ &         &         &   \\
033 & 276.608510 & $-33.775723$ & 11.064 & 10.240 & 10.041 & 3953 & $ +46.289$ & $-0.25$ & $-0.02$ & $-0.15$ &   \\
034 & 276.530697 & $-33.808578$ & 11.077 & 10.243 & 10.056 & 3959 & $ +72.605$ & $-0.58$ & $-0.60$ & $+0.25$ &   \\
035 & 276.734398 & $-33.905167$ & 11.112 & 10.288 & 10.008 & 3775 & $ +37.840$ & $-0.17$ & $-0.06$ & $+0.16$ &   \\
036 & 276.693013 & $-33.893147$ & 11.119 & 10.247 & 10.060 & 3874 & $+110.060$ & $-0.54$ & $-0.34$ & $+0.04$ &   \\
037 & 276.455229 & $-33.789265$ & 11.126 & 10.329 & 10.132 & 4017 & $ +25.919$ & $-0.06$ & $-0.44$ & $+0.29$ &   \\
038 & 276.531016 & $-33.587563$ & 11.157 & 10.323 & 10.122 & 3933 & $-105.342$ & $-0.38$ & $-0.03$ & $+0.22$ &   \\
039 & 276.476724 & $-33.652061$ & 11.191 & 10.461 & 10.223 & 4074 & $ -99.668$ & $+0.03$ & $-0.15$ & $+0.13$ &   \\
040 & 276.632713 & $-33.608341$ & 11.204 & 10.397 & 10.238 & 4075 & $-213.279$ & $-0.82$ & $-1.06$ & $+0.38$ &   \\
041 & 276.611649 & $-33.623035$ & 11.212 & 10.433 & 10.228 & 4037 & $  +7.052$ & $-0.40$ & $-0.55$ & $+0.20$ &   \\
042 & 276.514543 & $-33.867790$ & 11.212 & 10.396 & 10.256 & 4096 & $  +3.906$ & $-0.65$ & $-0.80$ & $+0.27$ &   \\
043 & 276.545896 & $-33.910049$ & 11.236 & 10.429 & 10.287 & 4109 & $ -60.971$ & $-0.41$ & $-0.55$ & $+0.20$ &   \\
044 & 276.501519 & $-33.712631$ & 11.270 & 10.456 & 10.228 & 3917 & $ -66.401$ & $-0.08$ & $+0.18$ & $+0.10$ &   \\
045 & 276.497877 & $-33.881786$ & 11.378 & 10.535 & 10.362 & 3969 & $+240.911$ & $-1.11$ & $-1.12$ & $+0.19$ &   \\
046 & 276.469519 & $-33.651062$ & 11.403 & 10.619 & 10.362 & 3921 & $  -1.719$ & $-0.54$ & $-0.34$ & $+0.16$ &   \\
047 & 276.718276 & $-33.890141$ & 11.407 & 10.593 & 10.388 & 3957 & $ +29.551$ & $+0.05$ &         &         &   \\
048 & 276.593535 & $-33.598366$ & 11.415 & 10.584 & 10.345 & 3857 & $  -9.884$ & $+0.14$ & $+0.23$ & $+0.33$ &   \\
049 & 276.751404 & $-33.686954$ & 11.428 & 10.661 & 10.459 & 4065 & $ +81.687$ & $-0.06$ & $-0.37$ & $+0.21$ &   \\
050 & 276.668148 & $-33.844818$ & 11.433 & 10.578 & 10.353 & 3829 & $ +29.158$ & $-0.44$ & $-0.12$ & $-0.10$ &   \\
051 & 276.897946 & $-33.802887$ & 11.446 & 10.612 & 10.445 & 3992 & $ -53.729$ & $-0.83$ & $-1.01$ & $+0.26$ &   \\
052 & 276.827522 & $-33.672203$ & 11.457 & 10.650 & 10.435 & 3952 & $ -28.859$ & $+0.00$ &         &         &   \\
053 & 276.598353 & $-33.763283$ & 11.464 & 10.658 & 10.412 & 3893 & $ +16.198$ & $-0.02$ & $+0.20$ & $+0.09$ &   \\
054 & 276.753347 & $-33.748871$ & 11.480 & 10.655 & 10.504 & 4049 & $ +21.622$ & $-0.74$ & $-0.92$ & $+0.23$ &   \\
058 & 276.763338 & $-33.726757$ & 11.555 & 10.775 & 10.634 & 4167 & $-133.305$ & $-0.97$ & $-1.07$ & $+0.26$ &   \\
060 & 276.814847 & $-33.799526$ & 11.592 & 10.814 & 10.653 & 4126 & $ +19.204$ & $-0.64$ & $-0.79$ & $+0.18$ &   \\
061 & 276.745468 & $-33.614552$ & 11.610 & 10.791 & 10.615 & 4011 & $-206.421$ & $-0.56$ & $-0.75$ & $+0.24$ &   \\
065 & 276.767964 & $-33.723507$ & 11.682 & 10.895 & 10.664 & 3961 & $ -59.417$ & $+0.60$ &         &         &   \\
067 & 276.764757 & $-33.714439$ & 11.695 & 10.914 & 10.713 & 4036 & $ -42.029$ & $-0.33$ & $-0.38$ & $+0.13$ &   \\
069 & 276.732436 & $-33.691238$ & 11.750 & 10.953 & 10.767 & 4035 & $ +28.948$ & $-0.30$ & $-0.43$ & $+0.20$ &   \\
070 & 276.809016 & $-33.594540$ & 11.763 & 10.955 & 10.743 & 3957 & $ -54.899$ & $-0.19$ & $-0.20$ & $+0.00$ &   \\
072 & 276.594486 & $-33.730743$ & 11.791 & 11.111 & 10.937 & 4325 & $ -52.206$ & $-0.52$ & $-0.59$ & $+0.20$ &   \\
073 & 276.836739 & $-33.717419$ & 11.816 & 11.071 & 10.883 & 4139 & $ -56.535$ & $-0.28$ & $-0.39$ & $+0.16$ &   \\
075 & 276.910260 & $-33.767544$ & 11.837 & 11.034 & 10.873 & 4070 & $ +45.375$ & $-0.45$ & $-0.59$ & $+0.16$ & * \\
080 & 276.663676 & $-33.680828$ & 11.956 & 11.185 & 11.002 & 4099 & $ +67.611$ & $-0.63$ & $-0.80$ & $+0.21$ &   \\
081 & 276.598738 & $-33.673889$ & 11.964 & 11.282 & 11.096 & 4294 & $-155.581$ & $-0.12$ & $-0.25$ & $+0.23$ &   \\
084 & 276.723840 & $-33.562702$ & 11.989 & 11.339 & 11.120 & 4290 & $ +29.832$ & $+0.49$ & $+0.21$ & $+0.16$ &   \\
085 & 276.663499 & $-33.799038$ & 12.006 & 11.199 & 11.069 & 4134 & $ +58.329$ & $-0.41$ & $-0.51$ & $+0.15$ & * \\
086 & 276.810910 & $-33.617214$ & 12.009 & 11.269 & 11.096 & 4185 & $  +9.035$ & $+0.60$ & $+0.22$ & $+0.23$ &   \\
088 & 276.843442 & $-33.804676$ & 12.058 & 11.340 & 11.215 & 4343 & $ +84.524$ & $-0.42$ & $-0.54$ & $+0.12$ &   \\
089 & 276.708411 & $-33.645660$ & 12.060 & 11.331 & 11.179 & 4261 & $ -58.044$ & $-0.14$ & $-0.30$ & $+0.21$ &   \\
096 & 276.728929 & $-33.625156$ & 12.172 & 11.457 & 11.252 & 4171 & $ +38.482$ & $-0.48$ & $-0.69$ & $+0.21$ &   \\
101 & 276.650041 & $-33.653530$ & 12.257 & 11.553 & 11.355 & 4213 & $-123.588$ & $-0.24$ & $-0.34$ & $+0.14$ &   \\
102 & 276.603598 & $-33.725780$ & 12.263 & 11.594 & 11.421 & 4352 & $ -37.926$ & $-0.61$ & $-0.74$ & $+0.25$ &   \\
104 & 276.764294 & $-33.611069$ & 12.270 & 11.528 & 11.326 & 4119 & $ +23.454$ & $+0.06$ & $-0.21$ & $+0.19$ &   \\
107 & 276.796933 & $-33.752850$ & 12.304 & 11.596 & 11.481 & 4392 & $ +16.073$ & $-1.03$ & $-1.17$ & $+0.35$ &   \\
108 & 276.818554 & $-33.721001$ & 12.315 & 11.546 & 11.354 & 4080 & $ -55.377$ & $+0.16$ & $-0.05$ & $+0.12$ &   \\
110 & 276.831116 & $-33.661152$ & 12.343 & 11.625 & 11.495 & 4334 & $-109.396$ & $-0.97$ & $-1.11$ & $+0.32$ &   \\
112 & 276.844085 & $-33.719276$ & 12.386 & 11.681 & 11.584 & 4444 & $  -3.301$ & $-0.69$ & $-0.89$ & $+0.32$ &   \\
115 & 276.876631 & $-33.669006$ & 12.398 & 11.613 & 11.454 & 4115 & $-172.477$ & $-0.36$ & $-0.51$ & $+0.29$ &   \\
116 & 276.886129 & $-33.817890$ & 12.401 & 11.716 & 11.560 & 4346 & $ -54.145$ & $-0.99$ & $-1.08$ & $+0.24$ & * \\
117 & 276.839217 & $-33.735954$ & 12.407 & 11.626 & 11.535 & 4277 & $-141.307$ & $-0.92$ & $-1.00$ & $+0.27$ &   \\
118 & 276.605428 & $-33.607613$ & 12.438 & 11.750 & 11.607 & 4380 & $ +31.103$ & $-0.99$ & $-1.18$ & $+0.38$ &   \\
121 & 276.850407 & $-33.727764$ & 12.470 & 11.850 & 11.667 & 4441 & $  -2.194$ & $-0.10$ & $-0.32$ & $+0.28$ &   \\
123 & 276.733228 & $-33.686802$ & 12.482 & 11.722 & 11.609 & 4278 & $ -85.450$ & $-0.76$ & $-0.82$ & $+0.23$ &   \\
128 & 276.674814 & $-33.794342$ & 12.520 & 11.823 & 11.717 & 4446 & $ +32.514$ & $-0.70$ & $-0.81$ & $+0.20$ &   \\
131 & 276.862902 & $-33.842964$ & 12.596 & 11.871 & 11.725 & 4277 & $ -28.810$ & $-0.37$ & $-0.55$ & $+0.22$ &   \\
132 & 276.529376 & $-33.635666$ & 12.607 & 11.953 & 11.827 & 4514 & $-159.866$ & $-0.74$ & $-1.06$ & $+0.49$ &   \\
133 & 276.754996 & $-33.640835$ & 12.630 & 11.933 & 11.784 & 4341 & $+136.734$ & $-0.22$ & $-0.46$ & $+0.20$ & * \\
136 & 276.670396 & $-33.678337$ & 12.694 & 11.991 & 11.848 & 4342 & $ -16.510$ & $-0.50$ & $-0.50$ & $+0.21$ &   \\
139 & 276.669257 & $-33.560745$ & 12.724 & 12.024 & 11.849 & 4278 & $  -8.143$ & $+0.06$ & $-0.16$ & $+0.30$ &   \\
142 & 276.592785 & $-33.609268$ & 12.796 & 12.171 & 11.994 & 4454 & $-120.627$ &         &         &         &   \\
143 & 276.720586 & $-33.791100$ & 12.801 & 12.219 & 11.977 & 4391 & $ -28.694$ &         &         &         & * \\
144 & 276.724865 & $-33.593685$ & 12.806 & 12.156 & 12.024 & 4504 & $ +47.662$ &         &         &         &   \\
147 & 276.605327 & $-33.599644$ & 12.834 & 12.130 & 12.014 & 4407 & $ -13.128$ &         &         &         &   \\
148 & 276.808190 & $-33.749828$ & 12.835 & 12.127 & 11.988 & 4335 & $ -40.138$ & $-0.14$ & $-0.36$ & $+0.08$ &   \\
154 & 276.884497 & $-33.711712$ & 12.903 & 12.231 & 12.130 & 4521 & $ -74.863$ & $-0.00$ & $-0.14$ & $+0.09$ &   \\
156 & 276.802245 & $-33.615543$ & 12.915 & 12.209 & 12.115 & 4453 & $  +5.379$ & $+0.17$ & $-0.04$ & $+0.20$ & * \\
158 & 276.607949 & $-33.626446$ & 12.936 & 12.308 & 12.179 & 4574 & $ +47.731$ & $-0.07$ & $-0.12$ & $+0.20$ &   \\
159 & 276.506642 & $-33.859463$ & 12.947 & 12.320 & 12.168 & 4514 & $ -67.344$ & $+0.55$ & $+0.32$ & $+0.18$ &   \\
167 & 276.696311 & $-33.687943$ & 12.992 & 12.354 & 12.170 & 4398 & $ +16.036$ & $-0.27$ & $-0.64$ & $+0.16$ &   \\
169 & 276.585971 & $-33.583626$ & 13.016 & 12.356 & 12.206 & 4434 & $ -21.109$ & $+0.02$ & $-0.37$ & $+0.28$ &   \\
172 & 276.674085 & $-33.722809$ & 13.036 & 12.399 & 12.232 & 4444 & $ +62.310$ & $+0.06$ & $-0.09$ & $+0.18$ &   \\
178 & 276.673132 & $-33.740860$ & 13.069 & 12.451 & 12.242 & 4385 & $ -14.891$ & $+0.46$ & $+0.23$ & $+0.18$ &   \\
180 & 276.715472 & $-33.633411$ & 13.078 & 12.449 & 12.266 & 4424 & $ +57.926$ & $+0.65$ & $+0.39$ & $+0.19$ & * \\
182 & 276.772886 & $-33.712326$ & 13.085 & 12.412 & 12.316 & 4535 & $ -44.163$ & $-0.87$ & $-1.00$ & $+0.30$ &   \\
183 & 276.791120 & $-33.629635$ & 13.088 & 12.478 & 12.337 & 4584 & $ -74.273$ & $+0.39$ & $+0.18$ & $+0.17$ &   \\
184 & 276.841313 & $-33.728474$ & 13.093 & 12.437 & 12.271 & 4393 & $ +13.471$ & $+0.64$ & $+0.37$ & $+0.15$ &   \\
186 & 276.721044 & $-33.780018$ & 13.110 & 12.489 & 12.371 & 4618 & $ +29.773$ & $-0.70$ & $-0.80$ & $+0.24$ &   \\
187 & 276.533345 & $-33.873089$ & 13.112 & 12.435 & 12.314 & 4462 & $-142.788$ & $+0.34$ & $+0.12$ & $+0.23$ &   \\
188 & 276.753476 & $-33.945801$ & 13.119 & 12.543 & 12.370 & 4585 & $ +19.007$ & $-0.20$ & $-0.41$ & $+0.18$ &   \\
190 & 276.538801 & $-33.644936$ & 13.134 & 12.484 & 12.360 & 4530 & $ +11.192$ & $+0.47$ & $+0.23$ & $+0.24$ &   \\
191 & 276.627097 & $-33.841202$ & 13.146 & 12.522 & 12.389 & 4569 & $-175.576$ & $-1.05$ & $-1.11$ & $+0.32$ &   \\
192 & 276.790756 & $-33.846645$ & 13.151 & 12.493 & 12.363 & 4481 & $  -8.106$ & $-0.67$ & $-0.83$ & $+0.22$ &   \\
193 & 276.717695 & $-33.835640$ & 13.153 & 12.548 & 12.410 & 4605 & $ +56.213$ & $+0.15$ & $-0.05$ & $+0.17$ &   \\
194 & 276.864580 & $-33.723576$ & 13.154 & 12.547 & 12.417 & 4620 & $ -33.430$ & $-0.59$ & $-0.74$ & $+0.24$ &   \\
195 & 276.694217 & $-33.601109$ & 13.163 & 12.508 & 12.393 & 4537 & $  -3.184$ & $-0.68$ & $-0.77$ & $+0.30$ &   \\
196 & 276.841761 & $-33.689163$ & 13.167 & 12.539 & 12.385 & 4499 & $ -44.337$ & $-0.37$ & $-0.52$ & $+0.20$ & * \\
197 & 276.749744 & $-33.856068$ & 13.171 & 12.572 & 12.406 & 4544 & $ -22.102$ & $-0.16$ & $-0.33$ & $+0.10$ &   \\
198 & 276.638867 & $-33.930489$ & 13.179 & 12.505 & 12.351 & 4381 & $ +89.036$ & $+0.23$ & $+0.02$ & $+0.08$ &   \\
199 & 276.601595 & $-33.643906$ & 13.190 & 12.594 & 12.394 & 4469 & $ +55.281$ & $+0.62$ & $+0.39$ & $+0.19$ &   \\
200 & 276.910018 & $-33.774948$ & 13.190 & 12.585 & 12.420 & 4527 & $ -23.265$ & $+0.20$ & $-0.01$ & $+0.20$ &   \\
201 & 276.655418 & $-33.667728$ & 13.192 & 12.617 & 12.458 & 4637 & $ -12.106$ & $+0.58$ & $+0.34$ & $+0.21$ &   \\
202 & 276.600953 & $-33.920620$ & 13.193 & 12.561 & 12.403 & 4480 & $ -72.007$ & $-0.65$ & $-0.77$ & $+0.14$ &   \\
203 & 276.718361 & $-33.917198$ & 13.194 & 12.505 & 12.362 & 4370 & $-112.531$ & $-0.35$ & $-0.50$ & $+0.19$ &   \\
204 & 276.651481 & $-33.734238$ & 13.196 & 12.496 & 12.354 & 4351 & $ -44.013$ & $-0.36$ & $-0.49$ & $+0.17$ &   \\
205 & 276.778831 & $-33.720421$ & 13.203 & 12.613 & 12.452 & 4583 & $ -66.094$ & $-0.23$ & $-0.32$ & $+0.08$ &   \\
206 & 276.529719 & $-33.594994$ & 13.210 & 12.613 & 12.466 & 4613 & $ +26.825$ & $+0.69$ & $+0.46$ & $+0.14$ &   \\
207 & 276.854131 & $-33.731907$ & 13.221 & 12.595 & 12.447 & 4519 & $ +62.408$ & $-0.78$ & $-0.97$ & $+0.31$ &   \\
208 & 276.628729 & $-33.743752$ & 13.226 & 12.621 & 12.494 & 4642 & $ -39.333$ & $+0.24$ & $+0.09$ & $+0.08$ &   \\
209 & 276.798530 & $-33.662918$ & 13.233 & 12.540 & 12.415 & 4405 & $ +95.619$ & $-0.75$ & $-0.86$ & $+0.21$ &   \\
210 & 276.836967 & $-33.769405$ & 13.243 & 12.643 & 12.456 & 4484 & $ +21.124$ & $+0.26$ & $+0.11$ & $+0.11$ &   \\
211 & 276.756279 & $-33.788357$ & 13.248 & 12.575 & 12.430 & 4404 & $ -34.348$ & $-0.54$ & $-0.70$ & $+0.19$ &   \\
212 & 276.892143 & $-33.753590$ & 13.248 & 12.670 & 12.487 & 4552 & $ +21.790$ & $+0.12$ & $-0.05$ & $+0.10$ & * \\
215 & 276.537753 & $-33.603622$ & 13.256 & 12.649 & 12.491 & 4555 & $-119.873$ & $-1.49$ &         &         & * \\
216 & 276.630558 & $-33.724087$ & 13.258 & 12.652 & 12.484 & 4526 & $ +38.871$ & $+0.36$ & $+0.18$ & $+0.10$ &   \\
217 & 276.611185 & $-33.719536$ & 13.258 & 12.650 & 12.506 & 4586 & $ +53.834$ & $+0.62$ & $+0.37$ & $+0.21$ &   \\
218 & 276.533293 & $-33.793808$ & 13.263 & 12.630 & 12.449 & 4421 & $ +25.411$ & $+0.58$ & $+0.33$ & $+0.16$ &   \\
219 & 276.490017 & $-33.791950$ & 13.267 & 12.645 & 12.524 & 4614 & $ +70.196$ & $-0.68$ & $-0.79$ & $+0.29$ &   \\
220 & 276.786346 & $-33.598869$ & 13.267 & 12.685 & 12.496 & 4531 & $ +36.370$ & $+0.40$ & $+0.20$ & $+0.18$ & * \\
221 & 276.488605 & $-33.676846$ & 13.277 & 12.662 & 12.508 & 4544 & $ -68.765$ & $-0.96$ & $-0.98$ & $+0.34$ &   \\
222 & 276.765113 & $-33.566185$ & 13.280 & 12.582 & 12.447 & 4371 & $ +89.637$ & $-0.84$ & $-0.94$ & $+0.11$ &   \\
223 & 276.689571 & $-33.850536$ & 13.283 & 12.596 & 12.483 & 4452 & $ +63.721$ & $+0.19$ & $+0.06$ & $+0.12$ &   \\
224 & 276.530160 & $-33.772533$ & 13.283 & 12.630 & 12.546 & 4630 & $-159.749$ & $+0.10$ & $-0.06$ & $+0.12$ & * \\
225 & 276.490013 & $-33.854095$ & 13.287 & 12.654 & 12.528 & 4568 & $  -2.227$ & $+0.47$ & $+0.28$ & $+0.10$ &   \\
226 & 276.441487 & $-33.833378$ & 13.287 & 12.601 & 12.440 & 4344 & $ +86.733$ & $-0.55$ & $-0.71$ & $+0.21$ &   \\
227 & 276.460830 & $-33.865948$ & 13.295 & 12.693 & 12.545 & 4593 & $ -30.132$ & $+0.42$ & $+0.20$ & $+0.12$ &   \\
228 & 276.567606 & $-33.881664$ & 13.305 & 12.729 & 12.573 & 4641 & $  +5.293$ & $+0.17$ & $-0.05$ & $+0.12$ &   \\
229 & 276.488238 & $-33.744823$ & 13.308 & 12.659 & 12.507 & 4457 & $ -93.491$ & $-0.90$ & $-0.89$ & $+0.02$ &   \\
230 & 276.701957 & $-33.724052$ & 13.317 & 12.670 & 12.492 & 4390 & $ -15.167$ & $+0.44$ & $+0.21$ & $+0.17$ & * \\
231 & 276.521504 & $-33.882675$ & 13.318 & 12.634 & 12.517 & 4454 & $ +87.725$ & $-0.85$ & $-0.94$ & $+0.20$ &   \\
232 & 276.701163 & $-33.812206$ & 13.321 & 12.791 & 12.593 & 4650 & $ +46.717$ & $-0.24$ & $-0.39$ & $+0.16$ &   \\
233 & 276.742927 & $-33.655430$ & 13.327 & 12.768 & 12.577 & 4588 & $ -33.817$ & $+0.42$ & $+0.20$ & $+0.20$ &   \\
234 & 276.817331 & $-33.599476$ & 13.328 & 12.681 & 12.582 & 4598 & $  +3.010$ & $-0.17$ & $-0.34$ & $+0.06$ &   \\
235 & 276.561797 & $-33.754608$ & 13.331 & 12.760 & 12.590 & 4618 & $  +2.381$ & $+0.36$ & $+0.18$ & $+0.14$ &   \\
236 & 276.811041 & $-33.747356$ & 13.339 & 12.698 & 12.552 & 4485 & $ -36.405$ & $-0.64$ & $-0.84$ & $+0.42$ & * \\
237 & 276.543824 & $-33.638008$ & 13.349 & 12.771 & 12.601 & 4601 & $-128.142$ & $-1.17$ & $-1.26$ & $+0.35$ &   \\
238 & 276.707179 & $-33.863255$ & 13.363 & 12.721 & 12.560 & 4443 & $ -48.485$ & $+0.26$ & $+0.09$ & $+0.15$ &   \\
239 & 276.514484 & $-33.893902$ & 13.367 & 12.706 & 12.596 & 4534 & $  -6.349$ & $-0.66$ & $-0.74$ & $+0.18$ &   \\
240 & 276.494178 & $-33.865170$ & 13.374 & 12.671 & 12.545 & 4384 & $ -45.410$ & $-0.45$ & $-0.61$ & $+0.22$ &   \\
241 & 276.691228 & $-33.652210$ & 13.381 & 12.740 & 12.621 & 4563 & $+163.130$ & $-0.75$ & $-0.82$ & $+0.21$ &   \\
242 & 276.769980 & $-33.622349$ & 13.400 & 12.785 & 12.587 & 4420 & $ +37.049$ & $+0.28$ & $+0.14$ & $+0.08$ &   \\
243 & 276.445539 & $-33.820004$ & 13.401 & 12.802 & 12.662 & 4626 & $-201.762$ & $-0.51$ & $-0.68$ & $+0.34$ & * \\
245 & 276.458400 & $-33.680145$ & 13.410 & 12.870 & 12.674 & 4637 & $  -0.726$ &         &         &         &   \\
246 & 276.732205 & $-33.735016$ & 13.413 & 12.722 & 12.633 & 4506 & $ -20.965$ & $-0.26$ & $-0.40$ & $+0.19$ &   \\
247 & 276.652760 & $-33.666142$ & 13.415 & 12.859 & 12.647 & 4542 & $ -12.734$ & $+0.63$ & $+0.34$ & $+0.22$ &   \\
248 & 276.663722 & $-33.563961$ & 13.423 & 12.782 & 12.703 & 4680 & $ -91.605$ & $-0.82$ & $-0.94$ & $+0.39$ &   \\
249 & 276.874535 & $-33.706005$ & 13.432 & 12.805 & 12.675 & 4564 & $ +20.002$ & $+0.22$ & $+0.05$ & $+0.05$ &   \\
251 & 276.814218 & $-33.891129$ & 13.436 & 12.813 & 12.703 & 4630 & $-100.233$ & $-0.18$ & $-0.29$ & $+0.09$ &   \\
252 & 276.590433 & $-33.824047$ & 13.447 & 12.815 & 12.625 & 4399 & $ +14.280$ & $-0.16$ & $-0.37$ & $+0.14$ &   \\
253 & 276.698884 & $-33.720577$ & 13.450 & 12.802 & 12.707 & 4608 & $ -12.354$ & $+0.28$ & $+0.14$ & $+0.12$ &   \\
254 & 276.560443 & $-33.786304$ & 13.457 & 12.905 & 12.659 & 4463 & $ +47.473$ & $+0.37$ & $+0.19$ & $+0.15$ &   \\
256 & 276.885368 & $-33.847267$ & 13.459 & 12.843 & 12.729 & 4637 & $ +47.665$ & $-0.61$ & $-0.74$ & $+0.28$ &   \\
257 & 276.673226 & $-33.739399$ & 13.464 & 12.896 & 12.744 & 4676 & $ +53.227$ & $+0.51$ & $+0.30$ & $+0.21$ &   \\
258 & 276.509126 & $-33.638287$ & 13.466 & 12.795 & 12.648 & 4414 & $ +20.249$ & $+0.22$ & $+0.01$ & $+0.25$ &   \\
259 & 276.903426 & $-33.794418$ & 13.475 & 12.812 & 12.641 & 4362 & $  +0.338$ & $+0.09$ & $-0.07$ & $+0.07$ &   \\
260 & 276.856050 & $-33.636570$ & 13.477 & 12.791 & 12.647 & 4375 & $ -99.944$ & $-0.31$ & $-0.39$ & $+0.09$ &   \\
261 & 276.597652 & $-33.706051$ & 13.478 & 12.889 & 12.705 & 4530 & $ -56.387$ & $+0.25$ & $+0.06$ & $+0.14$ &   \\
262 & 276.627210 & $-33.776421$ & 13.481 & 12.854 & 12.717 & 4552 & $ -31.979$ & $-0.74$ & $-0.91$ & $+0.47$ &   \\
263 & 276.659752 & $-33.951805$ & 13.489 & 12.884 & 12.727 & 4553 & $  -5.578$ & $+0.04$ & $-0.12$ & $+0.09$ &   \\
264 & 276.545369 & $-33.932869$ & 13.493 & 12.880 & 12.781 & 4701 & $ -61.467$ & $-0.25$ & $-0.38$ & $+0.17$ &   \\
266 & 276.533760 & $-33.787659$ & 13.505 & 12.874 & 12.744 & 4562 & $ -68.790$ & $-0.75$ & $-0.83$ & $+0.21$ &   \\
267 & 276.616984 & $-33.586750$ & 13.506 & 12.954 & 12.792 & 4699 & $-101.202$ & $+0.59$ & $+0.35$ & $+0.18$ &   \\
268 & 276.494876 & $-33.647800$ & 13.515 & 12.919 & 12.794 & 4681 & $ +71.564$ & $+0.08$ & $-0.07$ & $+0.18$ & * \\
269 & 276.764459 & $-33.589680$ & 13.518 & 12.940 & 12.784 & 4635 & $ +69.289$ & $-0.04$ & $-0.15$ & $+0.14$ &   \\
270 & 276.614926 & $-33.938892$ & 13.526 & 12.948 & 12.808 & 4680 & $ +65.396$ & $-0.69$ & $-0.78$ & $+0.23$ &   \\
271 & 276.465499 & $-33.869881$ & 13.526 & 12.857 & 12.692 & 4372 & $ +29.413$ & $+0.66$ & $+0.39$ & $+0.06$ &   \\
272 & 276.827432 & $-33.753338$ & 13.529 & 12.833 & 12.770 & 4559 & $-104.732$ & $-1.17$ & $-1.37$ & $+0.38$ &   \\
273 & 276.785331 & $-33.659843$ & 13.538 & 12.873 & 12.710 & 4381 & $+223.568$ & $-0.55$ & $-0.66$ & $+0.12$ &   \\
274 & 276.842104 & $-33.732456$ & 13.541 & 12.852 & 12.714 & 4381 & $ -21.443$ & $-0.28$ & $-0.51$ & $+0.10$ &   \\
275 & 276.765440 & $-33.581825$ & 13.554 & 12.999 & 12.817 & 4626 & $ +68.660$ & $-0.57$ & $-0.67$ & $+0.20$ &   \\
276 & 276.728640 & $-33.949757$ & 13.557 & 12.951 & 12.756 & 4446 & $-121.711$ & $-0.71$ & $-0.85$ & $+0.22$ &   \\
277 & 276.505794 & $-33.774723$ & 13.561 & 12.941 & 12.782 & 4515 & $ -51.159$ & $+0.33$ & $+0.05$ & $+0.16$ &   \\
278 & 276.714377 & $-33.724106$ & 13.581 & 13.007 & 12.860 & 4672 & $ +16.094$ & $+0.29$ & $+0.09$ & $+0.12$ &   \\
279 & 276.619384 & $-33.610786$ & 13.591 & 12.968 & 12.842 & 4596 & $ +50.290$ & $-0.65$ & $-0.73$ & $+0.19$ &   \\
280 & 276.699742 & $-33.801800$ & 13.600 & 13.064 & 12.896 & 4722 & $ -22.286$ & $-0.43$ & $-0.62$ & $+0.37$ &   \\
281 & 276.632739 & $-33.920406$ & 13.601 & 12.931 & 12.843 & 4565 & $ -74.670$ & $+0.19$ & $-0.02$ & $+0.09$ &   \\
282 & 276.603736 & $-33.792931$ & 13.606 & 13.010 & 12.886 & 4678 & $ -59.064$ & $-0.70$ & $-0.79$ & $+0.23$ &   \\
283 & 276.839527 & $-33.652874$ & 13.614 & 12.951 & 12.909 & 4717 & $ -47.148$ & $-0.57$ & $-0.73$ & $+0.36$ & * \\
284 & 276.724473 & $-33.882050$ & 13.619 & 12.995 & 12.858 & 4555 & $  -6.884$ & $-0.38$ & $-0.58$ & $+0.19$ &   \\
285 & 276.810256 & $-33.899017$ & 13.632 & 12.985 & 12.875 & 4562 & $ +24.234$ & $-0.80$ & $-0.90$ & $+0.25$ & * \\
286 & 276.629095 & $-33.587997$ & 13.646 & 13.024 & 12.914 & 4645 & $-104.407$ & $-0.89$ & $-0.93$ & $+0.21$ &   \\
287 & 276.860156 & $-33.865948$ & 13.646 & 13.033 & 12.895 & 4578 & $ +75.530$ & $-0.57$ & $-0.64$ & $+0.19$ &   \\
288 & 276.589341 & $-33.604183$ & 13.659 & 13.078 & 12.939 & 4682 & $ -63.114$ & $-0.35$ & $-0.48$ & $+0.17$ &   \\
291 & 276.787583 & $-33.871380$ & 13.673 & 13.065 & 12.954 & 4673 & $ -22.181$ & $-0.53$ & $-0.64$ & $+0.14$ &   \\
292 & 276.501441 & $-33.769341$ & 13.689 & 13.075 & 12.971 & 4688 & $ -60.436$ & $-0.64$ & $-0.75$ & $+0.29$ &   \\
294 & 276.807204 & $-33.744389$ & 13.705 & 13.002 & 12.912 & 4469 & $ -92.853$ & $+0.49$ & $+0.14$ & $-0.03$ &   \\
295 & 276.671937 & $-33.722317$ & 13.712 & 13.142 & 13.001 & 4704 & $  -7.833$ & $+0.54$ & $+0.32$ & $+0.19$ &   \\
296 & 276.472514 & $-33.777775$ & 13.719 & 13.094 & 12.963 & 4578 & $+121.918$ & $-0.44$ & $-0.50$ & $+0.19$ &   \\
297 & 276.686566 & $-33.655930$ & 13.744 & 13.148 & 13.021 & 4669 & $-110.334$ & $-0.88$ & $-0.92$ & $+0.23$ &   \\
298 & 276.665050 & $-33.818905$ & 13.746 & 13.132 & 13.029 & 4684 & $ +87.153$ & $-0.58$ & $-0.67$ & $+0.27$ &   \\
299 & 276.859868 & $-33.871136$ & 13.748 & 13.085 & 12.995 & 4572 & $-105.991$ & $-0.39$ & $-0.50$ & $+0.15$ &   \\
300 & 276.442637 & $-33.752930$ & 13.750 & 13.103 & 13.048 & 4737 & $ -23.401$ &         &         &         & * \\
301 & 276.695536 & $-33.764545$ & 13.751 & 13.160 & 13.029 & 4669 & $ -68.258$ & $-1.46$ &         &         &   \\
302 & 276.749584 & $-33.841690$ & 13.752 & 13.130 & 13.059 & 4752 & $-165.133$ & $-0.54$ & $-0.60$ & $+0.19$ &   \\
304 & 276.677892 & $-33.890118$ & 13.758 & 13.194 & 12.984 & 4521 & $ +54.041$ & $-0.37$ & $-0.50$ & $+0.05$ &   \\
305 & 276.676672 & $-33.567745$ & 13.771 & 13.127 & 13.063 & 4715 & $ -54.614$ & $-1.62$ &         &         &   \\
306 & 276.686629 & $-33.707558$ & 13.777 & 13.146 & 13.024 & 4581 & $ -57.854$ & $-1.04$ & $-1.13$ & $+0.28$ &   \\
307 & 276.712714 & $-33.944599$ & 13.782 & 13.237 & 13.092 & 4760 & $ -73.813$ & $-0.01$ & $-0.17$ & $+0.12$ & * \\
308 & 276.542295 & $-33.749485$ & 13.790 & 13.240 & 13.092 & 4746 & $-103.250$ & $-1.28$ & $-1.31$ & $+0.35$ &   \\
309 & 276.845057 & $-33.643093$ & 13.797 & 13.196 & 13.031 & 4542 & $  -7.766$ & $-0.05$ & $-0.20$ & $+0.11$ &   \\
310 & 276.737301 & $-33.623081$ & 13.801 & 13.204 & 13.097 & 4724 & $ +68.140$ & $-0.59$ & $-0.78$ & $+0.28$ &   \\
311 & 276.855065 & $-33.892948$ & 13.813 & 13.230 & 13.106 & 4706 & $ -10.062$ & $-0.57$ & $-0.63$ & $+0.18$ &   \\
312 & 276.805007 & $-33.749069$ & 13.817 & 13.247 & 13.105 & 4695 & $ -18.932$ & $-1.12$ & $-1.21$ & $+0.29$ &   \\
313 & 276.803927 & $-33.834248$ & 13.821 & 13.261 & 13.126 & 4744 & $ +29.782$ & $-0.51$ & $-0.65$ & $+0.25$ & * \\
314 & 276.622601 & $-33.794971$ & 13.822 & 13.150 & 13.024 & 4460 & $ -60.428$ & $+0.05$ & $-0.20$ & $+0.07$ &   \\
315 & 276.821083 & $-33.773975$ & 13.822 & 13.218 & 13.084 & 4618 & $ +96.113$ & $-0.59$ & $-0.72$ & $+0.31$ & * \\
316 & 276.689193 & $-33.710064$ & 13.845 & 13.207 & 13.090 & 4575 & $  +0.896$ & $-0.07$ & $-0.19$ & $+0.10$ & * \\
317 & 276.527656 & $-33.798317$ & 13.846 & 13.287 & 13.138 & 4716 & $ -20.465$ & $+0.31$ & $+0.16$ & $+0.13$ &   \\
318 & 276.605035 & $-33.901566$ & 13.846 & 13.198 & 13.094 & 4583 & $  -8.413$ & $+0.20$ & $+0.06$ & $+0.10$ &   \\
319 & 276.676819 & $-33.581738$ & 13.852 & 13.225 & 13.053 & 4460 & $ -58.955$ & $+0.02$ & $-0.20$ & $+0.16$ &   \\
320 & 276.826131 & $-33.769146$ & 13.853 & 13.137 & 13.051 & 4444 & $ -41.389$ & $-0.48$ & $-0.62$ & $+0.12$ & * \\
321 & 276.547129 & $-33.646622$ & 13.857 & 13.271 & 13.155 & 4736 & $ +11.187$ & $+0.17$ & $+0.05$ & $+0.12$ &   \\
322 & 276.625298 & $-33.608067$ & 13.858 & 13.294 & 13.140 & 4687 & $ -54.255$ & $+0.13$ & $-0.01$ & $+0.09$ &   \\
323 & 276.850420 & $-33.753113$ & 13.859 & 13.172 & 13.081 & 4508 & $ -52.017$ & $-0.66$ & $-0.78$ & $+0.04$ & * \\
324 & 276.648919 & $-33.838467$ & 13.868 & 13.295 & 13.114 & 4577 & $ +29.848$ & $+0.54$ & $+0.30$ & $+0.08$ & * \\
325 & 276.887809 & $-33.831329$ & 13.871 & 13.297 & 13.117 & 4569 & $ -84.110$ & $-0.59$ & $-0.69$ & $+0.19$ & * \\
326 & 276.520313 & $-33.606983$ & 13.871 & 13.248 & 13.174 & 4753 & $ -13.705$ & $-0.46$ & $-0.47$ & $+0.43$ &   \\
327 & 276.743296 & $-33.606194$ & 13.874 & 13.244 & 13.065 & 4431 & $ +20.670$ & $-0.05$ & $-0.23$ & $-0.16$ &   \\
328 & 276.748212 & $-33.826157$ & 13.875 & 13.305 & 13.185 & 4761 & $ -29.200$ & $-0.47$ & $-0.55$ & $+0.15$ &   \\
329 & 276.494197 & $-33.811489$ & 13.876 & 13.340 & 13.159 & 4690 & $ +10.821$ & $-0.72$ & $-0.83$ & $+0.28$ &   \\
330 & 276.905290 & $-33.690044$ & 13.879 & 13.243 & 13.114 & 4542 & $ -21.543$ & $-0.51$ & $-0.64$ & $+0.32$ &   \\
331 & 276.661029 & $-33.805126$ & 13.882 & 13.245 & 13.169 & 4696 & $ -85.801$ & $-0.24$ & $-0.37$ & $+0.19$ &   \\
332 & 276.725178 & $-33.922646$ & 13.894 & 13.301 & 13.118 & 4514 & $ +74.659$ & $-0.84$ & $-0.94$ & $+0.18$ &   \\
333 & 276.803651 & $-33.785221$ & 13.895 & 13.274 & 13.161 & 4630 & $-160.010$ & $-0.93$ & $-1.08$ & $+0.32$ &   \\
334 & 276.573211 & $-33.613113$ & 13.897 & 13.384 & 13.189 & 4718 & $ +30.030$ & $-0.77$ & $-0.72$ & $+0.43$ & * \\
335 & 276.754422 & $-33.812862$ & 13.899 & 13.284 & 13.177 & 4666 & $  -3.150$ & $-0.25$ & $-0.39$ & $+0.18$ &   \\
336 & 276.775634 & $-33.868172$ & 13.902 & 13.205 & 13.179 & 4661 & $ -19.149$ & $-0.55$ & $-0.68$ & $+0.35$ & * \\
337 & 276.609719 & $-33.899956$ & 13.910 & 13.284 & 13.186 & 4663 & $ +86.522$ & $-0.65$ & $-0.76$ & $+0.25$ &   \\
338 & 276.802770 & $-33.607288$ & 13.910 & 13.270 & 13.182 & 4651 & $ -31.855$ & $+0.26$ & $+0.02$ & $+0.14$ & * \\
339 & 276.599163 & $-33.751358$ & 13.916 & 13.369 & 13.197 & 4682 & $ +21.543$ & $+0.52$ & $+0.31$ & $+0.17$ &   \\
340 & 276.862496 & $-33.881008$ & 13.916 & 13.287 & 13.203 & 4688 & $  -7.419$ & $+0.15$ & $+0.02$ & $+0.10$ &   \\
342 & 276.619675 & $-33.905045$ & 13.925 & 13.343 & 13.216 & 4708 & $+123.234$ & $-0.91$ & $-1.02$ & $+0.25$ &   \\
343 & 276.694856 & $-33.562214$ & 13.926 & 13.257 & 13.149 & 4519 & $ +69.723$ &         &         &         &   \\
344 & 276.586781 & $-33.832466$ & 13.931 & 13.301 & 13.158 & 4528 & $+214.266$ & $-0.81$ & $-0.93$ & $+0.26$ &   \\
347 & 276.736184 & $-33.875690$ & 13.949 & 13.316 & 13.213 & 4624 & $ -45.041$ & $-0.06$ & $-0.25$ & $+0.15$ &   \\
348 & 276.578552 & $-33.619892$ & 13.950 & 13.352 & 13.202 & 4600 & $  +5.935$ & $+0.41$ & $+0.21$ & $+0.15$ &   \\
349 & 276.737919 & $-33.695145$ & 13.956 & 13.350 & 13.207 & 4590 & $ -65.841$ & $+0.70$ & $+0.45$ & $+0.20$ &   \\
350 & 276.456610 & $-33.796585$ & 13.957 & 13.347 & 13.225 & 4647 & $-120.618$ & $-0.51$ & $-0.64$ & $+0.26$ &   \\
351 & 276.564966 & $-33.635128$ & 13.960 & 13.423 & 13.248 & 4706 & $+128.623$ & $-0.51$ & $-0.71$ & $+0.29$ &   \\
353 & 276.861593 & $-33.801018$ & 13.974 & 13.438 & 13.274 & 4728 & $ +22.706$ & $-0.73$ & $-0.83$ & $+0.29$ &   \\
354 & 276.722013 & $-33.592499$ & 13.974 & 13.376 & 13.238 & 4630 & $ +46.922$ & $-0.91$ & $-1.07$ & $+0.23$ &   \\
355 & 276.749304 & $-33.665966$ & 13.975 & 13.357 & 13.210 & 4547 & $ -58.417$ & $-0.44$ & $-0.53$ & $+0.17$ &   \\
356 & 276.762247 & $-33.786091$ & 13.982 & 13.460 & 13.272 & 4702 & $ +20.393$ & $-0.41$ & $-0.52$ & $+0.26$ &   \\
357 & 276.785991 & $-33.896404$ & 13.985 & 13.341 & 13.224 & 4553 & $ -84.393$ & $-0.71$ & $-0.82$ & $+0.23$ &   \\
358 & 276.833052 & $-33.776608$ & 13.989 & 13.368 & 13.235 & 4572 & $  -0.947$ & $-0.70$ & $-0.88$ & $+0.28$ &   \\
359 & 276.433122 & $-33.740566$ & 13.989 & 13.411 & 13.292 & 4753 & $ -32.267$ &         &         &         &   \\
360 & 276.691675 & $-33.846764$ & 13.993 & 13.347 & 13.275 & 4679 & $ -38.991$ & $-0.64$ & $-0.80$ & $+0.30$ &   \\
361 & 276.705482 & $-33.740154$ & 13.996 & 13.411 & 13.280 & 4687 & $ +37.993$ & $-0.65$ & $-0.71$ & $+0.21$ &   \\
362 & 276.617111 & $-33.760021$ & 13.996 & 13.410 & 13.278 & 4684 & $  +7.110$ & $-0.37$ & $-0.54$ & $+0.23$ &   \\
363 & 276.493794 & $-33.799362$ & 13.996 & 13.417 & 13.299 & 4750 & $ +30.066$ & $-0.89$ & $-1.01$ & $+0.31$ &   \\
364 & 276.697108 & $-33.596195$ & 13.996 & 13.460 & 13.297 & 4741 & $ +49.761$ & $-1.39$ & $-1.41$ & $+0.28$ &   \\
365 & 276.575708 & $-33.577957$ & 14.000 & 13.428 & 13.271 & 4656 & $  +0.930$ & $-0.85$ & $-0.94$ & $+0.31$ &   \\
366 & 276.663677 & $-33.647522$ & 14.002 & 13.400 & 13.306 & 4750 & $ +12.679$ & $-0.34$ & $-0.45$ & $+0.22$ &   \\
367 & 276.779362 & $-33.567799$ & 14.002 & 13.382 & 13.217 & 4495 & $+188.004$ & $-0.16$ & $-0.24$ & $+0.00$ &   \\
368 & 276.871279 & $-33.792786$ & 14.006 & 13.428 & 13.318 & 4764 & $ -35.508$ & $-0.59$ & $-0.76$ & $+0.40$ &   \\
369 & 276.580553 & $-33.632629$ & 14.007 & 13.438 & 13.271 & 4634 & $ -80.578$ & $-0.47$ & $-0.57$ & $+0.23$ &   \\
370 & 276.711834 & $-33.703365$ & 14.013 & 13.394 & 13.306 & 4714 & $ +48.722$ & $-0.08$ & $-0.23$ & $+0.12$ &   \\
371 & 276.743577 & $-33.898552$ & 14.014 & 13.444 & 13.263 & 4581 & $ -93.685$ & $+0.07$ & $-0.14$ & $+0.03$ & * \\
373 & 276.441719 & $-33.814545$ & 14.024 & 13.424 & 13.276 & 4601 & $  +2.274$ & $+0.28$ & $-0.05$ & $-0.04$ &   \\
374 & 276.570897 & $-33.627617$ & 14.025 & 13.431 & 13.228 & 4468 & $ -16.193$ & $+0.11$ & $-0.24$ & $+0.14$ &   \\
376 & 276.747908 & $-33.773846$ & 14.027 & 13.360 & 13.322 & 4718 & $ -31.917$ & $-0.26$ & $-0.34$ & $+0.18$ &   \\
377 & 276.641959 & $-33.642696$ & 14.027 & 13.439 & 13.293 & 4638 & $  +9.039$ & $+0.14$ & $-0.00$ & $+0.14$ & * \\
378 & 276.686341 & $-33.926971$ & 14.032 & 13.484 & 13.341 & 4758 & $+119.348$ & $-0.73$ & $-0.87$ & $+0.25$ &   \\
379 & 276.867112 & $-33.799599$ & 14.035 & 13.406 & 13.344 & 4755 & $ +36.952$ & $-1.07$ & $-1.14$ & $+0.23$ &   \\
380 & 276.797940 & $-33.867813$ & 14.041 & 13.329 & 13.276 & 4542 & $ +10.755$ & $-0.80$ & $-0.87$ & $+0.24$ &   \\
381 & 276.615896 & $-33.679054$ & 14.044 & 13.455 & 13.337 & 4718 & $ -41.469$ & $+0.57$ & $+0.33$ & $+0.20$ &   \\
382 & 276.642599 & $-33.707905$ & 14.051 & 13.414 & 13.319 & 4642 & $ +29.732$ & $-0.47$ & $-0.59$ & $+0.25$ & * \\
383 & 276.552809 & $-33.793800$ & 14.052 & 13.441 & 13.273 & 4514 & $ +58.636$ & $-0.30$ & $-0.37$ & $+0.14$ &   \\
384 & 276.767664 & $-33.790257$ & 14.054 & 13.476 & 13.288 & 4542 & $ -40.420$ & $+0.44$ & $+0.19$ & $+0.15$ &   \\
385 & 276.519644 & $-33.920509$ & 14.060 & 13.538 & 13.278 & 4504 & $ -29.684$ & $+0.12$ & $-0.09$ & $+0.16$ & * \\
386 & 276.778487 & $-33.904800$ & 14.063 & 13.482 & 13.386 & 4799 & $  +7.124$ & $-0.91$ & $-1.01$ & $+0.27$ &   \\
387 & 276.738181 & $-33.939156$ & 14.068 & 13.462 & 13.390 & 4797 & $ +93.169$ & $-0.47$ & $-0.57$ & $+0.20$ & * \\
388 & 276.529173 & $-33.642719$ & 14.071 & 13.505 & 13.331 & 4624 & $-107.163$ &         &         &         &   \\
389 & 276.805911 & $-33.751480$ & 14.078 & 13.399 & 13.380 & 4737 & $+179.763$ & $-0.69$ & $-0.79$ & $+0.31$ & * \\
390 & 276.654480 & $-33.594334$ & 14.081 & 13.422 & 13.341 & 4621 & $  -8.853$ & $-0.70$ & $-1.03$ & $+0.27$ &   \\
392 & 276.628139 & $-33.834789$ & 14.097 & 13.540 & 13.394 & 4726 & $  +7.430$ & $-0.44$ & $-0.60$ & $+0.24$ & * \\
393 & 276.710695 & $-33.846008$ & 14.100 & 13.400 & 13.312 & 4484 & $ +84.567$ & $-0.73$ & $-0.93$ & $+0.25$ &   \\
394 & 276.689541 & $-33.778378$ & 14.107 & 13.531 & 13.367 & 4616 & $ +28.170$ & $-0.26$ & $-0.54$ & $-0.01$ &   \\
395 & 276.514699 & $-33.757675$ & 14.110 & 13.464 & 13.352 & 4572 & $+143.624$ & $-1.63$ &         &         &   \\
396 & 276.694978 & $-33.943527$ & 14.117 & 13.546 & 13.403 & 4689 & $ +90.426$ & $-0.85$ & $-0.96$ & $+0.22$ &   \\
397 & 276.828208 & $-33.699005$ & 14.118 & 13.473 & 13.418 & 4731 & $ +31.582$ & $-0.75$ & $-0.83$ & $+0.21$ &   \\
398 & 276.515313 & $-33.784389$ & 14.119 & 13.550 & 13.332 & 4493 & $ -78.551$ & $+0.20$ & $+0.02$ & $-0.10$ &   \\
399 & 276.634575 & $-33.901833$ & 14.126 & 13.579 & 13.433 & 4755 & $ -52.671$ & $-0.38$ & $-0.50$ & $+0.26$ &   \\
400 & 276.829987 & $-33.806454$ & 14.127 & 13.557 & 13.385 & 4605 & $ -47.000$ & $-0.69$ & $-0.78$ & $+0.24$ &   \\
401 & 276.744083 & $-33.730270$ & 14.129 & 13.516 & 13.358 & 4530 & $ -64.671$ & $+0.45$ & $+0.21$ & $+0.18$ &   \\
402 & 276.879971 & $-33.675831$ & 14.138 & 13.436 & 13.357 & 4500 & $ +13.679$ & $-0.65$ & $-0.84$ & $+0.31$ &   \\
403 & 276.873784 & $-33.655762$ & 14.143 & 13.553 & 13.461 & 4785 & $+181.139$ & $-1.00$ & $-1.13$ & $+0.25$ &   \\
404 & 276.750878 & $-33.678814$ & 14.145 & 13.559 & 13.397 & 4593 & $ -16.703$ & $+0.50$ & $+0.32$ & $+0.20$ &   \\
406 & 276.773448 & $-33.816742$ & 14.159 & 13.602 & 13.453 & 4713 & $ +89.280$ & $-0.89$ & $-1.03$ & $+0.34$ &   \\
407 & 276.538950 & $-33.857552$ & 14.167 & 13.626 & 13.426 & 4617 & $ -59.382$ & $-0.94$ & $-1.02$ & $+0.26$ &   \\
408 & 276.702749 & $-33.817944$ & 14.168 & 13.557 & 13.489 & 4797 & $ -91.236$ & $-0.23$ & $-0.31$ & $+0.16$ &   \\
409 & 276.578957 & $-33.705822$ & 14.171 & 13.622 & 13.446 & 4665 & $  +0.475$ & $+0.07$ & $-0.04$ & $+0.14$ &   \\
410 & 276.753490 & $-33.738415$ & 14.173 & 13.552 & 13.448 & 4659 & $ +24.100$ & $+0.07$ & $-0.28$ & $+0.23$ &   \\
411 & 276.836391 & $-33.686768$ & 14.174 & 13.593 & 13.453 & 4669 & $ +21.788$ & $-0.30$ & $-0.40$ & $+0.15$ &   \\
412 & 276.501238 & $-33.637436$ & 14.174 & 13.595 & 13.412 & 4563 & $-161.418$ & $-1.13$ & $-1.27$ & $+0.41$ &   \\
413 & 276.662487 & $-33.626492$ & 14.179 & 13.651 & 13.492 & 4778 & $ +50.701$ & $-0.16$ & $-0.33$ & $+0.38$ &   \\
414 & 276.628673 & $-33.566936$ & 14.185 & 13.516 & 13.421 & 4555 & $  -5.715$ & $-0.88$ & $-0.99$ & $+0.23$ &   \\
415 & 276.534514 & $-33.915627$ & 14.195 & 13.583 & 13.469 & 4660 & $ +58.092$ & $-0.59$ & $-0.68$ & $+0.16$ &   \\
416 & 276.673953 & $-33.712120$ & 14.197 & 13.610 & 13.502 & 4751 & $-111.533$ & $-0.37$ & $-0.48$ & $+0.19$ &   \\
417 & 276.561145 & $-33.898586$ & 14.197 & 13.540 & 13.448 & 4592 & $ +47.080$ & $+0.41$ & $+0.17$ & $+0.18$ &   \\
419 & 276.578991 & $-33.573586$ & 14.206 & 13.587 & 13.472 & 4641 & $ -41.636$ & $-0.03$ & $-0.32$ & $-0.09$ &   \\
420 & 276.687936 & $-33.576183$ & 14.212 & 13.551 & 13.441 & 4535 & $ -69.762$ & $+0.25$ & $+0.03$ & $+0.16$ & * \\
422 & 276.852338 & $-33.874542$ & 14.215 & 13.578 & 13.489 & 4650 & $ -20.986$ & $-0.46$ & $-0.58$ & $+0.26$ &   \\
423 & 276.852706 & $-33.851723$ & 14.232 & 13.667 & 13.550 & 4782 & $ +45.436$ & $-0.50$ & $-0.58$ & $+0.13$ &   \\
424 & 276.886130 & $-33.787621$ & 14.234 & 13.670 & 13.553 & 4785 & $+101.421$ & $-1.66$ &         &         &   \\
425 & 276.661265 & $-33.711018$ & 14.238 & 13.657 & 13.487 & 4587 & $ +23.665$ & $-0.55$ & $-0.67$ & $+0.16$ &   \\
426 & 276.616342 & $-33.801567$ & 14.238 & 13.734 & 13.527 & 4704 & $-107.099$ & $-0.74$ & $-0.84$ & $+0.29$ &   \\
427 & 276.858769 & $-33.890724$ & 14.238 & 13.579 & 13.486 & 4575 & $-152.559$ & $-0.36$ & $-0.54$ & $+0.21$ &   \\
428 & 276.683750 & $-33.836472$ & 14.242 & 13.685 & 13.528 & 4692 & $ -70.446$ & $-0.51$ & $-0.62$ & $+0.19$ &   \\
429 & 276.794492 & $-33.592255$ & 14.245 & 13.644 & 13.565 & 4796 & $  +3.931$ & $-1.73$ &         &         &   \\
430 & 276.767692 & $-33.907009$ & 14.247 & 13.673 & 13.511 & 4622 & $-154.454$ & $-0.58$ & $-0.67$ & $+0.21$ &   \\
431 & 276.561013 & $-33.621979$ & 14.248 & 13.682 & 13.531 & 4692 & $ +82.186$ & $-0.34$ & $-0.60$ & $+0.12$ & * \\
433 & 276.801658 & $-33.837830$ & 14.256 & 13.632 & 13.536 & 4670 & $ -60.959$ & $-0.57$ & $-0.62$ & $+0.21$ & * \\
434 & 276.665024 & $-33.724117$ & 14.258 & 13.665 & 13.550 & 4713 & $ -22.264$ & $-0.23$ & $-0.44$ & $+0.47$ & * \\
435 & 276.768679 & $-33.857334$ & 14.259 & 13.664 & 13.559 & 4730 & $ +99.095$ & $-0.77$ & $-0.93$ & $+0.19$ &   \\
436 & 276.493553 & $-33.646214$ & 14.263 & 13.713 & 13.588 & 4822 & $ -50.196$ &         &         &         & * \\
437 & 276.827726 & $-33.647717$ & 14.268 & 13.643 & 13.542 & 4655 & $ -52.388$ & $-0.68$ & $-0.73$ & $+0.22$ &   \\
438 & 276.594804 & $-33.845116$ & 14.271 & 13.742 & 13.515 & 4573 & $  +5.719$ & $+0.40$ & $+0.20$ & $+0.08$ &   \\
440 & 276.631808 & $-33.564571$ & 14.278 & 13.711 & 13.603 & 4819 & $ +14.329$ & $-0.84$ & $-0.96$ & $+0.31$ &   \\
441 & 276.700830 & $-33.637337$ & 14.281 & 13.799 & 13.534 & 4598 & $ -18.383$ & $-0.91$ & $-1.12$ & $+0.31$ &   \\
443 & 276.727795 & $-33.730831$ & 14.297 & 13.769 & 13.573 & 4663 & $ -32.104$ & $-0.85$ & $-0.95$ & $+0.26$ &   \\
445 & 276.472846 & $-33.808666$ & 14.310 & 13.670 & 13.574 & 4634 & $ +73.543$ & $-0.12$ & $-0.33$ & $+0.33$ &   \\
446 & 276.438307 & $-33.774590$ & 14.311 & 13.721 & 13.542 & 4544 & $  -8.678$ & $+0.65$ & $+0.36$ &         &   \\
447 & 276.777476 & $-33.775871$ & 14.316 & 13.739 & 13.634 & 4786 & $ -81.331$ & $-0.98$ & $-1.02$ & $+0.24$ &   \\
448 & 276.706230 & $-33.843826$ & 14.316 & 13.735 & 13.588 & 4649 & $ -20.233$ & $+0.14$ & $-0.03$ & $+0.10$ &   \\
449 & 276.798371 & $-33.735714$ & 14.320 & 13.855 & 13.631 & 4765 & $  +5.772$ & $-0.74$ & $-0.79$ & $+0.20$ &   \\
451 & 276.887413 & $-33.852146$ & 14.324 & 13.738 & 13.636 & 4762 & $-184.313$ & $-0.85$ & $-1.06$ & $+0.38$ &   \\
452 & 276.661034 & $-33.891323$ & 14.334 & 13.821 & 13.654 & 4794 & $ -76.626$ & $-0.79$ & $-0.87$ & $+0.26$ &   \\
453 & 276.690616 & $-33.808006$ & 14.336 & 13.754 & 13.596 & 4615 & $ -48.148$ & $-0.30$ & $-0.43$ & $+0.20$ &   \\
454 & 276.698507 & $-33.939789$ & 14.343 & 13.800 & 13.657 & 4773 & $+115.364$ & $-0.98$ & $-1.07$ & $+0.24$ &   \\
455 & 276.845583 & $-33.661392$ & 14.344 & 13.746 & 13.667 & 4802 & $ +65.842$ & $-0.46$ & $-0.47$ & $+0.22$ &   \\
456 & 276.593340 & $-33.672127$ & 14.349 & 13.759 & 13.642 & 4719 & $ -23.056$ & $-0.32$ & $-0.47$ & $+0.31$ &   \\
457 & 276.833423 & $-33.746063$ & 14.353 & 13.767 & 13.581 & 4524 & $ -23.375$ & $-0.22$ & $-0.36$ & $+0.15$ &   \\
459 & 276.615328 & $-33.576347$ & 14.368 & 13.740 & 13.606 & 4561 & $ +49.302$ & $-0.79$ & $-0.92$ & $+0.30$ &   \\
460 & 276.722855 & $-33.943268$ & 14.391 & 13.838 & 13.708 & 4781 & $ -73.249$ & $-0.95$ & $-0.99$ & $+0.23$ &   \\
461 & 276.834942 & $-33.635342$ & 14.396 & 13.863 & 13.666 & 4643 & $ -19.355$ & $-1.33$ & $-1.41$ & $+0.32$ &   \\
462 & 276.789630 & $-33.841022$ & 14.400 & 13.855 & 13.727 & 4813 & $  +4.985$ & $-1.51$ &         &         &   \\
463 & 276.467867 & $-33.749195$ & 14.400 & 13.783 & 13.650 & 4595 & $-172.922$ & $-0.57$ & $-0.67$ & $+0.31$ &   \\
464 & 276.657623 & $-33.651539$ & 14.406 & 13.786 & 13.632 & 4526 & $ -67.065$ &         &         &         &   \\
465 & 276.631871 & $-33.704411$ & 14.417 & 13.879 & 13.739 & 4806 & $-121.778$ & $-0.75$ & $-0.92$ & $+0.33$ &   \\
466 & 276.861515 & $-33.729511$ & 14.417 & 13.903 & 13.742 & 4806 & $ -59.983$ & $-1.71$ &         &         &   \\
467 & 276.742712 & $-33.819698$ & 14.422 & 13.889 & 13.745 & 4802 & $  -9.504$ & $+0.04$ & $-0.15$ & $+0.20$ &   \\
468 & 276.511268 & $-33.792122$ & 14.422 & 13.791 & 13.707 & 4696 & $ +56.092$ & $-0.91$ & $-1.00$ & $+0.28$ &   \\
469 & 276.581345 & $-33.660347$ & 14.425 & 13.834 & 13.711 & 4699 & $ -44.438$ &         &         &         &   \\
471 & 276.439088 & $-33.761520$ & 14.447 & 13.839 & 13.738 & 4717 & $ +35.018$ &         &         &         &   \\
472 & 276.823834 & $-33.629124$ & 14.464 & 13.925 & 13.731 & 4635 & $+131.626$ & $-0.46$ & $-0.56$ & $+0.09$ &   \\
473 & 276.670917 & $-33.612740$ & 14.466 & 13.942 & 13.796 & 4833 & $ +24.364$ & $-0.49$ & $-0.64$ & $+0.32$ & * \\
474 & 276.547004 & $-33.669544$ & 14.475 & 13.910 & 13.772 & 4733 & $ -17.630$ & $-0.28$ & $-0.39$ & $+0.18$ &   \\
476 & 276.661846 & $-33.655987$ & 14.491 & 13.900 & 13.829 & 4859 & $ -29.627$ & $-0.29$ & $-0.41$ & $+0.24$ &   \\
477 & 276.853016 & $-33.748569$ & 14.496 & 13.867 & 13.809 & 4768 & $ -75.148$ & $-0.89$ & $-0.95$ & $+0.31$ &   \\
478 & 276.533779 & $-33.846905$ & 14.501 & 13.982 & 13.812 & 4772 & $ -38.002$ & $-0.16$ & $-0.35$ & $+0.42$ &   \\
481 & 276.777363 & $-33.671219$ & 14.502 & 13.896 & 13.736 & 4543 & $ -11.567$ & $+0.39$ & $+0.18$ & $+0.09$ & * \\
482 & 276.573957 & $-33.858578$ & 14.505 & 14.030 & 13.800 & 4722 & $-154.032$ & $-0.57$ & $-0.72$ & $+0.27$ &   \\
483 & 276.603613 & $-33.823669$ & 14.506 & 13.932 & 13.766 & 4618 & $ +31.359$ & $-0.45$ & $-0.65$ & $+0.23$ &   \\
484 & 276.594676 & $-33.571365$ & 14.507 & 13.849 & 13.829 & 4810 & $  -6.386$ &         &         &         & * \\
485 & 276.530138 & $-33.766701$ & 14.511 & 13.876 & 13.792 & 4684 & $+129.335$ & $-0.47$ & $-0.58$ & $+0.30$ &   \\
486 & 276.865712 & $-33.679306$ & 14.515 & 13.996 & 13.838 & 4801 & $ -56.181$ & $-0.52$ & $-0.62$ & $+0.16$ &   \\
487 & 276.849420 & $-33.771736$ & 14.519 & 13.847 & 13.786 & 4631 & $-141.982$ & $-0.55$ & $-0.72$ & $+0.27$ &   \\
488 & 276.598639 & $-33.744186$ & 14.521 & 13.967 & 13.832 & 4771 & $  -3.129$ & $-0.04$ & $-0.17$ & $+0.16$ &   \\
489 & 276.875366 & $-33.742172$ & 14.521 & 13.810 & 13.769 & 4577 & $ +73.034$ &         &         &         &   \\
490 & 276.712242 & $-33.794239$ & 14.531 & 13.928 & 13.856 & 4810 & $  -2.514$ & $-0.40$ & $-0.56$ & $+0.23$ &   \\
491 & 276.751951 & $-33.710484$ & 14.534 & 13.883 & 13.786 & 4593 & $ -47.811$ & $+0.59$ & $+0.31$ & $+0.27$ &   \\
493 & 276.631945 & $-33.662464$ & 14.538 & 13.956 & 13.862 & 4813 & $ +54.638$ & $-0.92$ & $-1.01$ & $+0.29$ & * \\
494 & 276.572343 & $-33.570671$ & 14.542 & 13.939 & 13.812 & 4653 & $ -28.906$ & $-0.15$ & $-0.28$ & $+0.12$ &   \\
495 & 276.548845 & $-33.833427$ & 14.549 & 13.924 & 13.793 & 4575 & $ +34.963$ & $-0.02$ & $-0.21$ & $+0.12$ &   \\
496 & 276.765063 & $-33.868202$ & 14.555 & 13.990 & 13.859 & 4742 & $ +77.859$ & $+0.03$ & $-0.18$ & $+0.16$ &   \\
497 & 276.830924 & $-33.896267$ & 14.555 & 13.988 & 13.816 & 4612 & $ -30.904$ & $-0.44$ & $-0.52$ & $+0.20$ &   \\
498 & 276.771491 & $-33.669971$ & 14.559 & 13.951 & 13.817 & 4610 & $ -23.933$ & $+0.31$ & $+0.13$ & $+0.21$ &   \\
499 & 276.633311 & $-33.762764$ & 14.567 & 13.899 & 13.838 & 4650 & $ +31.087$ & $-0.96$ & $-1.14$ & $+0.51$ &   \\
500 & 276.602649 & $-33.731819$ & 14.567 & 14.041 & 13.883 & 4787 & $  -1.571$ & $-0.24$ & $-0.35$ & $+0.22$ &   \\
501 & 276.858813 & $-33.772926$ & 14.568 & 13.883 & 13.878 & 4759 & $+174.776$ & $-1.23$ & $-1.31$ & $+0.32$ &   \\
503 & 276.436893 & $-33.808731$ & 14.580 & 13.968 & 13.852 & 4659 & $ -36.423$ & $-0.48$ & $-0.78$ & $+0.36$ &   \\
504 & 276.481768 & $-33.625542$ & 14.581 & 13.971 & 13.872 & 4718 & $ -67.245$ &         &         &         &   \\
505 & 276.537031 & $-33.778503$ & 14.584 & 14.036 & 13.893 & 4767 & $  -9.908$ & $-0.23$ & $-0.49$ & $+0.47$ &   \\
506 & 276.502711 & $-33.686172$ & 14.585 & 13.967 & 13.838 & 4604 & $ -10.462$ &         &         &         &   \\
507 & 276.736185 & $-33.691521$ & 14.591 & 13.961 & 13.833 & 4566 & $ -87.751$ & $-0.77$ & $-0.91$ & $+0.38$ &   \\
508 & 276.761504 & $-33.739830$ & 14.594 & 13.971 & 13.862 & 4638 & $-231.533$ & $-1.07$ & $-1.26$ & $+0.43$ &   \\
509 & 276.856836 & $-33.670521$ & 14.596 & 13.998 & 13.898 & 4737 & $-150.306$ & $-0.56$ & $-0.76$ & $+0.20$ &   \\
511 & 276.586662 & $-33.710934$ & 14.606 & 13.974 & 13.869 & 4629 & $+106.628$ & $+0.43$ & $+0.24$ & $+0.21$ &   \\
512 & 276.620755 & $-33.646839$ & 14.620 & 14.124 & 13.932 & 4776 & $ -19.865$ &         &         &         &   \\
513 & 276.868502 & $-33.816536$ & 14.620 & 14.136 & 13.959 & 4851 & $ +39.173$ & $-0.23$ & $-0.51$ & $+0.29$ &   \\
514 & 276.727030 & $-33.799290$ & 14.624 & 14.014 & 13.922 & 4727 & $ +42.665$ & $-0.70$ & $-0.86$ & $+0.30$ & * \\
\end{longtable}
\tablefoot{Meaning of the columns: (1): Target number in this programme; (2):
Right ascension from 2MASS; (3): Declination from 2MASS; (4), (5), and (6):
2MASS $J$-, $H$-, and $K$-magnitude; (8): Effective temperature; (7):
Heliocentric radial velocity; (9): General metallicity [M/H]; (10): Iron
abundance [Fe/H]; (11): $\alpha$-element abundance [$\alpha$/Fe]; (12):
Flag for foreground star candidates.}
}

\end{appendix}

\end{document}